%% file: paper.tex
\documentclass[10pt,journal,compsoc]{IEEEtran}

\ifCLASSOPTIONcompsoc
  \usepackage[nocompress]{cite}
\else
  \usepackage{cite}
\fi

\usepackage{amsmath}
\usepackage{amssymb}
\usepackage{amsfonts}
\usepackage[usenames]{color}

\usepackage{enumitem}
\usepackage{graphicx}
\usepackage{listings}
\usepackage{multirow}
\usepackage{subfig}

\usepackage{fancybox}

\usepackage{soul}  % for highlighting without wrapping problems of package 'color'
\usepackage{url}
\usepackage{algorithm}
\usepackage{algorithmicx}
\usepackage[noend]{algpseudocode}
\usepackage{algpseudocode}
\makeatletter
\renewcommand{\ALG@beginalgorithmic}{\small}
\makeatother
\makeatletter
\newcommand\fs@ruled@notop{\def\@fs@cfont{\bfseries}\let\@fs@capt\floatc@ruled
  \def\@fs@pre{\kern4pt\hrule height.5pt depth0pt \kern1pt}% <----removed
  \def\@fs@post{\kern1pt\hrule\relax}%
  \def\@fs@mid{\kern1pt\hrule\kern1pt}%
  \let\@fs@iftopcapt\iftrue}
\renewcommand\fst@algorithm{\fs@ruled@notop}
\makeatother

\usepackage{lipsum}
\setlist{nosep}

\renewcommand\[{\begin{dmath}}
\renewcommand\]{\end{dmath}}
\newlength\myindent
\newcommand{\diver}{\textsc{Diver}}
\newcommand{\distia}{\textsc{DistIA}}
\newcommand{\distea}{\textsc{D$^2$Abs}}
\newcommand{\tool}{\distea}
\newcommand{\tech}{\tool}

\newcommand{\EAS}{\textsc{EAS}}
\newcommand{\PI}{\textsc{PathImpact}}

\newcommand{\LTS}{\textsc{LTS}}

\def\denseitems{
  \itemsep1pt plus1pt minus1pt
  \parsep0pt plus0pt
  \parskip0pt\topsep0pt}

\usepackage{color}

\usepackage{mdframed}
\usepackage{breqn}

\usepackage{parcolumns}
\usepackage{pifont,calc}
\definecolor{verylightgray}{gray}{0.55}
\newlength\lsthorizontalpadding
\newcommand*\lstnumberstyle{\ttfamily\tiny}
\newlength\lstnumbersep
\newlength\lstnumberwidth
\usepackage{cite}
\usepackage[table]{xcolor}% http://ctan.org/pkg/xcolor
\usepackage{colortbl}
\usepackage{tcolorbox}

\usepackage{times}
\usepackage{textcomp}
\usepackage{balance}

\usepackage{comment}

\usepackage{hyperref}

\usepackage{tikz}
\newcommand*\circled[1]{\tikz[baseline=(char.base)]{
             \node[shape=circle,draw=red!0!green!32!blue!96!,fill=red!0!green!32!blue!96!,inner sep=.5pt] (char) {\textcolor{white}{#1}};}}

\usepackage{pifont}% http://ctan.org/pkg/pifont
\newcommand{\cmark}{\ding{51}}%
\newcommand{\xmark}{\ding{55}}%

\definecolor{Gray}{gray}{0.85}
\definecolor{LightCyan}{rgb}{0.88,1,1}

\newcolumntype{g}{>{\columncolor{Gray}}c}
\newcolumntype{w}{>{\columncolor{white}}c}

\newcommand{\haipengemail}{haipeng.cai@wsu.edu}

\newfont{\ttlsc}{phvb8t at 15pt}
\newfont{\autt}{pcrr8t at 11pt}

\definecolor{Gray}{gray}{0.85}
\definecolor{LightCyan}{rgb}{0.88,1,1}

\newcolumntype{g}{>{\columncolor{Gray}}c}
\newcolumntype{w}{>{\columncolor{white}}c}

\begin{document}
\pagenumbering{arabic}
  \pagestyle{plain}
\setlength{\parskip}{0pt}

% ============= Mandatory Copyright Information Begin ==============================
%
% --- Author Metadata here ---
%\conferenceinfo{ASE}{'16, September 3--7, 2016, Singapore}
%\CopyrightYear{2016} % Allows default copyright year (20XX) to be over-ridden - IF NEED BE.
%\crdata{0-12345-67-8/90/01}  % Allows default copyright data (0-89791-88-6/97/05) to be over-ridden - IF NEED BE.
% --- End of Author Metadata ---
%\title{DistIA: Dynamic Impact Analysis of Distributed Systems}
%

%\title{Dynamic Dependence Abstraction for Impact Analysis and Measurement of Distributed Programs}
\title{{\tech}: A Framework for Dynamic Dependence Abstraction of Distributed Programs}

\author{\IEEEauthorblockN{Haipeng Cai}
\IEEEauthorblockA{%School of Electrical Engineering and Computer Science\\
Washington State University, Pullman, USA\\
haipeng.cai@wsu.edu}
\and
\IEEEauthorblockN{Xiaoqin Fu}
\IEEEauthorblockA{%Department of Computer Science\\
Washington State University, Pullman, USA\\
xiaoqin.fu@wsu.edu}
}

\author{Haipeng~Cai~%~\IEEEmembership{Member,~IEEE,}
        and~Xiaoqin~Fu%,~\IEEEmembership{Life~Fellow,~IEEE}% <-this % stops a space
\IEEEcompsocitemizethanks{\IEEEcompsocthanksitem Haipeng Cai and Xiaoqin Fu are with the
School of Electrical Engineering and Computer Science, Washington State University, Pullman, WA. %USA.\protect\\
%Haipeng Cai is the corresponding author.
E-mail: haipeng.cai@wsu.edu, xiaoqin.fu@wsu.edu
%\IEEEcompsocthanksitem Barbara Ryder is with the Department of Computer Science, Virginia Tech, Blacksburg, VA. %\protect\\
%E-mail: ryder@cs.vt.edu
}% <-this % stops an unwanted space
\thanks{Manuscript received April 1, 2018; revised August 26, 2015.}}

\markboth{Journal of \LaTeX\ Class Files,~Vol.~14, No.~8, August~2015}%
{Cai \MakeLowercase{\textit{et al.}}: \papertitle}

 %===================== Mandatory Copyright Information End ==============================
%\numberofauthors{1}
%\author{
%\alignauthor Haipeng Cai, Douglas Thain, and Raul Santelices \\
%\affaddr{Department of Computer Science and Engineering, University of Notre Dame} \\
%\{\href{mailto:\haipengemail}{hcai},\href{mailto:\dougemail}{dthain},\href{mailto:\raulemail}{rsanteli}\}@nd.edu
%}

\maketitle
%\thispagestyle{plain}
%\begin{abstract}
%\end{abstract}
\input{abstract}

\IEEEpeerreviewmaketitle

%\category{D.2.7}{Software Engineering}{Distribution, Maintenance, and Enhancement}
%%\category{F.3.2}{Logic and Meaning of Programs}{Program Analysis}
%\terms{Reliability, Experimentation}
%\keywords{Dynamic impact analysis; distributed program; non-blocking} % NOT required for Proceedings
%
%\begin{CCSXML}
%<ccs2012>
%<concept>
%<concept_id>10011007.10011074.10011111.10011113</concept_id>
%<concept_desc>Software and its engineering~Software evolution</concept_desc>
%<concept_significance>500</concept_significance>
%</concept>
%%<concept>
%%<concept_id>10011007.10011074.10011111.10011696</concept_id>
%%<concept_desc>Software and its engineering~Maintaining software</concept_desc>
%%<concept_significance>500</concept_significance>
%%</concept>
%</ccs2012>
%\end{CCSXML}
%
%%\ccsdesc[500]{Software and its engineering~Maintaining software}
%\ccsdesc[500]{Software and its engineering~Software evolution}
%\printccsdesc
%\keywords{Impact analysis; distributed systems; dynamic partial ordering} % NOT required for Proceedings

\input{intro}
\input{example}
\input{motivation}

\input{overview}
\input{design}

\input{technique}

\input{analysisalgorithms}
\input{implementation}
\input{evaluation}

\input{discussion}

\input{related}

\input{concl}
\renewcommand{\baselinestretch}{0.96}

%\newpage
%{
%\baselineskip=12pt
%\setstretch{0.9}
%\emergencystretch=\maxdimen
%\raggedright
%\bibliographystyle{latex8}
%\balance
\bibliographystyle{IEEEtran}
\bibliography{paper}

\end{document}

%% file: abstract.tex
%!TEX root = paper.tex

\begin{abstract}
As modern software systems are increasingly developed for running in distributed environments,
it is crucial to provide fundamental techniques such as dependence analysis
for checking, diagnosing, and evolving those systems.
However, traditional dependence analysis is either inapplicable or of very limited utility for
distributed programs due to the decoupled components of these programs that run in concurrent processes at
physically separated machines.

Motivated by the need for dependence analysis of distributed software and the diverse cost-effectiveness needs of dependence-based applications, this paper
presents {\tech}, a framework of dynamic dependence abstraction for %multiple-process,
distributed programs.
By partial-ordering distributed method-execution events and inferring causality from
the ordered events, {\tech} abstracts method-level dependencies both within and
across process boundaries.
Further, by exploiting message-passing semantics across processes, and incorporating static dependencies and statement coverage within
individual components, we present three additional instantiations of {\tech} that trade efficiency
for better precision.
We present the design of the {\tech} framework %in the context of its application in impact analysis,
and evaluate the four instantiations of {\tech} on distributed systems of various architectures
and scales using our implementation for Java.
Our empirical results show that {\distea} is significantly more effective than
existing options while offering varied levels of cost-effectiveness tradeoffs.
As our framework essentially computes whole-system run-time dependencies, it
naturally empowers a range of other dependence-based applications. % other than impact analysis.
\end{abstract}

%% file: intro.tex
%!TEX root = paper.tex

\section{Introduction}\label{sec:intro}
%\vspace{-2pt}
%% big context: why target distributed software systems concerning reliability and security
In response to scientific and societal demands on the scalability of data storage and computation,
a rising number of modern software systems are %becoming increasingly
{\em distributed} by
design~\cite{Coulouris2011DSC} to leverage decentralized, high-performance %vastly scalable
computing infrastructure and resources. In fact, most critical software and services
today, such as financial systems, web search, airline services, and medical networks, are all distributed systems in nature~\cite{dean2009designs}.
The quality (e.g., reliability and security, among other quality factors) of these systems is thus of paramount importance.
Significant advances in these regards have been made in the areas of parallel and distributed computing, yet mainly from {\em coarse}, {\em system-level} perspectives such as those of architecture, networking,
and resource management.
In contrast, {\em code-level} quality via %characteristics enabled by
%more {\em fine-grained} and
deeper analyses of
programs in distributed systems has not been much studied, and there is a lack of
tool support for code quality assurance for distributed software.

%%% why dependency analysis
%
%Historically, dependence analysis has underlaid a wide range of code-based software engineering tasks and
%associated techniques~\cite{podgurski90sep,horwitz1992use}. In particular, a large body and variety of dependence-based
%approaches have been proposed over the past few decades to support development and maintenance in general ~\cite{gallagher1991using,loyall1993using,Li2013ASC,orso04apr,petrenko2009variable}, and fault diagnosis~\cite{baah10jul,bates93jan,faultlocsurvey2016} and security defense in particular~\cite{dalton2007raksha,attariyan2010automating,kemerlis2012libdft}.
%For instance, for maintaining and evolving a software system, it is essential to assess the influence of given program entries of interest on the rest of the program~\cite{bohner96jun,Li2013ASC} for change planning (deciding whether to realize changes at
%candidate change locations) and fulfillment (deciding where to realize the changes)~\cite{bohner96jun,rovegard2008empirical,tao2012software}.
To attain and sustain various quality factors of distributed systems, %, including reliability and security,
it is crucial to
model and reason about the complex %, dynamic
%relationships and
interactions among code entities via
{\em dependencies} in these systems.
Historically, dependence analysis has been a foundational methodology that underlies a wide range of code-based
software engineering tasks and associated techniques~\cite{podgurski90sep,horwitz1992use}.
Generally, a dependence analysis can be static or dynamic (either purely dynamic or mixed with some static analysis).

As it attempts to produce results that hold for all possible execution scenarios, {\em static} dependence analysis is
known to be imprecise in general.
For distributed systems, it has to be even more conservative, hence is subject to even greater imprecision,
due to the {\em implicit} interactions among the components of these systems that are decoupled by networking facilities~\cite{murphy1996lightweight}.
As a result, there are no {\em explicit} references or invocations among code entities across those components~\cite{popescu2012impact,Tragatschnig2014IAE},
on which existing dependence analysis approaches typically rely.
In consequence, existing approaches %to dependence analysis
are largely limited to centralized software which
%and inapplicable for distributed programs.
mostly runs in a single process (whether the program is single- or multi-threaded) hence has explicit interactions
(via invocations or references) among code entities.
There is also no trivial adaptation of existing static
dependence analyses to distributed programs~\cite{jayaram2011program,garcia2013identifying,beschastnikh2014inferring}.
%
%The primary reason is because they typically rely on the existence of
%{\em explicit} interactions (invocations or references) among code entities.
%These explicit interactions are commonly present in a centralized program, which typically runs in a single process, whether the program is single- or multi-threaded.
%Yet in a common distributed system, components are decoupled by networking facilities, thus
%there are no explicit references or invocations among those components~\cite{popescu2012impact,Tragatschnig2014IAE}.
%Due to the implicity of the inter-component interactions, there is also no trivial adaptation of existing
%dependence analysis to distributed programs~\cite{jayaram2011program,garcia2013identifying,beschastnikh2014inferring}.
%%A primary blockade lies in the lack of the explicit relationships between distributed program
%%components that are decoupled by networking facilities~\cite{popescu2012impact,Tragatschnig2014IAE}.

% why dynamic dependence analysis
In contrast, {\em dynamic} dependence analysis has two advantages.
First, it has the potential to overcome the {\em implicity challenge} as it may infer
dependence relationships among code entities exercised from the observed concrete executions.
The second merit lies in its greater precision than static approaches~\cite{cai2018hybrid}, because
it is only concerned about dependence relations
%among code entities (e.g., statements or methods) between a given location ({\em query}) and the set of influenced entities ({\em dependency set}) of the program,
with respect to specific program executions~\cite{cai2016diapro} instead of considering
all possible ones. For developers working with these concrete executions, dynamic dependence analysis can be
a powerful technique to assist them with development and maintenance tasks, as it narrows down
the search space of complex dependency relations to the focused operational profiles of a program.
%
%This specific context (the profiles of interest) enables dynamic dependence analysis to
%achieve much greater precision than static approaches~\cite{cai2018hybrid} while
%%Also, dynamic analysis may overcome
%overcoming the lack of explicit invocations/references in distributed software.
%These merits, however, come at the cost of generally much
%greater overheads.
%%A dynamic analysis could be rapid but would generate merely very coarse results, or it
Therefore, in this paper we focus on {\em dynamic} dependence analysis of distributed systems.

% problems with existing peer approaches
%Dynamic code analyses of distributed software systems~\cite{korel1992dynamic,kamkar1995dynamic,duesterwald1993distributed}
%have been explored, including recent efforts which focus on detailed analysis of program dependencies~\cite{mohapatra2006distributed,barpanda2011dynamic,pani2012slicing}.
%
Unfortunately, the majority of current approaches
to the dynamic analysis of distributed program dependencies (e.g.,~\cite{mohapatra2006distributed,barpanda2011dynamic,pani2012slicing})
were designed only for procedural programs~\cite{barpanda2011dynamic}.
For distributed object-oriented programs, \emph{backward} dynamic slicing algorithms have been developed,
yet it is still unclear whether they can work with real-world systems~\cite{mohapatra2006distributed,barpanda2011dynamic}.
Also, for many maintenance and evolution tasks (e.g., impact analysis), \emph{forward} dependence analysis would be
needed instead (e.g., given a query, entities that transitively depend on the query are considered potentially impacted).
Moreover, fine-grained (statement-level) analysis of slicing would be excessively heavyweight for those tasks, which compromises
scalability and impedes their practical adoption.

% additional challenges
%
%However, so much as static analysis is known to be subject to imprecision challenges, dynamic analysis commonly suffers
%from its mostly substantial overheads which bring efficiency and scalability challenges.
In addition, dynamic dependence analysis is subject to a common, fundamental {\em challenge of balancing the analysis
cost and effectiveness}, evidenced by relevant prior works on centralized programs~\cite{cai2016method,cai2016diapro,cai2018hybrid}.
To deal with its substantial analysis
overheads, dynamic analysis has been mostly studied in two
directions. In one direction, efficient analysis often provides fast answers to dependence queries~\cite{law03may,apiwattanapong05may}, yet
with coarse results that require tremendous efforts for post-analysis inspection~\cite{tao2012software,cai15jss,cai14diver}.
%(e.g., checking the resulting impact set during impact analysis).
In the other direction, highly precise analysis saves developers' inspection efforts by producing few false positives, %greater precision,
yet incurring analysis overheads that may not be practically affordable~\cite{acharya11may,zhang03may}.
Frameworks offering better cost-effectiveness tradeoffs along with more such tradeoff options also
have been studied recently~\cite{cai15diverplus,cai2018hybrid}. However, the frameworks were developed for
centralized, single-process software, without a straightforward path to extending them for distributed programs~\cite{cai2016distia}.
%Thus, in essence, the challenge boils down to difficult balancing between analysis cost and effectiveness, evidenced by relevant prior works on centralized programs~\cite{cai2016method,cai2016diapro,cai2018hybrid}.
%
%Given their typically
%large size, great complexity, and long executions, achieving good cost-effectiveness balances in dynamic dependence analysis of distributed programs is even more challenging. Well-known non-determinism along with the variety of, and uncertainties in, runtime environments of distributed systems further exacerbate the challenges.
Given their typically
large size, great complexity, and long executions, distributed systems represent a software domain where
achieving good cost-effectiveness balances in dynamic dependence analysis is even more challenging.
Aiming to address these challenges, we develop {\tech} (\underline{D}istributed program \underline{D}ependence \underline{ABS}traction),
a framework of dynamic dependence analysis for the most commonly deployed kind
of distributed systems---where components %have no explicit invocations or references to each other but
communicate via message passing via standard socket facilities and/or their encapsulations.
%
%\footnote{We define a {\em component} as the code that runs in a separate process.}
%
%---we distinguish components as such that each runs in a separate process. %}% and use both terms interchangeably.}%network I/Os. %from other parts of the system.
By exploiting %the partial order of, hence
the happens-before relation~\cite{lamport1978time} between method-execution events and the semantics of message passing among
distributed components,
the framework %predicts impacts of one method on others of a given system
computes abstract dynamic dependencies to method level for a given system and its executions
%methods of the system
both within and across its %components and
concurrent processes.
%Akin to the execute-after-sequences ({\EAS}) technique~\cite{apiwattanapong05may},
Our approach offers \emph{rapid} results that are safe~\cite{jackson2000software} relative to (i.e., guaranteed to hold for) the analyzed executions, % (as safe as {\EAS}),
while relying on neither well-defined inter-component interfaces nor message-type specifications
needed by peer approaches (e.g., those for distributed event-based systems---DEBS~\cite{jayaram2011program,popescu2012impact,garcia2013identifying}).

Further, utilizing intra-component {\em static} dependence abstraction~\cite{cai2015abstracting,cai2016method} that
incorporates threading-induced and exception-driven dependencies,
and whole-system {\em dynamic} statement coverage data,
{\tech} trades efficiency for precision by pruning false-positive dependencies.
As a whole, the framework unifies four dynamic dependence abstractions each
providing a distinct level of cost-effectiveness balance.
%By design, {\tech} works with programs %centralized programs
%running in single processes as well, either concurrent or sequentially.
{\tech} naturally empowers dynamic impact analysis of distributed programs that accommodates different
usage scenarios with varying demands for precision and budgets for analysis and inspection costs.
%
%Using this framework, we further explore dynamic measurement of distributed systems concerning the
%coupling between the distributed processes.
%conformance of their design to core principles
%
%As such, our framework offers a flexible set of alternative cost-effectiveness options that provide
%higher precision at reasonably greater costs.

%%%% ================ highlights on implementation, evaluation, and results =====================
We evaluate {\tech} on eight distributed Java software, including six
enterprise-scale systems, and demonstrate that it is able to work with large, complicated distributed systems
using blocking and/or non-blocking (e.g., selector-based~\cite{artho2013software}) communication.
%We also evaluated our technique on these subjects with respect of its effectiveness and performance.
%
In the absence of more advanced prior peer techniques,
%
%directly comparable to ours,
%
%we use a coverage-based approach~\cite{orso03sep}, which
%reports as impacted all methods covered in the utilized executions, as a safe baseline alternative, and
%measure the effectiveness of our three abstraction techniques against it. %in terms of impact size ratios.
%
we use our coarse dynamic dependence abstraction purely based on control flows~\cite{cai2016distea} as
the baseline, and
measure the effectiveness of the three more precise abstraction techniques against it in
the application context of dynamic impact analysis.
%in terms of impact size ratios.
We also gauge the costs of each technique, including those for different phases of our framework
(static analysis, run-time profiling, and post-processing for dependence querying), and compare
against the baseline.

Our results show that, without using static dependencies or dynamic coverage data, {\tech} %as a dynamic impact analysis
can reduce
%impact-inspection effort,
the size of baseline impact (dependence) sets to be inspected
by over 15\% on average.
%(based on coverage), %existing alternative options
%, with highly promising scalability: overall, {\distea}
%reports only x\% of impacts that would have to be all inspected with existing options, and the average analysis time
%of {\distea} is only a few seconds.
%The results also support the high efficiency {\distea} is highly efficient, finishing the entire analysis within one minute
%and answering a query in three seconds for an average case.
The average cost was 73 seconds to finish the one-time instrumentation and 61 milliseconds for answering a query, with run-time overhead of 8\%. Incorporating static dependence abstraction in our framework largely further reduced baseline impact (dependence) sets
by 41\% on average, at the cost of reasonable additional overheads for most of our subject systems.
Exploiting whole-system statement coverage additionally even further enhanced the precision of our analysis (54\% impact-set reduction of baseline results), which causes generally negligible increases in total analysis overheads compared to just incorporating static dependencies.
%
%
% motivation: introduce what we are gonna do
%Nevertheless, when making incremental code changes in distributed systems, developers need to
%understand the impacts of those changes, including the impacts \emph{across component boundaries}.
%Despite of the known imprecision, impact analysis based on execute-after relation (e.g.,EAS~\cite{apiwattanapong05may}) gives
%a safe impact-set approximation very efficiently~\cite{orso04may,cai14diver} thanks to its lightweight and conservative computation:
%for a query method $c$, every method that executes after $c$ executes is potentially affected by,
%or any changes in, the queried method $c$.
%However, although such analysis handles multithreaded programs, it does not apply to distributed multiprocess executions
%where method events in different components often occur with respect to different systems of physical clock~\cite{beschastnikh2014inferring,lou2010mining}. The main challenge comes from the unsynchronized timing of events at
%different hosts, which makes it difficult to infer the execute-after relation among methods across all system components
%(e.g., server and its clients).

The development of {\tech} started with our preliminary work {\distia}~\cite{cai2016distia}, where
we presented the first two levels of abstraction (i.e., based on method-level control flows with/without
message-passing semantics exploited to enhance effectiveness). Technically, this paper represents a substantial expansion
of that earlier work by developing two more levels of abstraction, using {\em method-level static} dependencies with/without
{\em run-time statement-level} coverage data. Experimentally, we expand the scale of empirical study by including two more real-world
subject distributed programs and examining an additional research question (on variable cost-effectiveness). We used the
basic abstraction in~\cite{cai2016distia} as the baseline (instead of using all methods covered in an execution as baseline impacts) in assessing the cost and effectiveness of these newly added levels of abstraction. We also conducted extensive statistical analyses to
understand how various forms of program information used in our framework contributes to its effectiveness.

The main contributions of this work include:
\begin{itemize}
\item A framework of dynamic dependence analysis for distributed programs that unifies four dependence abstractions, each providing a distinct level of cost-effectiveness tradeoff.
%\item Two clients of the framework, dynamic impact analysis and inter-process coupling measurement, that demonstrate its
%    applications in the analysis of distributed software.
\item An extensive evaluation of our framework for dynamic impact analysis against subject programs of varied sizes and domains that shows its promising effectiveness and scalability, as well as flexible cost-effectiveness tradeoffs.
\item An empirical investigation of how the use of various forms of data in the framework affects its cost-effectiveness.
\item An {\em open-source} implementation of the framework that works with diverse, large enterprise distributed systems
    with either or both of blocking and non-blocking communication.
\end{itemize}

\vspace{4pt}
\noindent
\textbf{Paper organization.}
The rest of this paper is organized as follows. We motivate our work while introducing a running example for illustration purposes in Section~\ref{sec:motive}. To present our technical approach, we start with an overview in Section~\ref{sec:overview}, followed by details on the design, instantiations, and implementation of the {\tech} framework presented in Sections~\ref{sec:design},~\ref{sec:tech}, and~\ref{sec:impl}, respectively.
We evaluate {\tech} in Section~\ref{sec:eval}, where we describe our experimental setup and methodology before discussing empirical results, and then
discuss additional issues with our evaluation and the use of our technical approach in Section~\ref{sec:discussion}.
Prior work related to {\tech} is discussed in Section~\ref{sec:related}, before we give concluding remarks in Section~\ref{sec:conclusion}.

%% file: example.tex
%!TEX root = paper.tex
%  \vspace{-6pt}
\begin{figure}[t]
 \centering
%\begin{tabular}{cc}
 %\subfloat[Component C: the client]{
   %\begin{minipage}[t]{.45\textwidth}
   %\begin{lstlisting}[firstnumber=1]
 \lstset{basicstyle=\small}
  \begin{minipage}[t]{.46\textwidth}
%  \begin{mdframed}
  \vspace{-0pt}
\begin{lstlisting}
public class S {
  Socket ssock = null;
  public S(int port) { ssock = new Socket(port); ssock.accept(); }
  char getMax(String s) {...}
  void serve() { String s = ssock.readLine();
      char r = getMax(s); ssock.writeChar(r); }
  public static int main(String[] a) {
      S s = new S(33); s.serve(); return 0; }}
public class C {
  Socket csock = null;
  public C(String host,int port) { csock = new Socket(host,port); }
  void shuffle(String s) {...}
  char compute(String s) { shuffle(s); csock.writeLine(s);
      return csock.readChar(); }
  public static int main(String[] a) { C c = new C('localhost',33);
      System.out.println( c.compute(a[0]) ); return 0; }}
\end{lstlisting}
    \vspace{-0pt}
% \end{mdframed}
\end{minipage}
    %return 0;
   %System.out.println('Done.');
   %\label{fig:csserver}
  %}
 %\end{tabular}
 \vspace{-8pt}
 \caption{An example distributed program $E$ consisting of two components: S (server component) and C (client component).}
 \label{fig:csexample}
  \vspace{-8pt}
\end{figure}
%// MySocket is a common library class
    %\end{lstlisting}
    %\end{minipage}
   %\label{fig:csclient} \hfill
   %}
 %\subfloat[Component S: the server]{
  %\begin{minipage}[t]{.5\textwidth}
  %  \begin{lstlisting}

%% file: motivation.tex
%!TEX root = paper.tex
\vspace{-4pt}
\section{Motivating and Working Example}\label{sec:motive}
%% first motivation: existing analyses are not applicable; and why that is
%In this section, we first present a usage scenario of dynamic impact analysis that motivates
%this work. Then, we give necessary background on techniques
%%execute-after (EA) as an approach to dynamic impact analysis
%underlying the design of {\distea}.
%This section presents a motivating usage scenario of dynamic impact analysis, followed by necessary background underlying the design of {\distea}.

%\subsection{Problem and Motivation}
%\subsection{Motivation}

%The evolution of modern software is
%%The new paradigm of software engineering is based on software evolution
%driven by incremental changes\cite{rajlich2006changing}.
In this section, we illustrate a motivating use scenario of dynamic dependence abstraction
with a simple distributed program as example, for its application in change impact analysis.

When maintaining and evolving a distributed program which consists of multiple components,
the developer needs to understand potential change effects not only in the component where the change
is proposed, but also in all other components.
%Particularly in the context of dynamic impact analysis for distributed programs, since the components
%usually execute in separate processes either on a single machine or over different hosts, impacts propagated
%within and across processes and hosts need both be inspected for safely planning changes in the system
%as a whole. Dynamic impact analysis has also found proposed applications in a number of other tasks ranging
%from testing and debugging to refactoring and re-engineering.
%
%An increasing number of today's software systems runs not only concurrently but in distributed manner.
%To achieve high computing performance and scalability, distributed systems are usually designed in
%such a way that its components are loosely coupled or entirely decoupled, encouraging implicit
%invocations among components through network-based message passing.
%These design strategies, however, %also bring new challenges to traditional program analysis techniques,
%render the usefulness of existing dynamic impact analysis to a very limited extent.
%
%To achieve better flexibility and scalability,
%
By design, the components that constitute a distributed system are %usually loosely coupled or entirely
decoupled as a result of
{\em implicit} invocations and/or references among them, realized via networking-based %socket-based
message passing. %, which,
This design paradigm, however, greatly reduces
the utility of existing dependence analysis and its client analysis techniques (e.g., impact analysis). %techniques.
%to a very limited extent.
%As for other types of software systems, understanding
%To achieve high flexibility and scalability, the design of distributed systems often encourages
%implicit invocations among components
%
Consider the program $E$ of Figure~\ref{fig:csexample}, which consists of a server and a client component, implemented in classes {\tt S} and {\tt C}, respectively.
%The program consists of two components, the
%client (class C) and server (class S).
%The class {\tt MySocket} is an external %a user-extended common-utility
%library class wrapping networking utilities % relevant facilities
%in the legacy Java SDK (e.g., Java Socket~\cite{javasocket} and/or Java NIO APIs~\cite{javanio}).
%%In an example scenario,
The client retrieves the largest character in a string by delegating the task to the server (line 13), which finishes the task and sends the result back to the client (line 6).

Suppose the developer proposes to change {\tt S::getMax} as part of an upgrade plan for the server, and thus needs to determine which other parts of the program may also have to be changed. Having an available set $I$ of inputs, the developer wants to perform a dynamic impact analysis to get a safe estimation of the potential impacts of the candidate changes with respect to the program executions against $I$. Note that static approaches would be largely disabled by the implicit communication between these two components (via message passing, which is in this case realized through a network socket).
%check which methods in $E$ would be influenced

To accomplish this task, method-level dynamic impact analyses of varying cost-effectiveness tradeoffs (e.g., ~\cite{apiwattanapong05may,cai14diver,cai15diverplus}) seem able to offer the developer with many options.
%Unfortunately, it soon turns out that those existing options have merely quite limited utility in this context.
However, since there are no explicit dependencies between {\tt S} and {\tt C}, existing approaches would
predict impacts within the \emph{local} component (i.e., where the changes are located; $S$ in this case) only.
In consequence, the developer would have to ignore impacts in \emph{remote} components ($C$ in this case),
or make a worst-case assumption that all methods in remote components will be impacted.
%mainly due to the lack of explicit information about interactions and dependencies among the decoupled components~\cite{jayaram2011program}.
%
%At first glance, it seems that the developer has many options (e.g., ~\cite{apiwattanapong05may,cai14diver,cai15diverplus})
%to accomplish this task. Unfortunately, it soon turns out that those existing options have
%merely quite limited utility in this
%context. Since there is no explicit dependencies between {\tt S} and {\tt C}, existing dynamic impact analysis would
%predict impacts within the \emph{local} component (i.e., where the changes are located; $S$ in this case) only.
%In consequence, the developer would have to ignore impacts in \emph{remote} components ($C$ in this case),
%or make a worst-case assumption that all methods in remote components are to be impacted.
%%mainly due to the lack of explicit information about interactions and dependencies among the decoupled components~\cite{jayaram2011program}.
%
%%%% === what has been done and why not enough: impact analysis for specialized distributed systems/languages =====
%
%Developing a cost-effective dynamic impact analysis for real-world distributed systems, however, faces
%multiple challenges.
%%Impact analysis is widely recognized as a key step during software development~\cite{rovegard2008empirical,tao2012software}.
%%While a rich body of impact analyses
%%exist for traditional centralized programs, very few is applicable to distributed systems~\cite{jayaram2011program,garcia2013identifying}.
%%

%Again,
One major difficulty here is the lack of explicit invocations or references among the two decoupled components~\cite{goswami2000dynamic,jayaram2011program,Tragatschnig2014IAE},
% as a result of inter-component decoupling
whereas traditional, %slicing-based
dependence-based approaches often rely on such explicit information to compute dependencies for impact prediction.
%
%
%
%among program entities.
%Lately, a few dependence-analysis for particular types like event-based distributed systems (DEBS)~\cite{muhl2006distributed}
%and those built on specialized languages~\cite{eugster2009eventjava} have received some attentions very recently~\cite{popescu2012impact,lou2010mining,garcia2013identifying,Tragatschnig2014IAE}, impact analysis for
%general distributed multiprocess systems that adopt neither well-defined inter-component interfaces nor message-typing specifications has not been addressed as yet.
%On the other hand, the growing demands for, and increasing deployments of, distributed systems today entail effective impact analyses to support their development and evolution.
Lately, various analyses that are not based on code dependencies %other than slicing
have also been proposed~\cite{popescu2012impact,garcia2013identifying,Tragatschnig2014IAE}.
While efficient for \emph{static} impact analysis, these approaches are limited to %special type of systems %application domains
systems of special types
such as distributed \emph{event-based} systems (DEBS)~\cite{muhl2006distributed}, or rely on specialized
language extensions like EventJava~\cite{eugster2009eventjava}.
Other approaches are potentially applicable in a wider scope yet depend on information that is not always available, such as execution logs of particular patterns~\cite{lou2010mining}, or suffer from overly-coarse
granularity (e.g., class-level)~\cite{popescu2012impact,lou2010mining,garcia2013identifying} and/or unsoundness~\cite{murphy1996lightweight}, in addition to imprecision, of analysis results.

%An additional challenge lies in the difficult balance between cost and effectiveness in impact analysis of
%realistic distribute software, which is characteristic of large size and great complexity.
%To deal with the potentially heavy analysis overheads, impact analysis has been explored in two radical
%directions. Efficient analysis often provides fast answers to impact queries~\cite{law03may,apiwattanapong05may}, yet
%with coarse results that require tremendous efforts for post-analysis inspection~\cite{tao2012software,cai15jss,cai14diver}.
%On the flip side, precise analysis saves the inspection efforts by developers through greater precision, yet
%at the cost of analysis overheads that are not practically affordable~\cite{acharya11may,zhang03may}.
%Frameworks offering better cost-effectiveness tradeoffs along with more such tradeoff options also
%have been studied recently~\cite{cai15diverplus,cai2016d}. However, the frameworks were developed for
%centralized, single-process software, without a straight path to extending them for distributed programs~\cite{cai2016distia}.

%\vspace{4pt}
%\noindent
%\textbf{Scope.}
%
%inter-component impacts.
%\input{background}
This example immediately
illustrates the need for a dynamic impact analysis of distributed programs at method level, which
is a specific application of dynamic dependence abstraction.
Also, developers would need varied %balances
tradeoffs between precision and overhead of such analyses in
different task scenarios~\cite{cai15diverplus,cai2016diapro}.
{\bf For example}, if the developer aims at a {\em quick, high-level} understanding about the system behavior (with respect to
how {\tt S::getMax} interacts with other methods across the system),
a fast analysis with relatively low precision would be desirable than a precise yet much slower analysis.
Yet, %if the system has complicated dependencies among its methods,
given a sufficient time budget, if the developer is tasked to fix a bug in {\tt S::getMax},
it would be more desirable to have higher precision so that the developer only needs to inspect a few methods
when deciding how to apply the bug fixes in {\tt S::getMax} itself and %potentially
necessary changes in other methods too.
%
%the rate of false positives resulted from a
%low-precision impact analysis could be overwhelmingly high, causing a post-analysis inspection cost that is too high to be affordable
%by the developer.
In fact, earlier studies suggested such diverse needs when performing impact analysis (e.g., due to various types of change requests~\cite{rovegard2008empirical}) and its application tasks (e.g., due to varied usage scenarios~\cite{de2008empirical}; 
Another common reason lies in varying resource (including time) budget constraints developers are subject to in 
conducting these tasks~\cite{rovegard2008empirical,acharya11may}. 

Thus, more generally, a framework of dynamic dependence abstraction that offers
various levels of cost-effectiveness would be required to support a wide range of (dependence-based) tasks in
developing and maintaining distributed software.
In the rest of this paper, we demonstrate how to address these challenges and needs with our framework {\tech}.
%As illustrated by $E$, the distributed systems {\tech} addresses are those in which
%components located at networked computers communicate and coordinate their actions \emph{only} by passing messages~\cite{Coulouris2011DSC}: The components run concurrently in multiple processes \emph{without a global clock}, making it hard to infer impacts across components.
%
%
%We marked the two components in this example as {\tt server} or {\tt client} (Table~\ref{tab:examplelts})
%only to facilitate illustration. {\tech} does not recognize functionality roles of components/processes, and
%it identifies probing points for communication events based on message-passing APIs.
%
%
%In the rest of this paper,
%
%Hereafter,
We will continue to use this motivating program as a working example to illustrate the inner workings of our framework.

%% file: overview.tex
\section{{\tech} Overview}
\label{sec:overview}
%\vspace{-2pt}
% summary of approach
%To achieve an efficient dynamic impact analysis for distributed programs, % at method level, we
%{\distea} utilizes only lightweight run-time information such as method execution order.
%%we build {\distea} based on the order of method executions as a first step.
%%To that end, {\distea} leverages lightweight instrumentations to capture those method-execution events and partially
%%order them across all processes of the program, and infers execute-after relations among methods according to
%%the partially-ordered events to compute potential impacts of any query.
%%
%We first %define the essential concepts and
%present the fundamentals underlying our approach, including
%the definition of method events used by {\distea} and its rationale for impact prediction.
%%from sequences of those events.
%Then, we give an overview and illustration of the inner workings, %of {\distea} %works as a whole.
%%Finally, it describes details of the analysis algorithms of {\distea}.
%followed by details on the analysis algorithms, of the {\distea} approach.
%

This section gives an overview of our {\tech} framework, focusing on its architecture, workflow, configuration, %design factors,
and application scope.

%The {\tech} framework is an unification of four dependence abstractions that provide
%varied tradeoffs between precision and efficiency by adopting different
%%impact-relation
%dependence inference schemes, as described in Section~\ref{sec:fundamentals}.
%We first give an overview of our framework along with an illustration of its inner workings,
%followed by details on the core analysis %(dynamic dependence abstraction)
%algorithms.

%\subsection{{\distea} Overview}
%\subsection{Overview}\label{subsec:overview}

\begin{figure}[t]
  \vspace{-0pt}
  \begin{center}
  \includegraphics[scale=0.8]{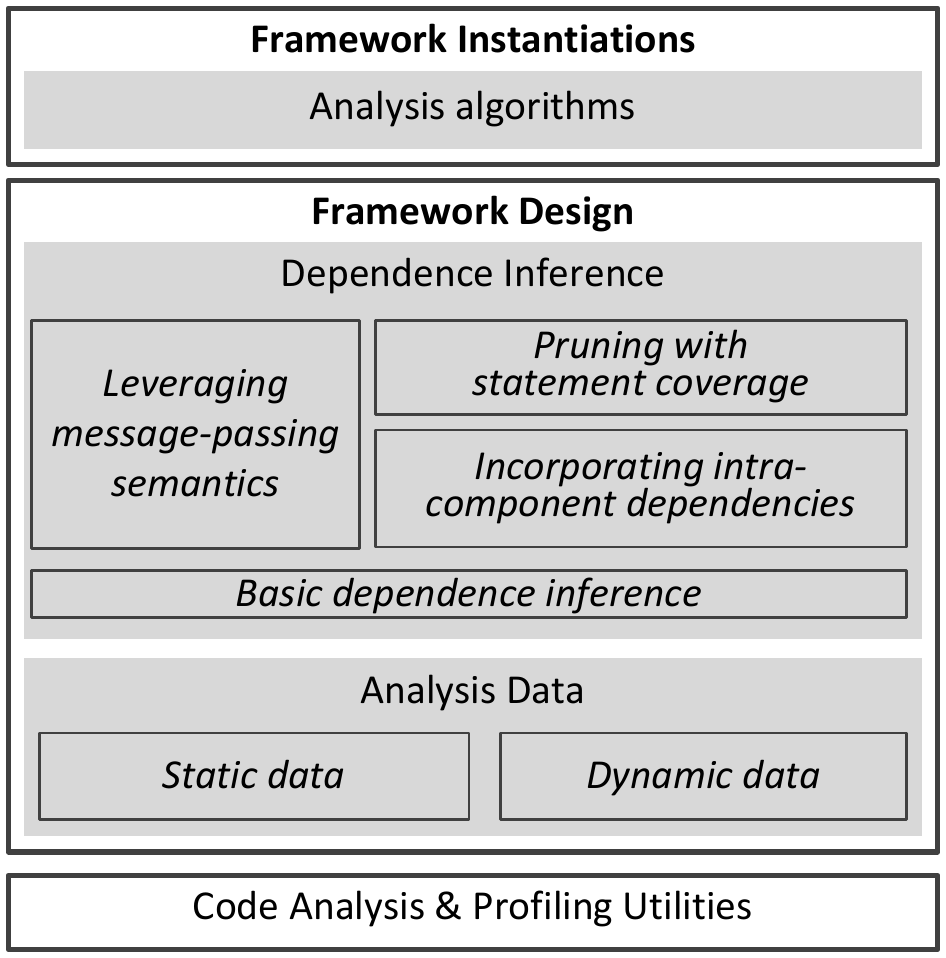}
  \vspace{-6pt}
  \caption{The architecture of {\tech} highlighting its key design elements, including {\em analysis data} used and the underlying {\em dependence inference} rules followed by various {\em instantiations} of the framework. Each layer relies on the layer immediately below it.}
  \label{fig:conceptual}
  \end{center}
  \vspace{-8pt}
\end{figure}

\subsection{{\tech} Architecture}\label{subsec:arch}
%\vspace{-2pt}
Figure~\ref{fig:conceptual} delineates the high-level conceptual (layered) architecture of our framework, elucidating its key design elements and their relationships.
The bottom layer consists of a set of static {\bf code analysis and profiling utilities} that enable the computation of, and actually purvey, the various forms of \underline{analysis data} utilized by the framework, including {\em static data} and {\em dynamic data} about the distributed program under analysis.
By using these analysis data in different, combinatorial ways, the framework infers dynamic dependencies through different \underline{dependence inference} rules.
Among these rules, {\em basic dependence inference} provides the basic form of dependence approximation underlying two other rules: {\em leveraging message-passing semantics} and {\em incorporating intra-component dependencies}; the latter further serves as a basis for another rule, {\em pruning with statement coverage}.
Such reliance relationships among the four dependence inference rules are consistent with the overall semantics of the spatial relationships among the layers in the architecture: beyond the bottom layer, every other layer relies on the layer that is immediately below it.
These analysis data and dependence inference rules constitute our {\bf framework design}.

Then, following these dependence inference rules, {\tech} spawns various {\bf framework instantiations} that each corresponds to a concrete dynamic dependence analysis for distributed programs. Importantly, each of these instantiations is expected to have a level of costs (e.g., time for producing the analysis data and for computing dependencies with the data) and a level of effectiveness (e.g., precision) that are both different from those offered by others.
On the other hand, these instantiations compute dynamic dependencies through a set of unified \underline{analysis algorithms}.
The abstraction of these analysis algorithms is reflected in the unified workflow of {\tech}, as described next (Section~\ref{subsec:workflow}).

%At a high level, Figure~\ref{fig:conceptual} depicts the two layers of {\tech}: the {\bf framework design}, and the framework instantiations which implement the design.
Details on the design of our framework are presented in Section~\ref{sec:design}, including \underline{analysis data} and \underline{dependence inference}, given in Sections~\ref{sec:data} and~\ref{sec:inference}, respectively.
Details on the instantiations of our framework are presented in Section~\ref{sec:tech}, including key \underline{analysis algorithms} given in Section~\ref{sec:depcomp}.
%
%Based on different dependence inference rules and using varied combinations of analysis data, {\tech} generalizes four different \underline{framework instantiations} (Section~\ref{sec:tech}) each providing a unique cost-effectiveness level, through unified analysis algorithms that work for any instantiation.

\subsection{{\tech} Workflow}\label{subsec:workflow}
\begin{figure}[t]
  \vspace{-0pt}
  \begin{center}
  \includegraphics[scale=0.68]{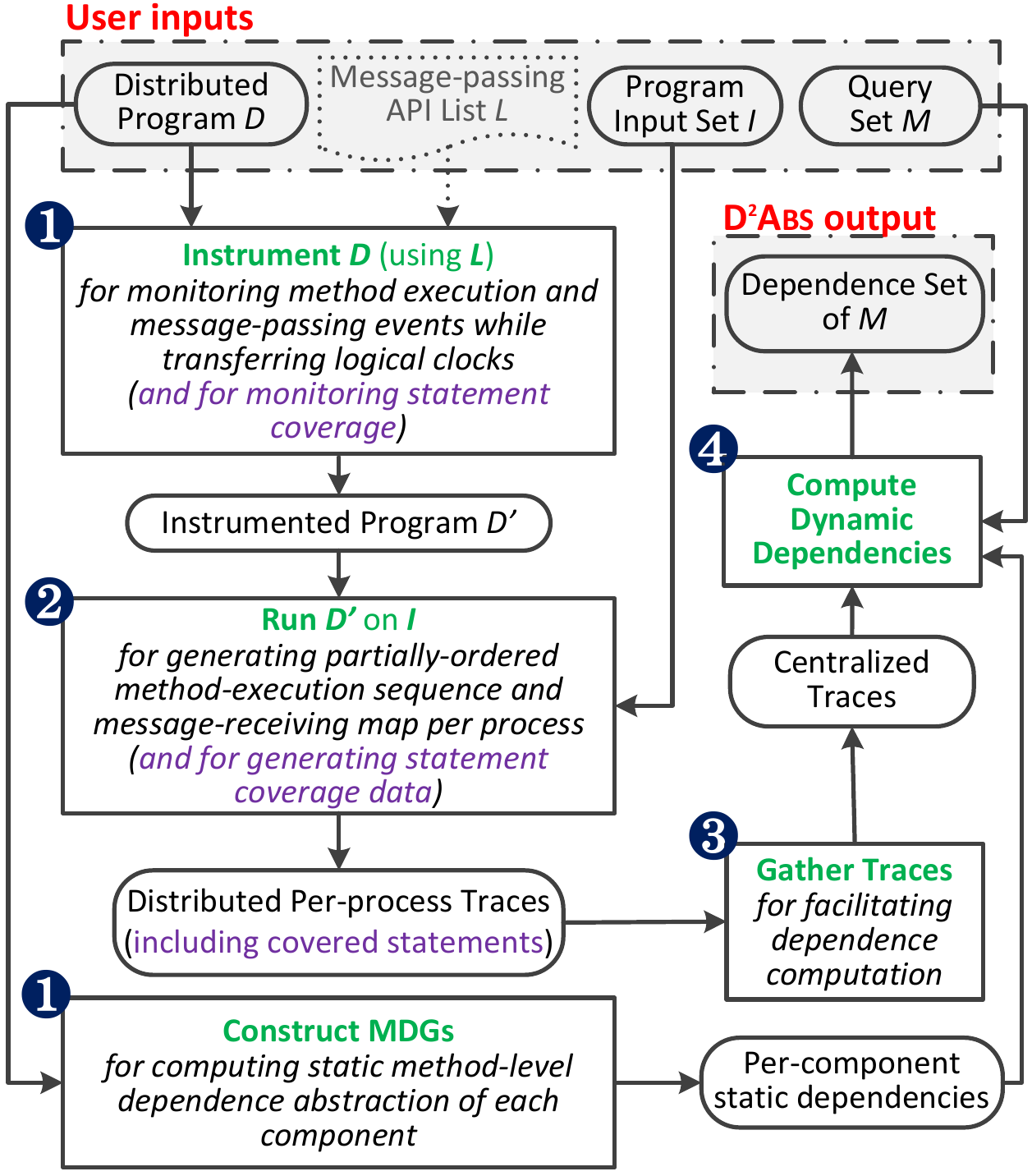}
  \vspace{-4pt}
  %\caption{{\distea} process flow, for an input system $D$ including $N$ distributed components \emph{C1} to \emph{CN}.}
  %\caption{The overall workflow of {\distea}, where the numbered steps are detailed in Section~\ref{subsec:processflow}.}
  \caption{The architecture of {\distea} and its overall workflow in four key steps (numbered), including its inputs and outputs.}
  \label{fig:disteaprocess}
  \end{center}
  \vspace{-12pt}
\end{figure}

%\subsubsection{Workflow}\label{subsec:processflow}
%\noindent
%\textbf{Workflow.}
The overall workflow of our framework is depicted in Figure~\ref{fig:disteaprocess}, where
the three primary {\bf user inputs} are the program $D$ under analysis,
a set $I$ of program inputs for $D$, and a query set $M$.
An {\em optional} input, a message-passing API list $L$ can also be specified to
help {\distea} identify program locations for profiling inter-process communications during the execution of $D$ driven by $I$.
In practice, typically this list would not be necessary since the framework implementation for a language (e.g., Java) could
handle most commonly used network I/Os for message passing (e.g., in Java distributed software) as a built-in feature. And this built-in feature
would suffice for precisely and completely capturing inter-process communications as required for the dynamic dependence abstraction in our framework.
%where probes for communication events should be instrumented (as detailed in Section~\ref{sec:impl}).
The %output of {\distea}
{\bf {\tech} output}
is a set of %potential
dependencies (impacts)
of $M$ (i.e., the set of methods dynamically dependent on any method in $M$)
computed from the given inputs in {\bf four steps} as annotated in the figure and described below.
% the overall working process of {\distea} consists of three phases

As mentioned earlier, the key motivation for {\tech} is to enable dynamic dependence abstraction for distributed programs while
offering varied tradeoffs between analysis precision and efficiency. To that end, {\tech} utilizes both static and dynamic
data about the program. These data are of different forms each coming with a different level of {\em cost} (i.e., the time cost incurred by generating and utilizing the data) and {\em benefit} (i.e., the data's contribution to the dependence analysis effectiveness).
At a high level, the four steps of {\tech} are about (1) {\em generating} these data ({\bf Step \circled{1},\circled{2},\circled{3}}) and then {\em utilizing} them for dependence computation ({\bf Step \circled{4}}).
The workflow of {\tech} as regards how these steps collaborate is elaborated below.

The \emph{first step} ({\bf Step \circled{1}}) %performs static analysis of the input program $D$ to instrument it
of {\tech} has two aims: (1) computing the only kind of {\em static data} (static dependencies per component) through a static analysis of $D$, and (2) preparing for collecting different forms of {\em dynamic data} (method-execution events, message-passing events, and statement coverage) through static instrumentation of $D$.
%The \emph{first step} instruments the input program $D$
%for both monitoring method-execution and message-passing events, and for synchronizing logical clocks among concurrent processes.
%In addition, if opted, statement coverage monitoring is also probed during the instrumentation.
%This step produces the instrumented version $D'$ of $D$.
%Optionally, this step also constructs the method dependence graph (MDG) for each component, which results in the
%per-component static dependencies represented by the MDG.
%%This step results in the , which will .
The results of this step are (1) the per-component static dependence graph and (2) the instrumented version $D'$ of $D$.

The \emph{second step} ({\bf Step \circled{2}}) aims to profile the dynamic data that is probed for in the previous step.
Specifically, this is done by executing $D'$ on the given input set $I$, during which
all the dynamic data instrumented are generated and collected (serialized into respective disk files).
%internal events are produced and time-stamped by means of communication events such that the partial order for all internal events is preserved.
An additional item of data, \emph{message-receiving map}, is derived from the message-passing events and serialized as part of
the resulting execution trace.
The map records for each process the timestamp of its receiving the first message from any other processes.
The tracing in each process is performed concurrently with that in others during the system execution.
%, while the processes of the system run concurrently.
%
%Meanwhile, a \emph{message-receiving map} is produced, where each entry indicates the
%sending process and time of a message-receiving event---for each unique sending process only the first such event is recorded.
%%all internal events.
%%with respect to the logic clocks exchanged among all processes during the runtime. The timestamps attached to
%%the method events implicitly preserve the partial ordering of those events.
%%
%If statement coverage monitoring is instrumented, the second step will include statement coverage data in the traces.

In the \emph{third step} ({\bf Step \circled{3}}), the execution traces separately generated in individual processes are gathered to the machine where the dependence computation is to be performed. This assisting step is necessary as the dependence analysis in {\tech} is
centralized, while the tracing is performed in a distributed manner.
%and merged to a holistic ordered sequence stored in either one or multiple traces. %(trace files). %in order to facilitate impact computation. As
%the processes are supposed to run on separated machines, {\distea} collects these traces to one machine before merging them.

In the \emph{fourth step} ({\bf Step \circled{4}}), {\tech} takes all the static and dynamic data as inputs to
compute the dependence set of the query set $M$, using a unified dependence abstraction algorithm. The algorithm automatically
chooses different combinations of the collected data and utilizes varied dependence inference rules accordingly, as per user specifications.
%
%the query set $M$ and  %merged event sequence
%
%the traces to compute the dependence set of $M$. %using the impact inference. %from the merged.

%\vspace{1pt}
%\noindent
%\textbf{Configuration.}
\subsection{{\tech} Configuration}\label{subsec:config}
A main merit of our framework lies in its offering variable and flexible cost-effectiveness tradeoffs in dependence analysis.
This merit enables {\tech} to accommodate varying developer needs (e.g., due to their varied precision expectations and cost budgets).
In essence, the flexibility of {\tech} is attributed to its customizability, realized via varied configurations.
As depicted in Figure~\ref{fig:disteaprocess} (as noted in parentheses), the framework allows users to choose which data to
%opt for or not
be utilized by the dependence analysis algorithm.
For example, the user may choose not to utilize statement coverage in the analysis. Then, accordingly, the instrumentation and profiling (in {\bf Steps \circled{1} and \circled{2}}, respectively) for statement coverage would be skipped.
Similarly, if the user does not want to use static dependencies in the dependence abstraction, the corresponding static analysis in {\bf Step \circled{1}} would be ignored.

%\vspace{1pt}
%\noindent
%\textbf{Configuration.}
%As shown, the framework can be instantiated at one of the four levels of dynamic dependence abstraction according to the
%analysis configuration in terms of how much data to be collected and used by the analysis algorithm.
%A key design motivation and principle of {\tech} is to provide flexible cost-effectiveness tradeoffs to accommodate
%varying developer needs (e.g., due to their varied precision expectations and cost budget).
%Thus, {\tech} allows users to choose different such tradeoffs through the analysis configuration.

% define distributed system, distributed program, component, process, thread, etc. as part of the scope
\subsection{Scope}\label{sec:scope}
{\tech} targets the most common type of distributed systems, as opposed to specialized ones such as DEBS and others relying on
special inter-component interfaces (e.g., RMI---remote method invocation~\cite{pitt2001java}) or message-type specifications (e.g., CORBA--common object request broker architecture~\cite{siegel2000corba}).
As exemplified by the running example program $E$ of Figure~\ref{fig:csexample},
a {\em common distributed system} has three characteristics according to the definition
in~\cite{Coulouris2011DSC}:
(1) the system consists of multiple components located at networked computers,
(2) these components communicate and coordinate their actions {only} by passing messages,
and (3) the components run concurrently in multiple processes {without a global clock}.
These systems represent the applicability scope of {\tech}, although
{\tech} only deals with the software part of such systems (as opposed to system-level concerns such as distributed computing infrastructure/platform and resource management), referred to as {\em distributed software}.
More specifically, as we focus on code analysis, the actual analysis scope of {\tech} is
the program (code) of distributed software, referred to as {\em distributed program}.

We further confine as a {\em component} the code entities that runs in a separate {\em process} from the rest of the system,
where each process may host one or multiple {\em threads}.
{\bf For example}, the two components in the example program $E$, {\tt server} and {\tt client}, %(Table~\ref{tab:examplelts})
each runs in a separate process and they communicate only through message passing (via a network socket).
These components are marked for illustration purposes---{\tech} does not recognize functionality roles of components/processes, and
it identifies probing points for inter-process communication (IPC) %(i.e., message-passing)
events based on the given list of message-passing APIs.

%the distributed systems {\tech} addresses are those in which components located at networked computers communicate and coordinate their actions \emph{only} by passing messages: The components run concurrently in multiple processes \emph{without a global clock}, making it hard to infer impacts across components.

%% file: design.tex
%!TEX root = paper.tex

\section{The {\tech} Framework Design}
\label{sec:design}
%\vspace{-2pt}
In this section, we elaborate the framework design of {\tech}, as shown in Figure~\ref{fig:conceptual}. 
The two major design elements that constitute the framework are the \underline{analysis data} utilized for dynamic dependence abstraction of distributed programs, and the \underline{dependence inference} rationales (rules) according to which dynamic dependencies are derived from the analysis data. 

\input{analysisdata}

\input{dependenceinference}

%% file: analysisdata.tex
%!TEX root = paper.tex
%  \vspace{-6pt}

\subsection{Analysis Data}\label{sec:data}
This section presents the various forms of program data utilized by {\tech} for its dependence abstraction
of variable cost-effectiveness.
As noted earlier, {\tech} uses one form of static and three forms of dynamic data in its analysis,
because of the different levels of cost and benefit of each of these forms of data which can contribute to
the variable cost-effectiveness {\tech} aims at.

\subsubsection{Static Data}
To trade analysis cost for better precision (than purely dynamic dependence analysis), {\tech} exploits
the static dependencies within each component of the distributed program under analysis (noted as {\em intra-component dependencies}).
Treating each component as a single-process program, traditional dependence analysis can be utilized for this purpose.
However, immediately adopting the traditional dependence model would compromise the scalability of {\tech}, given
the known heavyweight nature of computing fine-grained (statement-level) dependencies~\cite{horwitz90jan,cai14diver,cai2018hybrid}.
Thus, we employ the method-level dependence abstraction algorithm developed
for single-process Java software~\cite{cai2015abstracting,cai2016method} to
compute the static dependencies among methods.
%{\tech} refers to these static method-level dependencies to prune false positives resulting otherwise from a purely dynamic dependence analysis.
%in the control-flow approximation of dynamic dependence (impact) relations between methods.
%
%derived from the basic inference (Section~\ref{subsec:basicinfer}).
Here we summarize the key ideas of this abstraction algorithm and then discuss its extension for including
threading-induced dependencies.

\vspace{3pt}
\noindent
\textbf{Static dependence abstraction.}
Specifically, for each component $c$, the {\em method dependence graph} $MDG_c$ is constructed to
represent the static dependencies among methods, with intraprocedural dependencies abstracted
with summaries~\cite{cai2015abstracting}. Each method of $c$ is represented as a node of $MDG_c$, where
each edge represents data or control dependence between two methods. The intraprocedural dependence
summaries for each method are represented by a mapping from incoming to outgoing flows, resulting from an
intraprocedural reachability analysis on the procedure dependence graph~\cite{ferrante87jul} of the method.
For the sake of scalability, interprocedural dependencies are computed in a flow-insensitive manner
(although at the cost of compromising precision).
Further, since we do not perform any analysis of static dependencies alone but use them as
contexts for dynamic dependence refinement, %pruning,
the computation of interprocedural dependencies in $MDG_c$ is
also context-insensitive. Discarding context- and flow-sensitivity largely enhances the scalability of the
static intra-component dependence abstraction~\cite{cai2016method}, hence potentially that of
{\tech} when incorporating the static dependence information in the dynamic dependence abstraction.

\vspace{3pt}
\noindent
\textbf{Threading-induced dependencies.}
Realistic distributed systems typically run multiple threads in each component (process). Therefore, {\tech}
also includes threading-induced dependencies in its static intra-component dependence computation.
In particular, two additional types of control dependencies, {\em synchronization dependencies} and
{\em ready dependencies}~\cite{giffhorn2009precise}, and an additional type of data dependence, {\em interference dependencies}~\cite{Nanda2006ISM}, are considered in our framework.
For Java programs, {\em synchronization dependencies} are induced by the use of synchronization blocks while
{\em ready dependencies} are due to the {\tt wait-notify} synchronization mechanism.
{Interference dependencies} are generally a result of data sharing between threads.
%Formal definitions
Each of these three types of dependencies lies between two different methods across two different threads.
For each component $c$, {\tech} adds these inter-thread dependencies to the $MDG_c$ computed for $c$ in
the static dependence abstraction step described above.

\vspace{3pt}
\noindent
{\bf Illustration.}
For example, in the program $E$ of Figure~\ref{fig:csexample}, with {\tt init} representing the constructor of a class, 
the $MDG$$_{S}$ includes the dependencies of
{\tt main} on {\tt init} (via return varaible $s$) and {\tt serve} (via instance field $ssock$), of
{\tt serve} on {\tt getMax} (via return variable $r$), and of {\tt getMax} on {\tt serve} (via parameter $s$).

\subsubsection{Dynamic Data} \label{subsec:eventsdef}
%\vspace{-8pt}\subsection{Motivating Example}\label{subsec:motiveexample}
\setlength{\tabcolsep}{9.0pt}
\renewcommand{\arraystretch}{0.95}
\begin{table}[tp]
\begingroup
\renewcommand*{\arraystretch}{0.95}
  \centering
  %\caption{EXECUTION TRACE OF EVENTS OF THE EXAMPLE PROGRAM E3. THE EVENTS ARE PARTIALLY ORDERED FROM TOP TO BOTTOM, SEPARATELY FOR THE TWO COMPONENTS SERVER AND CLIENT; THE LTS COLUMN SHOWS THE LTS FOR EACH INTERNAL EVENT}
  %
  %\caption{\hspace{-0pt}{A full sequence of method-execution events for an example program $E$}}
  \caption{\hspace{-0pt}{A full sequence of (i.e., instance-level) method-execution (and message-passing) events of the program $E$}}
  \vspace{-8pt}
    \begin{tabular}{|l|r||l|r|}
    \hline
    \multicolumn{2}{|l||}{\textbf{Server process}} & \multicolumn{2}{l|}{\textbf{Client process}} \\
    \hline
    Event & Timestamp & Event & Timestamp \\
    \hline
    {\tt S::main}$_e$ & 0 & {\tt C::main}$_e$ & 0 \\
    \hline
    {\tt S::init}$_e$ & 1 & {\tt C::init}$_e$ & 1 \\
    \hline
    {\tt S::init}$_i$ & 2 & {\tt C::init}$_i$ & 2 \\
    \hline
    {\tt S::init}$_r$ & 3 & {\tt C::init}$_r$ & 3 \\
    \hline
    {\tt S::main}$_i$ & 4 & {\tt C::main}$_i$ & 4 \\
    \hline
    {\tt S::serve}$_e$ & 5 & {\tt C::compute}$_e$ & 5 \\
    \hline
    $E_c$(C,S) & - & {\tt C::shuffle}$_e$ & 6 \\
    \hline
    {\tt S::getMax}$_e$ & \textbf{10} & {\tt C::shuffle}$_i$ & 7 \\
    \hline
    {\tt S::getMax}$_i$ & 11 & {\tt C::shuffle}$_r$ & 8 \\
    \hline
    {\tt S::getMax}$_r$ & 12 & {\tt C::compute}$_i$ & 9 \\
    \hline
    {\tt S::serve}$_i$ & 13 & $E_c$(C,S) & - \\
    \hline
    $E_c$(S,C) & - & $E_c$(S,C) & - \\
    \hline
    {\tt S::serve}$_r$ & 14 & {\tt C::compute}$_r$ & \textbf{14} \\
    \hline
    {\tt S::main}$_i$ & 15 & {\tt C::main}$_i$ & 15 \\
    \hline
    {\tt S::main}$_r$ & 16 & {\tt C::main}$_r$ & 16 \\
    \hline
    \end{tabular}%
  \label{tab:examplelts}%
  \vspace{-14pt}
  \endgroup
\end{table}%

Our framework utilizes only lightweight run-time (i.e., dynamic) information, at method and statement levels.
In particular, %two forms of such as method execution order and statement coverage data.
{\tech} uses three forms of dynamic data as mentioned earlier:
{\em method-execution events} and {\em message-passing events} record the occurrence and timing of method calls
and IPCs, respectively, while {\em statement coverage} records which statements are exercised during the execution.
%
%We first describe the definition of method events used by {\tech} (Section~\ref{subsec:eventsdef}), and the
%basic inference of %impact (dependence)
%dependence relations among the methods (Section~\ref{subsec:basicinfer}).
%We then describe three ways to refine the inference so as to improve the
%precision of the dynamic dependence abstraction in {\tech} at additional costs:
%(i) leveraging message-passing semantics (Section~\ref{subsec:messagesemantics}),
%(ii) incorporating intra-component static dependencies (Section~\ref{subsec:intradep}),
%and (iii) pruning spurious dynamic dependencies using statement coverage (Section~\ref{subsec:stmtcov}).
%%
%In this paper, we use dynamic impact analysis as an illustrating application (client analysis) of {\tech}, for which the dependence relation between two methods derived by {\tech} immediately corresponds to the impact relation between the two methods:
%a method $m2$ is considered to be impacted by a method $m1$ (i.e., $m1$ impacts $m2$) if $m2$ is considered to depend on $m1$.
%%Hereafter, we use impact and depend
%Thus, for our dependence-based dynamic impact analysis application, we may use {\em be impacted by} and {\em depend on}" exchangeably hereafter.
In the general context of distributed systems, an {\em event} is defined as any happening of interest observable from
within a computer~\cite{lamport1978time}. More specifically, %~\cite{muhl2006distributed}
%, a particular type of distributed systems, however,
events in a DEBS are often expressed as messages transferred among system components and defined by a set of attributes~\cite{muhl2006distributed,garcia2013identifying}.
%While working for a broad scope of multiprocess programs with respect to
%both definitions of events,
While it also deals with message passing in distributed systems, currently {\tech} neither makes
any assumption nor reasons about the structure or content of the messages (doing so would compromise its applicability).

%its key components.
%\vspace{-2pt}
%\subsubsection{Method-Execution Events}
\vspace{3pt}
\noindent
\textbf{Method-execution events.}
As the very basic form of dynamic data for its dynamic dependence analysis at method level, {\tech}
%
%Concretely, for dynamic impact analysis, {\distea}
monitors and utilizes %two major classes of events as defined below:
%communication events and internal events, as informally defined as follows.
%As we focus on analyzing multiprocess programs, we distinguish components in a distributed system as
%such that each runs in a separate process.
method-execution events.
We define a \underline{method-execution event} $E_I$ as an occurrence of method execution within a component $c$, denoted as $E_I$($c$).
Further, we differentiate three subcategories of such events: %at method level:
\emph{entering a method}, \emph{returning from a method}, and \emph{returning into a method}, denoted as $m_e$, $m_r$, and $m_i$, respectively, for the relevant method $m$.

Note that {\tech} captures both the return (from) %($m_r$)
and returned-into %($m_i$)
 events for each method. %$m$.
{\bf For example}, for a call to a method $g$ in a method $f$, a {\em return from} event, associated with $g$, indicates the event that
the program control gets out of the scope of $g$, marking the end of one execution instance of $g$.
Differently, the corresponding {\em return into} event, associated with $f$, indicates the event that the program control gets
back into the scope of $f$, marking the continuation (after the callsite targeting $g$) of one execution instance of $f$.
However, we distinguish them during instrumentation %static analysis %(instrumentation)
only and treat them equally in the %online
%monitoring algorithm (Section~\ref{subsec:easalgo}).
analysis algorithms of {\tech}.
%of runtime monitoring.
% and tracing.
%The reason is to deal with interleaving method-execution sequences in the case of multithreaded  executions,
In sequential program executions, a method $m$ is potentially affected by any
changes in the query $q$ if $m$, or part of it, executes after $q$, and
monitoring method entry and returned-into events suffices for retrieving
such execute-after relations; in the case of concurrent (single-process) programs, however,
$m$ is potentially affected by the changes also if $m$, or part of it, executes
in parallel with $q$,
and method return events need to be monitored as well to correctly identify
%the impact
dependence relations from interleaving method executions in multiple threads~\cite{apiwattanapong05may}.
\vspace{3pt}
\noindent
\textbf{Message-passing events.}
%Different from the communication events, with which the event of passing a message is represented by a method call to a relevant message-passing API, message-passing events in our framework record.
%%% to do: okay, categorize the communication events above as message-passing events... or change this to first message-passing map
%
Besides method-execution events, {\tech} monitors message-passing events, used to reason about the effects of
%inter-process communications (IPCs)
IPCs
in the distributed system~\cite{xqfu19icpc} on the dynamic dependencies among methods across all the processes of
the system's execution.
We define a \underline{message-passing event} $E_C$ as the occurrence of a message transfer between two components $c1$ and $c2$, denoted as $E_C$($c1$,$c2$) if $c1$ initiates $E_C$ which attempts to reach $c2$. Further, according to the direction of message flow, we distinguish two major subcategories of such events: \emph{sending a message} to a component and \emph{receiving a message} from a component. %, %In a client/server scenario, for instance, the socket API invocations of {\tt send} and {\tt recv} are the \emph{sending} and \emph{receiving} events, respectively.
We refer to the process running the component that sends and receives the message as the \emph{sender process} and \emph{receiver process}, respectively.

%\vspace{-2pt}
%\subsubsection{Statement Coverage}
\vspace{3pt}
\noindent
\textbf{Statement coverage.}
For containing analysis costs, most kinds of program information {\tech} uses are at method level. To provide cost-effectiveness options with more emphasis on precision, {\tech} also considers using statement coverage as a form of statement-level data. More specifically, it records which statements are covered during the system execution, separately for each component of the system. This fine-grained form of data is used to refine the per-component (method-level) static dependencies so that they can contribute to an even more precise dynamic dependence abstraction.

\vspace{3pt}
\noindent
{\bf Illustration.}
As an example, Table~\ref{tab:examplelts} shows the method-execution events (e.g., {\tt S::main}$_e$, {\tt C::init}$_r$, and {\tt S::getMax}$_i$) and message-passing events (e.g., $E_c$(C,S)) captured during the execution of the program $E$ of Figure~\ref{fig:csexample}. These events are recorded separately for each process and are listed in the order of their timestamp.
In this simple program, there are no branches. As a result, during the system execution, in accordance with the events of Table~\ref{tab:examplelts}, all statements of $E$ are covered.

%% file: dependenceinference.tex
%  \vspace{-6pt}
\subsection{Dependence Inference}\label{sec:inference}
In this section, we present the varied dependence inference rules underlying the {\tech} %dependence abstraction
framework. Each rule utilizes the static and/or dynamic data as described earlier (Section~\ref{sec:data}) in different ways, implying different cost-benefit
tradeoffs for the framework.
We first describe the
basic inference of %impact (dependence)
dependence relations among executed methods (Section~\ref{subsec:basicinfer}).
We then discuss three ways to refine the basic inference so as to improve its
precision %of the dynamic dependence abstraction in {\tech}
at additional costs:
(i) leveraging message-passing semantics (Section~\ref{subsec:messagesemantics}),
(ii) incorporating intra-component static dependencies (Section~\ref{subsec:intradep}),
and (iii) pruning spurious dynamic dependencies using statement coverage (Section~\ref{subsec:stmtcov}).

%In this paper,
We use dynamic impact analysis as an illustrating application (client analysis) of {\tech}, for which the dependence relation between two methods derived by {\tech} immediately corresponds to the impact relation between the two methods:
a method $m2$ is considered to be impacted by a method $m1$ (i.e., $m1$ impacts $m2$) if $m2$ is considered to depend on $m1$.
%Hereafter, we use impact and depend
Thus, for our dependence-based dynamic impact analysis application, we may use ``{\em be impacted by}" and ``{\em depend on}" exchangeably hereafter.

%% comment on novelty
Note that while our dependence inference rules may seem intuitive and straightforward, they appear to be so after we have
revealed them---these rules have not been presented before, especially in the context of dynamic analysis for common distributed programs.
In particular, the partial-ordering algorithm underlying the basic inference was originally proposed for event synchronization in
distributed systems~\cite{lamport1978time}. Yet it has not been exploited for modeling code-based dynamic %method-level
dependencies.
Thus, as opposed to computational challenges, exploring how to infer dependencies between two code entities (e.g., methods) in distributed programs comes more with methodological challenges---it was not known (1) which information, among various possible kinds, to utilize, (2) how to combine varied kinds of information to derive dependencies, and furthermore (3) how to infer the dependencies in different yet cost-effective ways.

\subsubsection{Basic Dependence Inference}\label{subsec:basicinfer}
%As mentioned earlier,
One challenge to developing {\tech} is to infer dependence relations based on execution order in the presence of asynchronous events over concurrent multiprocess executions. Fortunately, %in the absence of synchronized physical timing for all processes,
maintaining a logical notion of time per process to discover just a partial ordering of %the causality relations among
method-execution events suffices for that inference. %required in {\tech}. %of execution-after order of associated methods that {\distea} needs.
%that is required by EA-based impact analysis.
%
The dependence relation between any two methods can be semantically
over-approximated by the {\em happens-before} relation between relevant execution events of corresponding methods;
%method-execution events can be transformed to.
and the partial ordering of the execution events reveals such happens-before relations~\cite{lamport1978time}.
%can be revealed from the partial ordering of the distributed method events
Formally, given two methods $m1$ and $m2$, we have
%
%Furthermore, to apply the partial ordering of events directly to the EA-based impact analysis by a unified inference,
%we map the EA relation between two method $m1$ and $m2$ as defined in~\cite{apiwattanapong05may} to the happens-before relation
% between relevant events associated with the two methods, as follows:
%
%Concretely, given two executed methods $m1$ and $m2$, we map the execute-after (EA) relation between them
% to the happens-before relation between their internal events, as follows:
%
\vspace{-5pt}
\begin{equation}\label{eq_ea}
\vspace{-5pt}
%EA($m1$,$m2$) $\Longleftrightarrow$ $m1_e$$\longrightarrow$$m2_r$ $\bigvee$ $m1_e$$\longrightarrow$$m2_i$
%m1_e \longrightarrow m2_r \bigvee m1_e \longrightarrow m2_i \Longrightarrow  EA(m1,m2)
%m1_e \prec m2_r \bigvee m1_e \prec m2_i \Longrightarrow m1~\emph{impacts}~m2
m1_e \prec m2_r \bigvee m1_e \prec m2_i \Longrightarrow m2~\emph{depends on}~m1
\end{equation}
%where the notion ``$\longrightarrow$" denotes the happens-before relation that signals causal relationship.
%Without loss of generality, $m1_e$$\longrightarrow$$m2_r$ (or $m1_e$$\prec$$m2_r$), $m1_e$$\longrightarrow$$m2_i$ (or $m1_e$$\prec$$m2_i$), and EA($m1$,$m2$) all equally imply that ``$m2$ executes after $m1$, thus $m2$ may be affected by $m1$ or any change to occur in $m1$.", hence the equivalence mapping between the execute-after relation and (internal) method-event partial ordering above.
%where ``$a$$\prec$$b$" denotes that $a$ \emph{happens-before} $b$, which signals the causal relationship that $a$ \emph{causes}
%$b$ in our context of impact analysis.
where $\prec$ denotes the \emph{happens-before} relation.
Without loss of generality, either of $m1_e$$\prec$$m2_r$ and $m1_e$$\prec$$m2_i$ implies that ``$m2$ executes after or in parallel with $m1$, thus $m2$ may be affected by (any changes in) $m1$", hence the dependence (impact) relation between $m1$ and $m2$.
%the equivalence mapping between the execute-after relation and (internal) method-event partial ordering above.
{\bf For example}, with respect to the example execution of program $E$ shown in Table~\ref{tab:examplelts},
{\tt C::init}$_e$$\prec${\tt C::main}$_i$ implies that {\tt C::main} depends on {\tt C::init},
and {\tt C::compute}$_e$$\prec${\tt C::shuffle}$_r$ implies that {\tt C::shuffle} depends on {\tt C::compute}.

Based on the above inference, for a given query $q$, computing the dependence set $DS$($q$) of $q$ %($IS$ for \emph{Impact Set})
is reduced to retrieving methods, from multiprocess method-execution event sequences, that satisfy the partial ordering of
%$c$ and those with respect to their
the execution events of candidate methods as follows:
\vspace{-5pt}
\begin{equation}\label{eq_is}
\vspace{-5pt}
%$IS$($q$)~=~\{m$\mid$$q_e$$\prec$$m_i$ $\vee$ $q_e$$\prec$$m_r$\}.
DS(q)~=~\{m\mid q_e\prec m_i \vee q_e \prec m_r\}
\end{equation}
Note that only %the three types of
method-execution events are directly used for the inference, whereas
message-passing events %, by contrast,
are utilized to maintain the partial ordering of method-execution events across processes.
%
%the former across processes.
%
%during multiprocess executions of the distributed system under analysis.
%Also, since the execution order between two methods can only \emph{over-approximate} the actual impact relation between them,
%the impact inference in {\distea} is conservative, but sound and lightweight in nature. %lead to false-positive impacts, but
%
%Also, while the dependence inference in {\tech} is conservative, it is sound~\cite{jackson2000software} and only requires lightweight dynamic analysis and computation.
%
While leading to potentially excessive imprecision, the conservative nature of this basic dependence inference in {\tech}
also implies that it provides safe results. It also only requires highly lightweight dynamic analysis and computation, which
implies great efficiency. Therefore, this basic inference still provides a useful cost-effectiveness option.

\subsubsection{Leveraging Message-Passing Semantics}\label{subsec:messagesemantics}
The above basic inference %based on happens-before relation
leads to a safe yet possibly \emph{overly} conservative approximation of dynamic dependencies (impacts),
because it is based purely on happens-before relations between method-execution events.
Recall that message passing is the only communication channel between processes in the
systems we address %(Section~\ref{sec:scope})
to propagate data flows or control decisions.
Thus, a method ${{P_1}^m}$ in process $P_1$ would not depend on %be impacted by
a method ${{P_2}^m}$ in
process $P_2$ if $P_1$ never received a message %directly, or transitively via other processes
from $P_2$ \emph{before} the last execution event of ${{P_1}^m}$.
This is true even if ${{P_1}^m}$ is %ever
executed after ${{P_2}^m}$ in the whole-system method-execution event partial ordering, since
no data or control would even flow between the two processes.
\begin{figure}[t]
  %\vspace{-5pt}
  \begin{center}
  \includegraphics[scale=0.75]{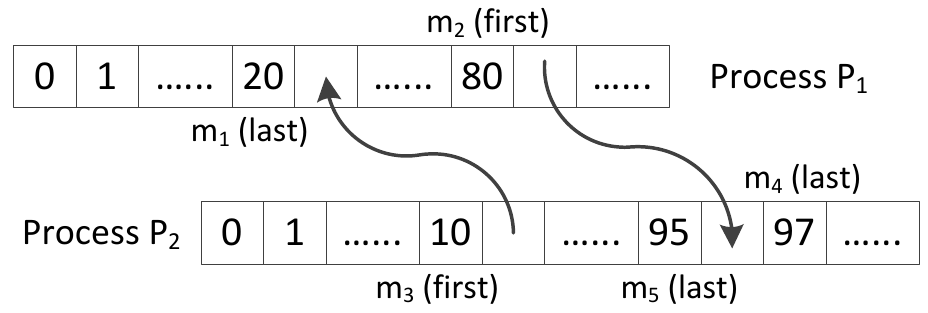}
  \vspace{-10pt}
  %\caption{{\distea} process flow, for an input system $D$ including $N$ distributed components \emph{C1} to \emph{CN}.}
  \caption{Inter-process impact constrained by message passing.}
  \label{fig:msgprune}
  \end{center}
  \vspace{-20pt}
\end{figure}

Figure~\ref{fig:msgprune} {\bf illustrates} how message passing could constrain the dependence %impact
relation defined by Equation~\ref{eq_ea}. Each numbered cell represents the first or last execution event
of a method with the number indicating the time (stamp) when that event occurs, and each empty cell represents a
message receiving or sending event.
The arrowed line on the left and right indicates the \emph{first} message passing from
$P_2$ to $P_1$ and $P_1$ to $P_2$, respectively.
As per the basic inference, $m_1$ in $P_1$ would be inferred as dependent on $m_3$ in $P_2$, because
$m_1$ was executed after $m_3$.
With the message-passing semantics considered, however, this dependence is deemed as spurious hence should be pruned.
The reason is because
$P_2$ sent the first message to $P_1$ after the last time $m_1$ was executed.
Similarly, a spurious dependence of $m_5$ on $m_2$ (as produced per the basic inference) should also be pruned.
On the other hand, $m_4$ is still considered as
dependent on %is potentially impacted by
$m_2$ even with message-passing semantics considered,
because not only was $m_4$ executed after $m_2$ was first executed, but also $P_2$ received a message from $P_1$ prior to the first execution of $m_2$.

Generally, considering the message-passing semantics,
${{P_i}^m}$ potentially depends on %is potentially impacted by
${{P_j}^{m'}}$ only if
(i) the first execution event of ${{P_j}^{m'}}$ happens before %has a smaller timestamp than
the last execution event of ${{P_i}^m}$---this essentially enforces the basic inference rule,
(ii) $P_j$ sends at least one message (directly or transitively) to $P_i$---this is a quick check against spurious dependencies, because if there is no message passing between two processes, two methods across the processes should not have dependence relationships,
and (iii) the last execution event of ${{P_i}^m}$ happens after %has a larger timestamp than
the first message receiving event in $P_i$ from $P_j$, and the message is sent after the first execution event of ${{P_j}^{m'}}$---this last condition ensures that the message, which represents interprocess data flow, can actually first influence the method in the sender process and then propagate the influence to the method in the receiver process.
Formally, let $T_F(m)$ and $T_L(m)$ denote the time (stamp) of the first and last execution event of a method $m$,
respectively, and $T_S(P_i,P_j)$ denote the time (stamp) of the event of $P_i$ receiving the first message from process $P_j$, we define %the constrained form of happens-before
a customized form of partial order
relation between two methods ${{P_i}^m}$ and ${{P_j}^{m'}}$ as follows ($T_S$ is the {\em message-receiving map}):
\vspace{-2pt}
\begin{equation}\label{eq_constraint}
\vspace{-2pt}
%{\scriptsize
{{P_j}^{m'}}\prec{{P_i}^m}:=%\Longleftrightarrow
\begin{cases}
    T_L({{P_i}^m})\geq T_F({{P_j}^{m'}}),~~~~~~~\text{if } i=j \\
    \\
    \begin{aligned}
    & T_S(P_i,P_j)\ne \text{null} \wedge T_L({{P_i}^m}) \geq \\
    & ~T_S(P_i,P_j)\geq T_F({{P_j}^{m'}}),~~\text{if } i\ne j
    %& ~~\text{max}(T_F({{P_j}^{m'}}), T_S(P_j)),~\text{if } i\ne j
    \end{aligned}
\end{cases}
\end{equation}
%Intuitively there would be another condition that is symmetric to (3) constraining the
% first execution event of ${{P_j}^{m'}}$ happens before the last message sending event in $P_j$ to $P_i$. However, as per our internal event synchronization algorithm, this additional constraint is subsumed in the given three---immediately after the first message receiving event in $P_i$ from $P_j$, the logic clock of $P_i$ will be updated to be greater than the timestamp of the first execution event of ${{P_j}^{m'}}$, which thus must be smaller than the timestamp of the last message sending event in $P_j$ to $P_i$.

%since the change impact originated
%in $m_2$ may propagate through the message passing
With this constrained definition of $\prec$, a more precise dependence %impact
inference is obtained from Equation~\ref{eq_is} since
the potential method-level dependencies %potentially impacted methods
identified by the basic inference that do not satisfy the additional constraints (ii) and
(iii) will be pruned. Given the scope of distributed systems we address, this pruning does not compromise the safety of
resulting dependence sets. %impact sets.

%To facilitate our discussion and evaluation later on, hereafter we refer to {\tech} with and without the
%pruning as its \emph{basic} and \emph{enhanced}
%version, respectively.

\subsubsection{Incorporating Intra-Component Dependencies}\label{subsec:intradep}
With the static dependencies of a component $c$ and method execution events
of the process $P_c$ that executes $c$, the dependence/impact relation between
two methods ${P_c}^{m1}$ and ${P_c}^{m2}$ in $P_c$ is inferred from both
the partial ordering of execution events associated with these two methods, and
the static dependencies between them. Formally, we define another constrained partial
order relation as follows:
\begin{equation}\label{eq_constraint2}
\vspace{-5pt}
%{\scriptsize
{{P_c}^{m1}}\prec{{P_c}^{m2}}:= %\Longleftrightarrow
    \begin{aligned}
    T_L({{P_c}^{m2}})\geq T_F({{P_c}^{m1}})~\wedge~\\
    %{{P_c}^{m1}}\overset{MDG_c}{\longrightarrow}{{P_c}^{m2}}
    {{P_c}^{m1}}\xrightarrow{\text{\footnotesize $MDG_c$}}{{P_c}^{m2}}
    \end{aligned}
\end{equation}
where $a\rightarrow~b$ denotes {\em dynamic} dependence of $b$ on $a$ with respect to the (static)
dependence graph $MDG_c$ and the method-execution events in $P_c$.
This dynamic dependence is not simply determined through reachability on the graph, though, to avoid
imprecision induced by straightforward transitive dependence computation.
Instead, the dependence is computed iteratively according to the semantics of method-execution events while traversing the graph (as described in Section~\ref{subsubsec:impactcomp}).
%We elaborate on Detail The algorithm for integrating
Applying this constrained partial order relation to Equation~\ref{eq_is} leads to a more
precise dependence inference through pruning false positives within individual processes, as detailed below.

Given a query method $q$ defined in component $c$ and executed in process $P_c$,
methods that execute after $q$ but are not dynamically dependent on $q$ will be pruned
from the trace of $P_c$.
{\bf For example}, in the client process of program $E$, {\tt C::shuffle} executed after {\tt C::compute}.
Yet, on the MDG for the client component of $E$, {\tt C::shuffle} does not depend on {\tt C::compute}. Thus,
{\tt C::shuffle} will be pruned from the (forward) dependence set of {\tt C::compute} that would
be produced as per the basic inference.
To apply similar pruning based on static dependencies for
every other component $c'$ and corresponding process $P_{c'}$, we
need to identify a method similar to $q$ that serves as the starting point of
the dynamic dependence/impact propagation (noted as {\em spark method}).
We safely choose the first method in $P_{c'}$ that executed after $P_{c'}$ receives a message %directly, or transitively via other processes
from $P_c$ as the spark method of $c'$ for query $q$.
{\bf For example}, if the query is in the client component of program $E$, {\tt S::getMax} would be the spark method for the server
component of $E$.
Then, methods that are partial-ordered after but not dynamically dependent on the spark
method will be pruned from the trace of $P_{c'}$.

\subsubsection{Pruning with Statement Coverage}\label{subsec:stmtcov}
Conceptually, the intra-component dependencies in {\tech} aim to directly model dependencies at method level.
However, in terms of the concrete (graph) representation of these dependencies,
there may be multiple edges (i.e., interprocedural dependencies) between two methods (e.g., data dependencies each
due to the passing of a different parameter from the caller to the callee).
These edges are not %simply
conflated into a single dependence edge.
The rationale is that, by doing so, {\tech} can avoid imprecision accumulation during transitive dependence
propagation across methods, threads, and processes.
Specifically, all the individual interprocedural dependencies computed at statement level are kept to
represent method-level dependencies.
Thus, for any two nodes of an $MDG_c$, there may be multiple edges. {\tech} further uses edge annotations to
denote the type of interprocedural dependence for each edge (e.g., {\em interference}, {\em synchronization},
{\em traditional control}, etc.). These edge annotations are used later in the precise transitive dynamic dependence
%computation (see Section~\ref{subsubsec:impactcomp}).
analysis algorithm of {\tech}.

Given this underlying representation of the static dependencies in {\tech}, our framework provides an additional option of
collecting statement coverage data and utilizing the data to further prune possible spurious dynamic dependencies.
A method $m2$ is considered to (dynamically) depend on a method $m1$ if there is at least one static interprocedural dependence
of $m2$ on $m1$ on the corresponding $MDG_c$ that has both statements associated with the dependence %(i.e., both ends of the dependence edge)
covered during the execution analyzed by our framework. Otherwise, the method-level dependence between $m1$ and $m2$ should be pruned.
This pruning rule can be formally defined as another constrained partial
order relation as follows:
\begin{equation}\label{eq_constraint3}
\vspace{-5pt}
%{\scriptsize
{{P_c}^{m1}}\prec{{P_c}^{m2}}:= %\Longleftrightarrow
    \begin{aligned}
    T_L({{P_c}^{m2}})\geq T_F({{P_c}^{m1}})~\wedge~\\
    %{{P_c}^{m1}}\overset{MDG_c}{\longrightarrow}{{P_c}^{m2}}
    {{P_c}^{m1}}\xrightarrow[\text{at lease one covered}]{\text{\footnotesize $MDG_c$}}{{P_c}^{m2}}
    \end{aligned}
\end{equation}
where the notations are the same as in Equation~\ref{eq_constraint2}. The only difference is that {\em at least one} of
the interprocedural dependencies between $m1$ and $m2$ on $MDG_c$, in addition to their existence,
must be exercised ({\em covered}) during the execution.
{\bf For example}, in the server process of the working example program $E$, {\tt S::getMax} executed after {\tt S::serve} and the MDG of the server component also indicates that {\tt S::getMax} (statically) depends on {\tt S::serve} (because the former uses variable $s$ which is defined in the latter). Now suppose the statement {\tt char r = getMax(s);} is not covered (e.g., suppose there is an unexercised predicate guarding this statement) during the analyzed execution.
In this case, {\tt S::getMax} would be pruned from the (forward) dynamic dependence set of {\tt S::serve} that would be produced as per the {\em incorporating intra-component dependencies} inference rule (i.e., Equation~\ref{eq_constraint2}).
Note that like the pruning based on static intra-component dependencies, the pruning rule here based on statement coverage only applies
within a single process (albeit possibly across multiple threads).
Also, similarly, applying this constrained partial order relation to Equation~\ref{eq_is} leads to yet another level of
precision of dynamic dependence abstraction in {\tech} due to the removal of possible false positives associated with
statements that are not covered in the analyzed execution.

%% file: technique.tex
%\section{Approach}
%\label{sec:tech}
%%\vspace{-2pt}
%% summary of approach
%To achieve an efficient dynamic impact analysis for distributed programs, % at method level, we
%{\distea} utilizes only lightweight run-time information such as method execution order.
%%we build {\distea} based on the order of method executions as a first step.
%%To that end, {\distea} leverages lightweight instrumentations to capture those method-execution events and partially
%%order them across all processes of the program, and infers execute-after relations among methods according to
%%the partially-ordered events to compute potential impacts of any query.
%%
%We first %define the essential concepts and
%present the fundamentals underlying our approach, including
%the definition of method events used by {\distea} and its rationale for impact prediction.
%%from sequences of those events.
%Then, we give an overview and illustration of the inner workings, %of {\distea} %works as a whole.
%%Finally, it describes details of the analysis algorithms of {\distea}.
%followed by details on the analysis algorithms, of the {\distea} approach.

%\subsection{Essential Concepts} %{Main Concepts} %{Preliminaries}
%\subsection{Fundamentals}
%\section{Preliminaries}
%\section{Fundamentals}\label{sec:fundamentals}

\section{Framework Instantiations}\label{sec:tech}
%\vspace{2pt}
%\noindent
%\textbf{Instantiations.}
As we discussed in Section~\ref{sec:motive}, users would need variable levels of cost-effectiveness balances to accommodate different dependence-based task scenarios. Now that {\tech} aims to provide a dynamic dependence analysis framework that empowers a range of dynamic-dependence-based applications, it should have the capabilities for offering dynamic dependencies at variable cost-effectiveness levels so as to meet the diverse application needs.
To streamline this vision, {\tech} provides flexible options for users to enable/disable certain parts of its analysis data and some steps of its analysis algorithms, hence to accommodate different usage scenarios that need varied cost-effectiveness tradeoffs (e.g., some application tasks prioritize precision over efficiency, while some others accept
relatively rough/low-precision dependencies in exchange for high efficiency/scalability).
In particular, currently {\tech} unifies four instantiations, each corresponding to a version of
{\tech} that offers a distinct level of cost-effectiveness.
By presenting these different instantiations, we do not intent to investigate all possible levels of such tradeoffs; instead, the goal here is to
demonstrate the capabilities of our framework design in enabling varying dynamic dependence analysis of distributed programs to offer
variable levels of cost-effectiveness.
%
%To accommodate different usage scenarios that need varied cost-effectiveness tradeoffs,
%{\tech} provides flexible options for users to enable/disable certain analysis steps.
%The framework unifies four instantiations, each corresponding to a version of
%{\tech} that offers a distinct level of cost-effectiveness.

\setlength{\tabcolsep}{1.8pt}
\begin{table}[tp]
  \centering
  \caption{The four {\tech} instantiations defined by analysis data and dependence inference rules used}\vspace{-20pt}
    \begin{tabular}{|p{3.2em}|p{3.4em}|p{12em}|c|c|c|c|}
\cline{4-7}    \multicolumn{1}{r}{} & \multicolumn{1}{r}{} &   & \multicolumn{4}{c|}{Instantiation} \\
\cline{4-7}    \multicolumn{1}{r}{} & \multicolumn{1}{r}{} &   & {\bf Basic} & {\bf Msg+} & {\bf Csd+} & {\bf Scov+} \\
    \hline
    \multicolumn{3}{|c|}{dependence inference rule (Equation)} & \ref{eq_ea}  & \ref{eq_constraint}  &  \ref{eq_constraint2} & \ref{eq_constraint3} \\
    \hline
    \multirow{4}[1]{*}{\shortstack{analysis\\data}} & static & static dependencies & \xmark & \xmark & \cmark & \cmark \\
\cline{2-7}      & \multirow{3}[1]{*}{dynamic} & method-execution events & \cmark & \cmark & \cmark & \cmark \\
\cline{3-7}      &   & message-passing events & \xmark & \cmark & \xmark & \xmark \\
\cline{3-7}      &   & statement coverage & \xmark & \xmark & \xmark & \cmark \\
    \hline
    \end{tabular}%
  \label{tab:instances}%
\end{table}%

Table~\ref{tab:instances} outlines the four instantiations (2nd row) of {\tech}, defining each
in terms of the analysis data (static and/or dynamic, 4th--7th rows) and dependence inference rule used (3rd row).
The rationale for having these four is two-fold. %as also illustrated by the table.
First, the {\bf Basic} version uses only one form of
data: method-execution events. While used to partial order these events, message-passing events are not further utilized in the analysis algorithm for this instantiation. Second, from the 5th to 7th columns, each instantiation
%monotonically adds one more form of static/dynamic data to the previous one.
aims to enhance the {\bf Basic} version in a different way, by adding message-passing events only, static dependencies only, and
statement coverage along with static dependencies.
Statement coverage needs to go with static dependencies, because it cannot be used along with
other forms of data in our framework---the static dependencies are the only other form of data with statement-level details.
There could be more instantiations of {\tech}. We study these four as they have intuitively more {\em differentiable} cost-benefit tradeoffs than other possible instantiations, and because these four would suffice for the purpose of demonstrating the capabilities of our framework in offering diverse cost-effectiveness levels. 

\begin{itemize}%[leftmargin=14pt]
\item{\textbf{Basic} version.} This is the simplest, most lightweight instantiation of {\tech}, which computes
    dynamic dependencies among methods simply using the \textbf{basic} inference (Section~\ref{subsec:basicinfer}). It only utilizes method-execution events and their global partial ordering, essentially a dependence approximation purely based on inter-process control flows.
\item{\textbf{Msg+} version.} This instantiation exploits the \textbf{m}e\textbf{s}sa\textbf{g}e-passing semantics to prune false-positive dynamic dependencies {\em across} processes in the {\bf Basic} version (Section~\ref{subsec:messagesemantics}). It utilizes method-execution events and per-process message-receiving maps, essentially a dependence approximation based on inter-process control {\em and} data flows. %and inter-process data flows.
\item {\textbf{Csd+} version.} This version of {\tech} enhances the {\bf Basic} version by incorporating intra-\textbf{c}omponent \textbf{s}tatic \textbf{d}ependencies to prune false-positive dynamic dependencies {\em within} individual processes (Section~\ref{subsec:intradep}). It utilizes method-execution events and per-component static dependence abstraction, essentially a dependence approximation based on {\em inter-process} control flows and {\em intra-component} (data and control) dependencies.
\item {\textbf{Scov+} version.} This is a further refinement of the {\bf Basic} version, incorporating statement coverage in {\bf Csd+} to prune false-positive dynamic dependencies {\em within} individual processes (Section~\ref{subsec:stmtcov}). In essence,
    {\bf Scov+} utilizes both method- (i.e., execution events and static dependencies) and statement-level data for a method-level dynamic dependence abstraction that is supposedly more precise than {\bf Csd+}.
\end{itemize}

\begin{table}[htbp]
  \centering
  \caption{Cost-effectiveness tradeoffs of {\tech} instantiations}
    \begin{tabular}{|c||c|c|c||c|}
    \hline
    \shortstack{{\tech}\\Instantiation} & \shortstack{static analysis\\time} & \shortstack{runtime\\overhead} & \shortstack{querying\\time} & \shortstack{effectiveness\\(precision)} \\
    \hline
    {\bf Basic} &       & \cmark     & \cmark     & \xmark \\
    \hline
    {\bf Msg+}  &       &       & \xmark     & \cmark \\
    \hline
    {\bf Csd+}  & \xmark     &       & \xmark     & \cmark \\
    \hline
    {\bf Scov+} & \xmark     & \xmark     & \xmark     & \cmark \\
    \hline
    \end{tabular}%
  \label{tab:instancecosteff}%
\end{table}%
To further justify our decision for having these different instantiations of our framework, Table~\ref{tab:instancecosteff} shows what the cost-effectiveness factors are that each instantiation attempts to trade for and off: \xmark ~indicates a factor that is traded off (i.e., compromised) while \cmark indicates a factor that is traded for (i.e., prioritized).
For instance, the {\bf Basic} version trades effectiveness (in terms of precision) for efficiency in terms of both runtime overhead and querying time,
while {\bf Scov+} trades all the efficiency factors for greater effectiveness.
Note that in general which factor is considered {\em traded off} versus which is considered {\em traded for} is a relative notion: for example, relative to {\bf Basic}, {\bf Msg+} trades efficiency for effectiveness, but relative to {\bf Csd+}, {\bf Msg+} trades effectiveness for efficiency.
Here in Table~\ref{tab:instancecosteff}, the tradeoff made by {\bf Basic} is determined based on its nature (suffering very low precision for achieving high efficiency), while the tradeoff made by the other three instantiations is determined all relative to {\bf Basic}.

\begin{figure}[t]
  \vspace{-0pt}
  \begin{center}
  \includegraphics[scale=0.8]{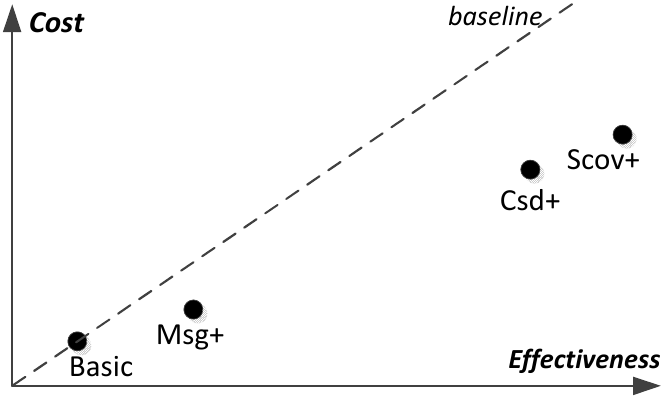}
  \vspace{-6pt}
  \caption{The cost-effectiveness space explored by {\tech} through its four instantiations. The coordinates of these instantiations are not based on empirical measurements but only provide estimated contrasts among them in terms of their cost-effectiveness levels.}
  \label{fig:designspace}
  \end{center}
  \vspace{-0pt}
\end{figure}
To complement Table~\ref{tab:instancecosteff} which gives the factors involved in the tradeoff made by each instantiation, Figure~\ref{fig:designspace} further gives estimated, relative comparisons of cost-effectiveness levels among the four instantiations---the coordinate for each instantiation does not correspond to its exact cost-effectiveness measure.
The dashed line ($y=x$) represents the baseline, indicating the level of cost-effectiveness achieved by the {\bf Basic} version. Accordingly, a point below this line means the associated analysis is better than the baseline, and the further the point is from the line the more cost-effective the associated analysis.
As depicted, through its varied instantiations each representing a distinct level of cost-effectiveness, our framework provides variable cost-effectiveness tradeoffs as a whole, meeting diverse application needs.

%% file: analysisalgorithms.tex
%!TEX root = paper.tex
%  \vspace{-6pt}
\subsection{Analysis Algorithms}\label{sec:depcomp}
%\vspace{-8pt}
%\subsection{Analysis Algorithms}
%\vspace{-2pt}
In this section, we present the unified dependence analysis (abstraction) algorithm in {\tech}.
It is referred to as {\em unified} because it generalizes all of the varied instantiations of the framework, using
different dependence inference rules according to the different selections of analysis data.
For obtaining each form of static/dynamic data, {\tech} has a dedicated algorithm as well. Among those algorithms,
we present in detail two online algorithms---one for monitoring message-passing events and the other for
partial ordering method-execution events, given their relative sophistication compared to other data-gathering algorithms.
%as other data-gathering algorithms are drawn from previous work as discussed in Section~\ref{sec:data}.

\subsubsection{Monitoring Message-Passing Events}\label{subsec:ltsalgo}
%At the core of {\distea} is the maintenance of causality among method events across distributed system components, and that causality analysis is based on the order in which those method events occur based on a system of logic clock.
%It is important to note that the order is only a partial ordering in distributed systems~\cite{lamport1978time}.
%Maintaining this order is realized through monitoring communication events in {\distea}.
%
%Preserving the partial ordering of internal events is at the core of {\distea}, for which
%two main options exist, both based on %for maintaining the partial ordering on top of
%a logical notion of time: the {\LTS} approach~\cite{lamport1978time} as described before, and vector clocks~\cite{fidge1988timestamps,mattern1989virtual}.
%In comparison, {\LTS} is lighter-weight as it just maintains a single counter as the logical clock for each process,
%while a vector clock keeps an array of clocks for all processes. %of information, such as process context which can be used for deriving concurrency relationships among communicating processes.
%%Yet, {\distea} does not use such information for the impact inference purpose, at least in its current design. Therefore, we adopt LTS in {\distea} for partially ordering method events in multiprocess
%%Since {\distea} does not use those extra information for impact inference (in its current design at least),
%Therefore, we adopt {\LTS} in this work since it suffices for the current {\distea} design. %for the ordering.
%
Preserving the partial ordering of method-execution events is at the core of {\distea}, since the partial-ordered
method-execution events are the default form of data used by any instantiation of the framework. %(see Table~\ref{tab:instances}).
For that purpose, we adopt the \emph{Lamport time-stamping} ({\LTS}) approach~\cite{lamport1978time} %as described before
based on a logical notion of time.
The idea is to maintain a synchronized logic clock across processes of the system execution analyzed, through
monitoring message-passing events. From these events, the {\em message-receiving map} is also derived, but only for the {\bf Msg+} version
 which uses the data to prune spurious dependencies resulting from the basic inference.
%Based on the original {\LTS} algorithm (Section~\ref{subsec:lts}), to observe the partial ordering {\distea} needs to
%maintain a logic clock for each process and exchange the logic clocks among processes when
%communicating data (messages) among them.
%
%
%We give more details on adapting this algorithm for impact analysis as part of the complete {\distea} algorithm in Section~\ref{algo}.
%the partial ordering maintenance consists of the following three steps, using the notations previously defined:
%\begin{enumerate}
%\item Create a logic clock $C_i$ for each process $P_i$ and initialize $C_i$ to 0.
%\item Upon the occurrence of each internal event $E_I$($q$), increment by 1 the logic clock for the hosting process of $q$.
%\item Upon the occurrence of each communication event $E_C$($c1$,$c2$), when sending the message to $c2$, $c1$ timestamps that message using the logic clock of its hosting process; when receiving the message from $c1$, $c2$ extracts the timestamp attached to that message, updates the timestamp of the hosting process of $c2$ to the greater of the two timestamps, and then further increments its local timestamp by 1.
%\end{enumerate}
%\subsection{Algorithms}
%
\setlength{\intextsep}{0pt}
\setlength{\textfloatsep}{0pt}% Remove \textfloatsep
%\vspace*{6pt}
\begin{algorithm}[tp]
\vspace{-0pt}
%\label{algo}
\scriptsize{
    %\caption{Transferring logic clocks}
    \caption{Monitoring message-passing events}
    %\footnotesize{let $C$ be the global counter used for tracking the internal events} \\
    \footnotesize{let $C$ be the logical clock of the current process} \\
    \footnotesize{remaining = 0 // remained length of data to read}
    \label{algo}
    \begin{algorithmic}[1]
        \Function{SendMessage}{$msg$}// on sending a message $msg$
            \State $sz$ = length of $sz$ + length of $C$ + length of $msg$
            \If{using the {\bf Msg+} version}
                \State $sz$ += length of the sender process id $sid$
                \State pack $sz$, $C$, $sid$, and $msg$, in order, to $d$
            \Else
                \State pack $sz$, $C$, and $msg$, in order, to $d$
            \EndIf
            \State write $d$
        \EndFunction
        \Function{RecvMessage}{$msg$}// on receiving a message $msg$
            \State read data of length $l$ into $d$ from $msg$
            \If{remaining $>$ 0}
                \State remaining -= $l$
                \State\Return $d$
            \EndIf
            \State retrieve and remove data length $k$ from $d$
            \State retrieve and remove logical clock $ts$ from $d$
            \State remaining = $k$ - length of $k$ - length of $ts$ - $l$
            \If{$ts$ $>$ $C$} %\Comment update local clock if the received clock is greater
                \State $C$ = $ts$
            \EndIf
            \State increment $C$ by 1
            \If{using the {\bf Msg+} version}
                \State retrieve and remove sender process id $sid$ from $d$
                \State add ($sid$, $C$) to message-receiving map if $sid$$\notin$its key set
                \State remaining -= length of $sid$
            \EndIf
            \State\Return $d$
        \EndFunction
    \end{algorithmic}
}
\vspace{-0pt}
\end{algorithm}

Algorithm~\ref{algo} summarizes %in pseudo code
our algorithm for %partially ordering internal events
monitoring message-passing events while carrying and synchronizing the logic clocks
based on
the original {\LTS} algorithm. %((Section~\ref{subsec:lts})).
The logical clock of the current process $C$ is initialized to 0 upon process start, as is the global
variable {\tt remaining}, %although initialized to 0,
which tracks the remaining length of data most recently
sent by the \emph{sender} process. The rest of this algorithm consists of two parts,
trigged upon the occurrence of IPC (i.e., message-passing) events during system executions.
%This is an online algorithm as its different parts will be trigged upon the occurrence of communication events during an execution of the system.
%in all communicating processes of the system

%corresponding to two runtime monitors instrumented at each point of the system where a communication event may occur.
The first part is the run-time monitor \Call{SendMessage}{} trigged online upon each message-sending event.
The monitor piggybacks (prepends) two extra data items to the original message: the total length $sz$ of the data to send,
and the present value of the local logical clock $C$ (of this \emph{sender} process) (lines 2--7); then, it
sends out the packed data (line 8). The sender process id is also packed in the {\bf Msg+} version (lines 3--5).

The second part is the other monitor \Call{RecvMessage}{}, which is trigged online upon each message-receiving event.
After reading the incoming message into a local buffer $d$ (line 10), the monitor decides whether to simply update the
size of remaining data and return (lines 11--13),
or to extract two more items of data first: the new total data length to read,
and the logical clock of the peer \emph{sender} process (lines 14--15).
In the latter case, the two items are retrieved, and then removed also, from the entire incoming message.
Next, the remaining data length is reduced by the length of the data already read in this event (line 16), and
 the local logical clock (of this \emph{receiver} process) is compared to the received one, updated to the greater, and incremented by 1 (lines 17--19). In the {\bf Msg+} version, the sender process id is retrieved as well (line 21) and
 a new entry is added to the message-receiving map if this is the first message received from that sender process (line 22).
 Lastly, the monitor returns the message as originally sent in the system
 (i.e., with the prepended data taken away).
 %%%% --- response to ase'16 reviewer#2's concerns --------
 %
 %what if l>remaining?
 %
 %Initially when remaining==0, l>remaining must be true but remaining will
 %be set at line 16. Later, each of the reads that consume different parts of the piece
 %of data sent by the sender in one write cannot receive data of length greater than
 %remaining, i.e., l>remaining cannot be true.
 %
 %what if k and ts did not fit into the buffer at the same
 %time? E.g. buffer size is 64 bits, same as the size of the data length
 %variable?
 %
 %we simply allocate a buffer of 16bytes to hold the two data-length (integer) variables.
 %

To avoid interfering with the message-passing semantics of the original system, {\distea}
keeps the length of remaining data (with the variable {\tt remaining} in the algorithm)
to determine the right timing for logical-clock (and sender id) retrieval.
In real-world distributed programs (e.g., Zookeeper~\cite{zookeeper}), %as we studied),
it is common that %between two communicating processes the
a \emph{receiver}
process may obtain, through several reads, the entire data sent in a single write by its peer \emph{sender} process.
For example, a first read just retrieves the data length so that an appropriate size of memory can be allocated to
take the actual data content retrieved in a second read.
%Since only one pair of the prepended data items is sent in that single write, %logic clock can be retrieved from the head of the entire message received,
%only the first read by the \emph{received} process could successfully extract that pair.
Therefore, not only is it unnecessary to attempt retrieving
the prepended data items (data length and logical clock) in the second read
since the first one should have already done so, but also such attempts can break the original network I/O protocols.
{\distea} addresses this issue by piggybacking the length of data to send and tracking the remaining length of data to
receive.

\subsubsection{Partial Ordering Method-Execution Events}\label{subsec:easalgo}
The {\bf Basic} dependence inference in {\distea} relies on the execution order of methods that is
deduced from the timestamps attached to all method-execution events.
%for which {\distea} monitors the occurrence of each internal event.
These timestamps are calculated based on each local logic clock, which is synchronized through {\tech}'s monitoring
message-passing events as described earlier.

%However, %, which can be easily retrieved based on the timestamps attached to those events.
As proved in~\cite{apiwattanapong05may}, recording the \emph{first} entry and \emph{last} returned-into
(or return) events only is equivalent to tracing the full sequence of those events for
 %capture the essential information for execute-after relation inference and
our dynamic dependence inference that is purely based on the method execution order.
%Similarly, this equivalence also applies in {\distea}.
%It is also worth emphasizing that {\distea} does not need to track the full sequence of events, but instead record the first and last counters per method only, which can be easily deduced from this table.
Thus, instead of keeping the timestamp for every method-execution-event occurrence (as shown in Table~\ref{tab:examplelts}),
for the {\bf Basic} version, {\distea} only %it suffices to bookkeep just
records two key timestamps for each method $m$: %in {\tech}:
the one for the first instance of $m_e$, and the one for the last instance of $m_i$ or $m_x$,
whichever occurs later. %, just as done by {\EAS} when considering multithreaded executions~\cite{apiwattanapong05may}.
As per Equation~\ref{eq_constraint}, this simplified tracing also suffices for the {\bf Msg+} version.
For the {\bf Csd+} and {\bf Scov+} versions, however, the full sequence of (i.e., instance-level) method-execution events is required for
effectively pruning false positives with intra-component %impact relations
dependencies~\cite{cai14diver,cai2016method}---the hybrid dependence analysis algorithm needs to traverse each
instance of every method-execution event to {\em precisely} recognize false positives.
Thus, the online monitor of method-execution events records either the two key timestamps for each method, or
the timestamp for every method-execution event, depending on which version is used. %chosen by the user.

%Accordingly, the online algorithm for monitoring internal events uses two counters %per method
%to record the two key timestamps for each method, similar to what {\EAS} does but different in
%that it does so \emph{in each process} rather than in just one process.
For each process, we use an integer counter for time-stamping, which
is updated using the per-process logical clock.
%Also, we use the per-process logical clock, instead of a global integer as used by {\EAS},
%to update the per-method counters during runtime.
Meanwhile, the logical clock $C_i$ of each process $P_i$ is maintained as follows:
%(1) It is initialized to 0 upon the start of current process;
%(2) It is increased by 1 upon each internal event;
%and (3) It is updated upon each communication event in the way as shown in Algorithm~\ref{algo}.
%We omit the pseudo code as it is straightforward.
\begin{itemize}[leftmargin=14pt]
\denseitems
\item Initialize $C_i$ to 0 upon the start of $P_i$.
\item Increase $C_i$ by 1 upon each method-execution event in $P_i$.
\item Update $C_i$ upon each message-passing event occurred in $P_i$ via the two online monitors shown in Algorithm~\ref{algo}.
\end{itemize}
Finally, for the offline dependence %impact
computation in {\distea}, the online algorithm here also dumps per-process method-execution-event sequences %(i.e., the two timestamps for each executed method)
as traces upon the %the event of
program termination (of each component).
Additionally, in the {\bf Msg+} version, the message-receiving maps are also serialized as part of the per-process traces.
The {\bf Scov+} version further carries statement-coverage information in the per-process traces.

%\subsubsection{Trace Processing and Impact Computation}
%
%\subsubsection{Impact Computation}\label{subsubsec:impactcomp}
\subsubsection{Dynamic Dependence Abstraction}\label{subsubsec:impactcomp}

\setlength{\textfloatsep}{0pt}% Remove \textfloatsep
\begin{algorithm}[!t]
%\label{algo}
\scriptsize{
    %\caption{Transferring logic clocks}
    %\caption{Computing dependence sets}
    \caption{Dependence abstraction}
    %\footnotesize{let $C$ be the global counter used for tracking the internal events} \\
    \footnotesize{let $P_1,P_2,...,P_n$ be the $n$ concurrent processes of the system} \\
    \footnotesize{let $q$ be the query method} \\
    \footnotesize{let $MDG[P_i]$ be the MDG of the component executed in $P_i$} \\
    \footnotesize{let $AM[P_i]$ be the spark method of $P_i$ with respect to $q$}
    \label{algo2}
    \begin{algorithmic}[1]
        \State{\emph{locDS} = $\emptyset$, \emph{extDS} = $\emptyset$, \emph{comDS} = $\emptyset$}
        \For{$i$=1 to $n$}
            %\State \emph{locIS} $\cup=$ computeLocalImpact($q$)
            %\If{using the csd+ version}~~$mdg$ = $MDG[P_i]$~~\textbf{else}~~$mdg$ = null
            %\EndIf
            \State $mdg$ = (using the {\bf Csd+} or {\bf Scov+} version)?$MDG[P_i]$:null
            \If{using the {\bf Scov+} version}
                \State $covStmts$ = obtainCovStmts(\emph{tr($P_i$)}, $MDG[P_i]$)
                \State updateMDG ($mdg$, $covStmts$)
            \EndIf
            %\State $scovdata$ = (using the scov+ version)?$MDG[P_i]$:null
            \State $ts_q$ = computeIntraDS($q$, \emph{locDS}, \emph{tr($P_i$)}, $mdg$)
            \If{$ts_q$==null}~~\textbf{continue}
            \EndIf
            \For{$j$=1 to $n$}
                \If{$i==j$}~~\textbf{continue}
                \EndIf
                \If{using the {\bf Csd+} or {\bf Scov+} version}
                    \If{using the {\bf Scov+} version}
                        \State $covStmts$ = obtainCovStmts(\emph{tr($P_j$)}, $MDG[P_j]$)
                        \State updateMDG ($MDG[P_j]$, $covStmts$)
                    \EndIf
                    \State computeIntraDS($AM[P_j]$, \emph{extDS}, \emph{tr($P_j$)}, $MDG[P_j]$)
                    \State \textbf{continue}
                \EndIf
                \If{using the {\bf Msg+} version~$\wedge$~\emph{S(tr($P_j$))}[$P_i$]==null}
                    \State \textbf{continue}
                \EndIf
                \For{{\bf each} method $m$$\in$ \emph{keyset(R(tr($P_j$)))}}
                    \If{\emph{R(tr($P_j$))}[$m$]~$\ge$~$ts_q$}
                        \If{using the {\bf Msg+} version}
                            \If{\emph{R(tr($P_j$))}[$m$]~$\ge$~\emph{S(tr($P_j$))}[$P_i$]}
                                \State \emph{extDS}~$\cup$=~\{$m$\}
                            \EndIf
                        \Else
                            \State \emph{extDS}~$\cup$=~\{$m$\}
                        \EndIf
                    \EndIf
                \EndFor
            \EndFor
        \EndFor
        \State \emph{comDS}~=~\emph{locDS}$~\cap~$\emph{extDS}
        \State\Return \emph{locDS}, \emph{extDS}, \emph{comDS}
    \end{algorithmic}
}
\vspace{-0pt}
\end{algorithm}
%{\distea} computes impacts based on execute-after relations between the queried method
%and each candidate method possibly impacted. These relations are identified from traces of internal events, which
%% collects the execution trace of internal events
%are produced by all components of the system in concurrent processes at runtime, and generally
%on spatially distributed machines.
%
During system executions, the online method-execution-event monitor
generates event traces concurrently (and typically on distributed machines).
Since it computes dependencies %impacts
offline, {\distea} gathers these traces after
%they are produced
their completion
to one machine, and computes dependence %impact
sets there as outlined in Algorithm~\ref{algo2}.
For the {\bf Csd+} version, the MDGs of individual components are also utilized; and
statement coverage is further computed and utilized on top of the MDGs for the {\bf Scov+} version.
For a detailed analysis, we refer to the process where the query is first executed as \emph{local process} versus all
other processes as \emph{remote process}, and dependencies (represented by the dependent methods) in local and remote processes as
\emph{local dependencies} and \emph{remote dependencies}, respectively.
For a given query $q$, {\tech} computes its dependence set
as three subsets: \emph{local dependence set}, \emph{remote dependence set}, and their intersection called \emph{common dependence set} (denoted as \emph{locDS}, \emph{extDS}, and \emph{comDS}, respectively, in Algorithm~\ref{algo2}).

The algorithm takes the query $q$, $n$ per-process traces, and $n$ per-component MDGs as inputs, and outputs the three subsets (line 27)
all initialized as empty sets (line 1). Per-process spark methods can be readily derived from the message-receiving maps.
The algorithm traverses the $n$ processes (loop 2--25) taking each as the local process (line 2) against
all others as remote processes (lines 9--10) to first compute the local dependence set (line 4).
Then,  the remote dependence set is computed (loop 11--25) based on Equation~\ref{eq_constraint2} if {\bf Csd+} or {\bf Scov+}
is used (lines 11--16), or on Equation~\ref{eq_constraint} if {\bf Msg+} is used (lines 17--25).
In particular, if the {\bf Scov+} is used, the static dependencies need to be pruned based on statement coverage before they are
used for local (lines 4--6) or remote (lines 12--14) dependence abstraction, as per Equation~\ref{eq_constraint3}.

The subroutine {\tt computeIntraDS} computes the dependence set (to return via the second argument) of the given query (taken as its first argument) {\em within} the given process (indicated by the third argument).
If the MDG of the component associated with this process is given as null (as the last argument), it will be simply ignored by the
subroutine, which will identify as dependants the methods whose last execution is not earlier than
the first execution of the query as in {\EAS}~\cite{apiwattanapong05may}.
Otherwise (i.e., in {\bf Csd+} or {\bf Scov+} version), the MDG will be utilized
along with the trace for the process (given as the third argument) to compute the dependence set using {\diver}~\cite{cai14diver}.
The key ideas of {\diver}'s hybrid dependence analysis are summarized below for self-containing purposes.
%we previously developed .

Generally, the {\diver} computation carefully checks whether a method-execution event $e$ in the trace leads the dependence (impact) originated in the query to propagate to the method associated with $e$ by referring to the MDG while using different propagation rules.
%The rules are based on (1) whether $e$ is an entry event or returned-into event, and (2)
The rationales underlying these rules are (1) through static dependencies induced by parameter or return-value passing,
impact can only propagate between two adjacently executed methods, and (2) through other kinds of static dependencies (those induced by definition-use relations %with respect to
between heap variables), the dependence can propagate from a method $a$ to any method that
executed after $a$. %Further details on these rules can be found in~\cite{cai2016method}.
The subroutine {\tt computeIntraDS} unionizes the newly computed dependence set with the one passed in, and
returns the first execution event time $ts_q$ of $q$ if $q$ is found in the input trace
and {\tt null} otherwise (line 8).

For the {\bf Scov+} version, subroutine {\tt obtainCovStmts} is needed to compute the covered statements (line 5)
from the coverage information contained in the trace of the given process (passed in as the first argument).
For the sake of efficiency, {\tech} instruments for and monitors covered branches in previous phases.
Now in the dependence abstraction, covered statements are derived according to the branch coverage and control dependencies
on the MDG for the component corresponding to the given process (passed in as the second argument).
Then, the third subroutine {\tt updateMDG} is invoked, which prunes dependencies with at least one statement not included in
the covered statements (line 6).

For the {\bf Csd+} or {\bf Scov+} version, the external dependence set is computed similarly (line 15), but using the spark method
of a remote process $P_j$ as the query (the first argument) and always taking the MDG of the component
that $P_j$ executes (the last argument). Computation of statement coverage and MDG pruning with the result for
the remote process (lines 13--14) are both similar to those for the computation of local dependencies.

Further, for the {\bf Msg+} version,
\emph{tr(P)} denotes the trace of process $P$, \emph{R(t)} denotes the
hashmap from each executed method to its last execution event time in trace $t$, \emph{S(t)} denotes the hashmap from
the id of each sender process to the event time when the remote process receives the first message from the sender, and
\emph{keyset(.)} returns the key set of a hashmap. If the (remote) process $P_j$ never received
any message from the (local) process $P_i$, no remote impacts would be found in $P_j$ (lines 17--18).

For a single test input, the dependence-computation algorithm computes the dependence set of one method at a time; for
multiple methods in the query set, the result is the union of all the one-method dependence sets computed separately (e.g., in parallel).
Similarly, the dependence set for multiple tests is the union of all per-test dependence sets. These treatments are similar to those adopted in union slicing~\cite{beszedes02oct}.

%With the enhanced version, we focus on safely reducing only the remote impact sets
%produced by the basic version.
%A complementary approach would be to reduce local impacts, such as through static-dependence-based pruning
%as in~\cite{cai14diver}. It is straightforward to incorporate such approaches in {\tech} to further
%improve its effectiveness, but that will largely increase the cost as well~\cite{cai15diverplus}.

\vspace{4pt}
\noindent
\textbf{Analysis soundness and result safety.}
The soundness of {\tech} relies on that of its static and dynamic analysis.
Our discussion and characterization on the analysis soundness and results safety of {\tech} refer to relevant definitions and discussions in~\cite{jackson2000software}: a sound static analysis produces information that holds for all possible program executions, while a
sound dynamic analysis produces information that holds for the analyzed execution alone.
As per these definitions, our static analysis is not sound, because of its inability to deal with dynamic language constructs in Java (e.g., reflection,
native code invocation via JNI, etc.)---the analysis does not see the relevant code at compile time thus its results may not hold for the executions involving such code.

In fact, sound static analysis for a language that allows for dynamic code constructs is generally rare~\cite{livshits2015defense}.
On the other hand, except for the dynamic code features that are widely known as statically unanalyzable, it is still important for a static analysis to be as sound as possible; thus, the notion of {\em soundiness} is introduced in~\cite{livshits2015defense}: a static analysis is soundy if it aims to be as sound as possible without excessively compromising precision and/or scalability; and soundy is the new (de facto) sound.
As per this definition, the static analysis in {\tech} is soundy because it does over-approximate all the common Java language features except for the known dynamic code constructs (reflection, JNI).

Also per the definition in~\cite{jackson2000software}, our dynamic analysis that only uses dynamic data is sound as it captures all the dynamic dependencies (at method level) exercised in the analyzed executions and does so always in a conservative manner: for instance, the {\bf Basic} version over-approximates
dynamic dependencies between two methods just according to their happens-before relationships.
However, in two instantiations of {\tech}, {\bf Csd+} and {\bf Scov+}, the dynamic analysis utilizes static dependencies also, which are produced by the static analysis that is soundy; thus, the resulting dynamic dependencies may not hold for some executions (i.e., those involving dynamic code constructs).
Therefore, considering all of its instantiations, {\tech} is soundy.
Note that while the advanced (more precise) versions of {\tech} (relative to the {\bf Basic} version) apply various pruning rules to improve precision, only dependencies that cannot possibly hold with respect to the analyzed executions are pruned---this conservative pruning maintains the soundiness of {\tech}.

Traditionally, the result of a sound analysis is considered {\em safe}.
Given soundy is the new sound~\cite{livshits2015defense}, we regard the resulting dependence sets of {\tech} as safe in light of the overall {\em soundiness} of our (hybrid) dynamic (dependence) analysis.
Note that here we use ``soundness/soundiness" as a property of an analysis while using ``safety" to characterize the results of the
analysis.
\vspace{4pt}
\noindent
\textbf{Illustrations.}
To illustrate %the inner workings of our framework with respect to these four instantiations,
%these analysis algorithms in {\tech},
the holistic analysis in {\tech},
consider $E$ of Figure~\ref{fig:csexample} as the program under analysis $D$.
Suppose the message-passing API list $L$ includes {\tt Socket::readLine}, {\tt Socket::writeLine}, {\tt Socket::readChar}, and {\tt Socket::writeChar}.
{\distea} first instruments $E$ and produces the instrumented server and client components $S'$ and $C'$.
For the {\bf Csd+} version, the MDGs for $S$ and $C$ are further constructed.
Then, suppose the components are deployed on two distributed machines, %hosts $A$ and $B$, respectively,
and $S'$ starts before $C'$. %before both processes execute concurrently,
%and that {\tt MySocket} we do not analyze library code,
When running concurrently (e.g., against an example input set $I$=\{``hello"\}), $S'$ and $C'$ generate two method-event sequences in two separate processes,
as listed \emph{in full}
in the first two and last two columns in Table~\ref{tab:examplelts}, respectively. After the last instance of event {\tt C::main}$_i$, the client prints 'o' (which is the maximum character in the input string).
%the event traces of this program will be what is shown in Table~\ref{tab:examplelts}.
As shown, logical clocks are updated upon message-passing events. For instance, the logical clock of the server
process is first updated to 10 upon the event $E_c$(C,S) originated in the client process, which is greater by 1 than the current logical clock of the client process. Later, the client logical clock is updated to 14 upon $E_c$(S,C). The method-execution events are time-stamped by these logical clocks while message-passing events are not, as marked by `-' (i.e., not applicable).
%
%Next, suppose the query set $M$=$\{${\tt S::getMax}$\}$, {\distea} merges event traces of the two processes and,
%by inferring impact relations from the time-stamped events, it gives
%%collected from both processes to the impact set will be
%$\{${\tt S::getMax}, {\tt S::serve}, {\tt S::main}, {\tt C::compute}, {\tt C::main}$\}$ as the impact set of $M$.
%As demonstrated, {\distea} can predict impacts across distributed components (processes). %of the input system.
%For example, if the developer plans for a change to method {\tt getMax} in the server, the methods {\tt compute} and {\tt main} in the client, in addition to the other two server methods, are potentially affected and, thus, need impact inspection by the developer before applying that change.
%%the change to the server code.

Next, suppose the query set $M$=$\{${\tt S::serve}$\}$, {\distea} gathers traces of the two processes and,
by inferring dynamic dependencies %impact relations
from the time-stamped events, the {\bf Basic} version gives
%collected from both processes to the impact set will be
$\{${\tt S::getMax}, {\tt S::serve}, {\tt S::main}, {\tt C::shuffle}, {\tt C::compute}, {\tt C::main}$\}$
as the dependence set of $M$. By exploiting message-passing semantics, the {\bf Msg+} version prunes {\tt C::shuffle} from this
dependence set according to Equation~\ref{eq_constraint}.
By utilizing the static dependencies of both components, the {\bf Csd+} version prunes
{\tt S::main} from the dependence set produced by the {\bf Basic} version according to Equation~\ref{eq_constraint2}.
In this simple example, since none of the methods contain branches, the fact that a method executed means that every statement in
that method is covered. Thus, statement coverage data did not lead to further reduction of the dependence set. As a result,
the {\bf Scov+} version produces the same dependence set as {\bf Csd+} does for the given query set.

Now let us respond to the challenge to the developer in the use scenario that motivated {\tech}.
As demonstrated, {\distea} can compute dynamic dependencies (hence predict impacts)
across distributed components (processes). %of the input system.
For an example application of these dependencies in impact analysis,
if the developer plans for a change to method {\tt serve} in the server, the methods {\tt compute} and {\tt main} in the client, in addition to the other server method ({\tt S::getMax}), are potentially affected and thus need impact inspection by the developer before applying that change.

%% file: implementation.tex
%!TEX root = paper.tex

\vspace{-3pt}
\section{Implementation}
 \label{sec:impl}\vspace{-3pt}
%{\distea} was implemented in Java
%\footnote{\vspace{-2pt}Download is available at \href{http://nd.edu/~hcai/distea}{\scriptsize \url{http://nd.edu/~hcai/distea}.}}
%
%{\tech}\footnote{\vspace{-2pt}Code and study results are  at~\href{https://chapering.github.io/distea/}{https://chapering.github.io/distea/}.}
%
%\footnote{\vspace{-2pt}Download link will be revealed in the camera-ready version.}
{\tech} consists of three main modules: a static analyzer, two sets of run-time monitors, and a post-processor.
%We discuss key implementation issues in each of these modules as follows in the context of the Java language.
%As the tool is currently implemented for Java programs, we limit our discussion to the context of the Java language.
%The entire {\distea} tool, including its library dependencies, is available online for download.
Careful treatments are crucial for a non-interfering implementation as recapped below.
We released the source code of {\tech} along with our empirical study results %are available
%at~\href{https://chapering.github.io/distea/}{https://chapering.github.io/distea/}.
at~\href{https://www.dropbox.com/s/4nldj8z6plzator/d2abs-artifact.zip?dl=0}{\underline{the {\tech} project archive}}.

%\subsection{Static Analyzer}
\vspace{4pt}
\noindent
\textbf{Static analyzer.}
%The primary role of the static analyzer is to identify proper locations in the input program
%and instrument there such that appropriate runtime monitors are to be triggered upon relevant method-execution events.
%%As another task of the static analyzer,
%Finding an accurate set of those locations for the instrumentation is crucial to the soundness and precision of our technique.
%In this static analysis, we utilized the Soot byte-code analysis framework~\cite{lam11oct} for both looking for instrumentation
%points and actually achieving the byte-code-level instrumentation.
%The primary role of the static analyzer is to
The static analyzer first instruments the input program such that
all relevant events are monitored accurately, which is crucial to the soundiness~\cite{jackson2000software,livshits2015defense} and precision of {\distea}.
We used Soot~\cite{lam11oct} %utilized the Soot static analysis framework~\cite{lam11oct}
for the instrumentation in two main steps.
%both looking for instrumentation
%points and actually achieving the byte-code-level instrumentation.
%
%The instrumentation consists of two major steps. %working at the level of Soot Jimple IR~\cite{lam11oct}.
First, {\distea} inserts probes for the three types of method-execution events in each method, for which we reused relevant
modules of {\diver}~\cite{cai14diver}, a hybrid impact analysis that is built on Soot and uses method-execution traces also.
%%, with
%%that for method-entrance events inserted at the method entry, and that for returned-into events
%%inserted after each call site and at the entry of each catch and finally block in the method.
%%those for method-entry and returned-into events inserted in the same way as {\EAS} did~\cite{apiwattanapong05may}, and
%%those for return events inserted before all
%as {\EAS} did~\cite{apiwattanapong05may}. Also, we continue to adopt the improvement for capturing
%return and returned-into events that the original {\EAS} algorithm would miss in the case of unhandled exceptions, as detailed
%in~\cite{cai15jss}.
%Moreover, we considered calls to {\tt System.exit} as an additional case of program return.
%
 %statements for a similar purpose.
The second step is to insert probes for message-passing events, for which {\distea} uses the
list $L$ of specified message-passing APIs to identify probe points: %based on string matching:
$L$ includes the %full
prototype of each API used in the input system for network I/Os.
%
%Among other user inputs, the message-passing API list $L$ is expected to give the full prototype of
%each unique API used in the input system for network-based I/O operations. This list facilitates the identification
%of instrumentation locations based on string matching, but it is optional.
If $L$ is not specified, a list of basic Java network I/O APIs is used
%by default, which %currently
covering two common means of blocking and non-blocking communication% , as well as our four experimental subject programs
: Java Socket I/O~\cite{javasocket} and Java NIO~\cite{javanio} (both are {\em socket-based}).
While not immediately addressing the {\em implicity challenge} to distributed program dependence analysis by itself, this API list is critical for
the effectiveness of dependence abstraction in our framework: it immediately affects whether {\tech} can completely and precisely capture
message-passing events hence how accurately it can model interprocess control and data flow.
Fortunately, it seemed to suffice that our implementation handles the most commonly used network I/O mechanisms
for message passing, at least in Java distributed software---for the diverse systems in our evaluation study, we did not need
to specify the list but just used the built-in support of our framework implementation.

%types of default list will be used.
%, which specifies the prototypes
%of APIs used in $D$ for message passing among its components (processes), is an optional input; if users do not specify $L$, a default list will be adopted. list of commonly used Java socket~\cite{javasocket} and NIO APIs~\cite{javanio}.
%
%monitoring method-execution events and partially ordering the distributed internal events.
%It takes either source code
%The rest of the static analysis is the same as {\EASc}~\cite{cai14mar}. Note that the first and last event counters per method are maintained separately from the logic clock per process, although both are integer counters. In essence,
%the first and last counters are updated upon internal events using the logic clock of the hosting process of the method associated with those per-method counters.
For the {\bf Csd+} version, the static analyzer proceeds with constructing the MDG for each component of
the input program, by reusing the algorithms for abstracting method-level dependencies
in single-process programs~\cite{cai2015abstracting,cai2016method}.
The MDG includes three classes of threading-induced dependencies: {\em ready}, {\em synchronization},
and {\em interfere} dependencies, computed partly in reference to the implementation of the
Indus project~\cite{giffhorn2009precise}.
The MDGs are serialized to disk, to be used by the post-processor.

%\subsection{Run-time Monitors}
\vspace{4pt}
\noindent
\textbf{Run-time monitors.}
%Corresponding to the probes inserted in the two steps of static analysis,
%two sets of runtime monitors are implemented accordingly.
%The first class is those for monitoring and timestamping internal events, which implement the algorithm described in
%Section~\ref{subsec:easalgo}.
%In addition, since some programs (e.g., service daemons) terminate only upon external kills,
%we hook in the monitor for program termination to ensure the output of method event sequences.
%%the serialization of method event sequences.
%The second class implements Algorithm~\ref{algo} for monitoring
% communication events and exchanging logic clocks among processes.
%%In particular, with respect to the cases in which reads do not mirror writes as discussed in Section~\ref{subsec:ltsalgo}.
The two sets of run-time monitors implement the two online algorithms: the first focuses on monitoring method-execution events
and the second is dedicated to preserving the partial ordering of them.
The first set again reuses relevant parts of {\diver}~\cite{cai14diver}.
For the second set, instead of invoking additional network I/O API calls to transfer logical clocks,
%transferring logic clocks via separate calls to the monitors in addition to the original network I/O API calls,
the monitors \emph{take over} the original message passing
so that they can %manipulate directly on the message (or data buffer) in order to
{piggyback} the three extra data items (i.e., the data length, logical clock, and sender id) to the original message.
%To that end, the probes for the monitors \emph{replace} the original network I/O API calls during the instrumentation.
%
In the message-passing event monitors, we carefully manage these extra data items with sophistication in order to support non-blocking
I/Os (e.g., Java NIO) and account for complications and variety in communication implementations
of real-world distributed systems (e.g., those that would cause interference with original communication semantics as discussed in Section~\ref{subsec:ltsalgo}).

\begin{comment}
%In other words,
%That is, the extra data items are carried on by the original message passing.
Our experience comparing different instrumentation schemes
% Our empirical tests, with our subject programs at least,
suggested that the piggyback strategy is more viable than inserting additional network I/O API calls, %to network I/O APIs,
especially when dealing with systems using selector-based non-blocking communications~\cite{artho2013software}.
%for non-blocking I/Os.
For instance, the ShiVector tool~\cite{abrahamson2014shedding,beschastnikh2014inferring}, which adopts the latter,
%additional I/O calls for vector-clock transferring,
%works with only two of our six subject programs (MultiChat and Voldemort).
did not work with most of our subject programs.
%to our subject programs, and with two of them (NioEcho and ZooKeeper)
%it did not work properly because of the interference caused by those additional calls.
One reason as we verified %with the two unworkable subjects
is that, for a pair of an original call and the corresponding additional call, the two messages may not be read in
the same order by the \emph{receiver} process as in which they are sent by the \emph{sender} process. As a result,
an original message-receiving call may encounter unexpected data in the message hence causing
%the original
network I/O protocol violations even system failures.
\end{comment}

%\subsection{Post-processor}
\vspace{4pt}
\noindent
\textbf{Post-processor.}
The post-processer is the module that actually answers %impact-set
dependence abstraction queries. %To that end,
It starts by
gathering distributed traces with a helper script which passes per-process traces to
the offline dependence-computation algorithm.
%The algorithm itself is implemented in Java, as are all other parts of {\distea}. Currently,
%Also, event traces are serialized after compression upon program termination. Thus, the post-processor decompresses
%per-process traces before merging them, and then computes the impact set.
To compute the dependence set following Algorithm~\ref{algo2}, the post-processor retrieves the
partial ordering of method-execution events by comparing the associated timestamps. %attached to those events.
For the {\bf Csd+} and {\bf Scov+} versions, the MDG is deserialized from the disk file dumped by the static analyzer.
Additionally for the {\bf Scov+} version, branches covered are retrieved from the traces as well and
covered statements are identified using control dependencies in the MDG: if a branch is covered, all
statements that are control dependent on that branch are all considered covered. Since our MDG construction
addresses exception-driven control flows as well, coverage of exception-handling constructs is handled
by our statement coverage computation as well.

%% file: evaluation.tex
\section{Evaluation}
 \label{sec:eval}
%%This section presents our empirical evaluation .....
%To evaluate our approach, we conducted an empirical study to answer
%%the utilized its implementation as discussed above, and attempted to seek
%%answers to
%the following
%three research questions:
We evaluate {\tech} in the context of its application for dynamic impact analysis,
with which an impact query corresponds to a dependence query (a dependence of
$m2$ on $m1$ implies $m1$ impacts $m2$). We seek to answer
%Our evaluation was guided by
the following research questions:

\begin{itemize}[leftmargin=14pt]
\denseitems
%\vspace{1pt}\noindent
%\item \textbf{RQ1} How effective is {\tech} in predicting impacts in comparison to existing options?
\vspace{0pt}
%\item \textbf{RQ1} Is {\tech} more effective than existing options?
\item \textbf{RQ1}: Are {\bf Msg+} and the two new levels of abstraction in {\tech} more effective than the basic {\distia}~\cite{cai2016distia}?
\item \textbf{RQ2}: Which kind of program information used by {\tech} contributes the most to its effectiveness?
%
%\vspace{1pt}\noindent
%\item \textbf{RQ2} How do impacts within processes compare with impacts across process boundaries?
%\item \textbf{RQ2} How are the impacts distributed across process boundaries in the impact sets given by {\tech}?
%
%\vspace{1pt}\noindent
\item \textbf{RQ3}: How efficient and scalable are the different versions of {\distea} in terms of various analysis overheads?
    %is {\tech} holistically as a framework? %in terms of the time and storage overheads it incurs?
%
%\item \textbf{RQ4} Can {\tech} be used for measuring the dynamic complexity of distributed software?
\item \textbf{RQ4}: Does {\tech} as a whole offer variable cost-effectiveness tradeoffs hence accommodate diverse use scenarios?
        %flexible options?
\end{itemize}
\vspace{-2pt}

%The main goal of this evaluation was to investigate the effectiveness (RQ1) and
%efficiency (RQ2) of {\distea}.
%We also intended to examine the composition of {\distea} impact sets %of impact sets given by {\distea}
%concerning how impacts propagate within and across constituent components (processes) in distributed systems (executions)
%%in distributed systems
%(RQ2).
%%to help understand the characteristics of impact analysis in distributed systems.
%%for which we formulated the following research questions
%%and applied to four distributed Java programs.

%% subject programs and inputs used in our study
\subsection{Experiment Setup}
%We studied four Java programs for the evaluation and their relevant characteristics
%are summarized in Table~\ref{tab:subjects}.
We evaluated {\distea} on eight distributed Java programs, as summarized in Table~\ref{tab:subjects}.
The size of each subject is measured by the number of
non-comment non-blank Java source lines of code (\emph{\#SLOC}) and number of
methods defined in the subject (\emph{\#Methods}) that we actually analyzed.
The last two columns list the type of test input used in our study (one test case per type)
%dynamic analysis, %including the type and size of each set,
and the number of methods executed at least once in the respective test (\emph{\#Cov.M.}).
We used each of these methods as a dependence %impact-set
query.
%The third and fifth columns together give the method-level coverage of each test (i.e., the fifth column divided by the third).
For each subject and input type, the ratio of the fifth column to the third column gives the method-level coverage of the test input (e.g., the method-level coverage of the integration test input for Thrift is 266/1,459).

\setlength{\tabcolsep}{3.5pt}
%\renewcommand{\arraystretch}{0.90}
%\begin{table}[tp]
%  \centering
%  \caption{Statistics of experimental subjects}
%    \begin{tabular}{|l|r|r|l|r|}
%    \hline
%    \multicolumn{1}{|l|}{\textbf{Subject (version)}} & \textbf{\#SLOC} & \textbf{\#Methods} & \textbf{Test type} & \textbf{\#Cov.M.} \\
%    \hline
%    MultiChat (r5) & 470 & 37 & integration & 25 \\
%    \hline
%    NIOEcho (r69) & 412 & 27 & integration & 26 \\
%    \hline
%    xSocket (v2.8.15) & 15,890 &   & integration & 391 \\
%    \hline
%    MINA (v2.0.16) & 47,419 &   & integration & 362 \\
%    \hline
%    Thrift (v0.11.0) & 12,366 &   & integration & 188 \\
%    \hline
%    Open Chord (v1.0.5) & 38,084 & 736 & integration & 354 \\
%    \hline
%    \multirow{3}[2]{*}{ZooKeeper (v3.4.11)} & \multirow{3}[2]{*}{62,450} & \multirow{3}[2]{*}{4,813} & integration & 749 \\
%\cline{4-5}      &   &   & system & 817 \\
%\cline{4-5}      &   &   & load & 798 \\
%    \hline
%    \end{tabular}%
%  \label{tab:subjects}%
%\end{table}%

\begin{table}[tp]
  \centering
  \caption{Statistics of experimental subjects}
    \begin{tabular}{|l|r|r|l|r|}
    \hline
    \multicolumn{1}{|l|}{\textbf{Subject (version)}} & \textbf{\#SLOC} & \textbf{\#Methods} & \textbf{Test type} & \textbf{\#Cov.M.} \\
    \hline
    MultiChat (r5) & 470 & 37 & integration & 25 \\
    \hline
    NIOEcho (r69) & 412 & 27 & integration & 26 \\
    \hline
    xSocket (v2.8.15) & 15,890 & 2,204  & integration & 391 \\
    %\hline
    %MINA (v2.0.16) & 47,419 & 3,608  & integration & 355 \\
    \hline
    Thrift (v0.11.0) & 12,366 & 1,459  & integration & 266 \\
    \hline
    Open Chord (v1.0.5) & 38,084 & 736 & integration & 354 \\
    \hline
    \multirow{3}[2]{*}{ZooKeeper (v3.4.11)} & \multirow{3}[2]{*}{62,450} & \multirow{3}[2]{*}{4,813} & integration & 749 \\
\cline{4-5}      &   &   & system & 817 \\
\cline{4-5}      &   &   & load & 798 \\
    \hline
    \multirow{3}[2]{*}{Voldemort (v1.9.6) } & \multirow{3}[2]{*}{163,601} & \multirow{3}[2]{*}{17,843} & integration & 2,048 \\
\cline{4-5}      &   &   & system & 1,242 \\
\cline{4-5}      &   &   & load & 1,323 \\
    \hline
    Freenet (v0.7.0) & 196,281 & 16,673 & integration & 2,477 \\
    \hline
    \end{tabular}%
  \label{tab:subjects}%
\end{table}%

\subsubsection{Subject Systems}
We chose these subjects %and their input sets
such that varied system scales and architectures, application domains, and uses of either and both of blocking and non-blocking I/Os are all considered.
\begin{itemize}
  \item MultiChat~\cite{multichat} is a chat application where multiple clients exchange
messages via a %central
server broadcasting the message sent by one client to all others.
%, using Java Socket I/O (blocking, stream-oriented) only.
  \item NioEcho~\cite{nioecho} is an %network
echo service %program
%using Java NIO only (non-blocking, buffer-oriented I/Os),
%where the server provides a simple echo service
via which the client just gets back the same message as it sends to the server.
  \item xSocket~\cite{xsocket} is an NIO-based library for building high-performance network applications.
%MINA is an Apache framework for network application development~\cite{mina}.
%Also an Apache project,
  \item Thrift~\cite{thrift} is a framework for scalable cross-language services development.
  \item Open Chord is a peer-to-peer lookup service based on distributed hash table~\cite{openchord}.
  \item ZooKeeper~\cite{zookeeper,hunt2010zookeeper} is a %high-performance
coordination service for distributed systems
to %easily
achieve consistency and synchronization.
  \item Voldemort~\cite{voldemort} is a distributed key-value storage system used at LinkedIn.
  \item Freenet~\cite{freenet} is a peer-to-peer data-sharing platform offering anonymous communication.
\end{itemize}
Some of these systems use Socket I/O or Java NIO only, while others use both mechanisms, for message passing
among their components.
%
%The latter two use both Socket I/O and Java NIO.
For all subjects, we checked out from their official repositories
the latest stable versions or revisions as shown in (the parentheses of) Table~\ref{tab:subjects}.

\subsubsection{System Executions}
We chose test inputs to cover different types of inputs when possible, including system test, integration test, and load test, which we assume exercise typical, overall system behaviour instead of a small specific area of the system. %source code.
The integration tests were created manually as elaborated below, while the other types of inputs come
with the subjects from their respective repositories.
%For unit tests, we used only those leading to multiprocess executions from the full original sets.
%: the ZooKeeper package includes only one system test ({\tt SimpleSysTest}) and one load test ({\tt GenerateLoad},
%for which we used two servers and two clients with request size of 10);
%for Voldemort, we randomly chose one system test ({\tt TestClientShutDown}) and one load test (End-to-end non-blocking
%population of key-value pairs with different keys) out of others of each type (we chose only one because of
%considerable setup effort for running such tests).

In each integration test, we started two to five server
and client nodes on different machines and performed client operations that
cover main system services---specially for peer-to-peer systems, we operated on all nodes, and
for ZooKeeper we started a container node in addition.
\begin{itemize}
  \item For MultiChat and NioEcho, the client requests were sending random text messages.
  \item For ZooKeeper, the ordered client operations were: create two nodes,
look up for them, check their attributes, change their data association, and
delete them.
  \item For Open Chord, the operations were in order: create an overlay network on machine (node) A, join the
network on machines B and C, insert a new data entry to the network on C, look up and then delete the data entry on A, and
list all data entries on B.
  \item For Voldemort, the operations were: add a key-value pair,
query the key for its value, delete the key, and retrieve the pair again.
  \item For Freenet, we first uploaded a file to the network with a note on node A, shared it to
nodes B and C, then on B and C accepted the sharing request and the note, followed by downloading the file and replying with a note.
  \item The remaining two %three
subjects, xSocket and Thrift, are frameworks/libraries, thus we needed to develop
user applications for exercising these subjects. Each of
the two applications consists of two components, a server and a client. %
For xSocket, one client sends a text message to the server, followed by the second client
sending a different message to the server.
%the application did something in the client, and the server xxxx;
%
%In the MINA application, the server sends a connection confirmation message to the first client connected to it, the client
%sends a reply to the server, the server sends confirmation message to the second client connected to it, and the
%second client sends a message to the server as response.
The Thrift application implements a calculator, for which the operations were
addition, subtraction, multiplication, and division of two numbers.
\end{itemize}

Our {\tech} implementation handled the varied system architectures and network I/O mechanisms represented by the chosen subject systems, including
blocking and non-blocking message passing among the distributed components of these systems.
Thus, no user-specified list of message-passing APIs was needed for our evaluation experiments.
Given the diversity of our study subjects, this implies that users of {\tech} would commonly not need to specify
the optional API list $L$ (Figure~\ref{fig:disteaprocess}).

%
%For ZooKeeper and Voldemort, the client operations are those listed on their
%start-guide pages for tutorial purposes.

\subsection{Experimental Methodology}\label{sec:methodology}

To answer our research questions, we evaluate {\tech} as a holistic framework through
assessing the cost and effectiveness of its four instantiations (Section~\ref{sec:tech}).
In particular, we compare the {\bf Msg+} version and the two newly added versions, together noted as {\em advanced versions}, %levels of abstraction
against the {\bf Basic} version (i.e., basic {\distia}~\cite{cai2016distea})
as the baseline, so as to understand the contribution of message-passing semantics, static dependencies, and coverage data to
effectiveness improvements and overhead increases in {\tech}.
%
%dynamic impact prediction (dependence abstraction). the cost-effectiveness of {\tech}.
%
%
We considered every method of each subject as a query, yet we report
results only for queries executed at least once in one process---only such queries
have a non-empty impact set. For a method executed in more than one
processes, we took it as executed in each process as a separate query.
%them as separate queries each per process.

\begin{figure*}[th!]
  \vspace{2pt}
  \centering
  %\includegraphics[scale=0.80]{graphics/allratios.pdf}
  %\includegraphics[scale=0.72]{graphics/allratios-imp.pdf}
  \includegraphics[width=0.98\textwidth]{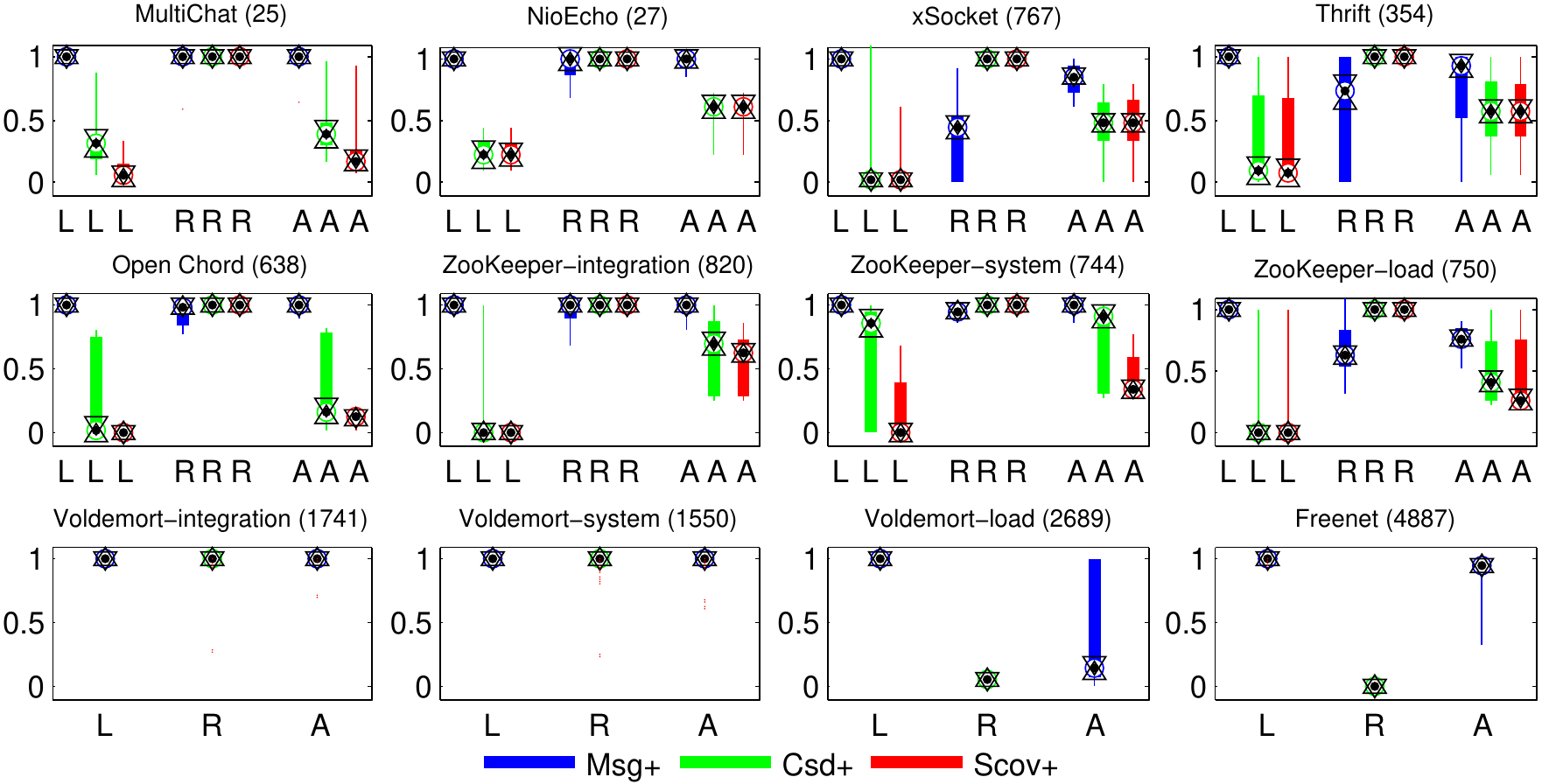}
  \vspace{-10pt}
  %\caption{Effectiveness of the basic {\distea} expressed by the ratios ($y$ axes, the lower the better) of its per-query impact-set sizes, including those of the local and remote subsets ($x$ axes), to the baseline, per subject and input type (with \#queries in parentheses).}
  \caption{Distribution of {\distea} effectiveness as the ratios ($y$ axes, the lower the better) of its per-query impact-set sizes ($A$), including those of the local ($L$) and remote ($R$) subsets ($x$ axes), to the baseline, per subject and input type (with \#queries in parentheses).}
  \label{fig:effdist}
  \vspace{-6pt}
\end{figure*}

For each query,
%For every method in each process as a query
we measure the effectiveness of {\distea} by comparing %the impacts it predicted to those given by \emph{MCov} in terms
it to the baseline in terms of
of %overall
impact-set (dependence-set) size ratios, and
%Further, we also
examine the composition of the impact set concerning its two
subsets: \emph{local impact set} (i.e., local dependence set) and \emph{remote impact set} (i.e., remote dependence set).
%
%while also analyzing %their intersection, referred to as
%the \emph{common impact set}.
%In our results,
For average-case analyses,
we added the common impacts (dependencies) into both subsets.
Accordingly, we measure the effectiveness of {\distea} with respect to two subsets (local and remote impact sets) relative to the corresponding %\emph{MCov}
baseline results as well.
In essence, we assess precision improvement through impact-set reduction as a {\em relative} measure.
We use the relative effectiveness measures for two reasons. First, computing the (precision and/or recall) measures in absolute terms requires ground-truth dependence sets of all possible queries, which are neither available for our subject systems and executions nor automatically computable (due to the lack of capable tools available to us).
Second, given that the baseline has been evaluated with respect to its effectiveness before~\cite{cai2016distea}, the main goal of this evaluation is to assess the relative improvements of the advanced versions over the baseline, for which the relative measures are sufficient.
%
%
%When computing the impact set for a query, for each type of inputs, we take the union of per-input impact sets of all
%inputs of that type %(e.g., impact sets of all unit tests are unionized for ZooKeeper and Voldemort)
%since each type %of inputs
%is usually intended to represent a different operational profile of the system under analysis.
Besides the impact-querying time, we report the static-analysis and run-time costs of {\distea}, and
storage costs, all as efficiency metrics.
%We also compare the effectiveness and efficiency of the basic versus the enhanced version.
%
%We also compare the three versions of {\tech} as distinct dependence abstraction techniques in terms of %effectiveness and efficiency, using \emph{MCov} as the common baseline.
%

For a fair comparison, all versions used the same test input for computing the impact set of each query. %in a subject.
This was realized by first recording our manual inputs per subject and test type and then replaying accordingly across
different {\tech} versions.
The machines used were all Linux workstations with an Intel i5-2400 3.10GHz CPU and 8GB DDR2 RAM.

\input{resultanalysis}

%% file: resultanalysis.tex
\subsection{Results and Analysis}
%This section presents the results of our study.
In this section, we report and discuss the empirical results of our evaluation on {\tech}, focusing on the effectiveness and
efficiency of {\tech} separately and both together as a cost-effectiveness measure, in order to answer the four research questions.
\vspace{-2pt}
\subsubsection{RQ1: Effectiveness}
Figure~\ref{fig:effdist} illustrates the effectiveness results of {\tech}, with one plot depicting the data distribution
%of all data points
for each subject and input type, shown as the plot title (hereafter, the input type is omitted for
 subjects for which only an integration test is available and utilized).
Each chart includes at most nine boxplots showing that data distribution for one of three categories (on $x$ axis):
the holistic impact set (\emph{A} for all) and its two subsets (\emph{L} for local and \emph{R} for remote), each produced
by one of the three advanced versions of {\tech} if applicable.
Two exceptions were Voldemort and Freenet, for which the {\bf Csd+} results are missing because
computing the static intra-component dependencies did not finish after ten hours (thus we killed the analysis).
The same was applied to {\bf Scov+} since it subsumes {\bf Csd+} in terms of the inclusion of the static dependence analysis.

Particularly in Figure~\ref{fig:effdist}, for each query, the common dependencies were removed from the two (i.e., local and remote) subsets of the impact set. This is done for a sanity check of the
pruning strategy of each advanced version: {\bf Msg+} only prunes remote dependencies while the other two only prune local dependencies.
%ran out of memory after eight hours.
%, and holistic (\emph{all}) impact sets to corresponding covered sets,
Each underlying data item of the chart indicates the effectiveness metric (i.e., impact-set size ratio to the baseline, as shown on $y$ axis) for one query.
%Thus, the total number of data points graphed by each plot is that of the queries, as given in Table~\ref{tab:subjects}, exercised on the corresponding subject and input set for obtaining those data.
The total numbers of queries involved are shown in the parentheses.
In each boxplot, the circled dot within the bar represents the median.
To facilitate the comparison of effectiveness differences among the three {\tech} versions, we also showed in each boxplot
a pair of triangular marks, which indicate the comparison interval of the median.
The statistical meaning of these intervals is: the medians of any two groups are significantly different
at the 5\% significance level if their intervals do not overlap.

The results indicate that all the three advanced versions are noticeably more effective than the baseline (basic {\distia}).
Incorporating the message-passing semantics reduced the baseline impact sets by 5\% up to over 20\% in most cases (according
to the 75\% quartiles).
Compared to the two smallest subjects, the large, real-world systems saw generally higher effectiveness improvements by
the {\bf Msg+} version. %instantiation of {\tech}.
Examining the two subsets of each impact set confirms that {\bf Msg+} only reduced remote impact sets only while not
changing the local impact sets relative to the baseline, consistent with the rationale of this technique.
Note that the ratios with respect to \emph{all} (rightmost three boxplots of each chart) impact sets are mostly higher than those with respect to
the {\em remote} subsets. This is because
%the two subsets of each impact set have substantial intersections,
each {\em all} impact set includes the corresponding common impact set as well,
while sizes of the common impact sets of {\bf Msg+} %the three {\tech} versions shown here
are consistently smaller than those of
the baseline (due to smaller remote impact sets).

The {\bf Csd+} and {\bf Scov+} versions of {\distea} further enhanced the baseline %{\bf Basic}-version
effectiveness by pruning only local impact sets, %and both local and remote impact sets, respectively,
according to the analysis algorithm (Algorithm~\ref{algo2}).
Just as {\bf Msg+} version did not change the local impact sets, the results confirmed that neither {\bf Csd+} nor {\bf Scov+}
changed the remote impact sets.
Since the static dependencies incorporated in {\bf Csd+}, and statement coverage in {\bf Scov+} additionally,
are both limited to individual components (Equations~\ref{eq_constraint2} and~\ref{eq_constraint3}),
the reduction of overall impact sets by {\bf Csd+} and {\bf Scov+} was ascribed to the decreases in
local impact-set sizes. Again, here we are looking at the two subsets with common dependencies removed; if we look at the
the complete local and remote impact sets, any of the three advanced versions would have reduced both local and remote impact sets
in most cases, because reducing one of the two subsets led to the reduction of the common set hence the reduction of the other subset.

% Table generated by Excel2LaTeX from sheet 'sizeRatios'
\setlength{\tabcolsep}{4.5pt}
\begin{table*}[th!]
  \centering
  \caption{Mean effectiveness improvement (in terms of impact-set size ratios) of the three new instances of {\tech} over the baseline}\vspace{-8pt}
    \begin{tabular}{|l||r|r|r||r|r|r||r|r|r||r|r|r|}
    \hline
    \multicolumn{1}{|c|}{\multirow{3}[1]{*}{\textbf{Subject \& test input}}} & \multicolumn{3}{c|}{\multirow{2}[0]{*}{\textbf{\shortstack{Average baseline\\effectiveness (impact-set sizes)}}}} & \multicolumn{9}{c|}{\textbf{Mean effectiveness improvement versus baseline}} \\
\cline{5-13}    \multicolumn{1}{|c|}{} & \multicolumn{3}{c|}{} & \multicolumn{3}{c|}{\textbf{{\bf Msg+} version}} & \multicolumn{3}{c|}{\textbf{{\bf Csd+} version}} & \multicolumn{3}{c|}{\textbf{{\bf Scov+} version}} \\
\cline{2-13}    \multicolumn{1}{|c|}{} & \multicolumn{1}{c|}{\textbf{Local}} & \multicolumn{1}{c|}{\textbf{Remote}} & \multicolumn{1}{c|}{\textbf{All}} & \multicolumn{1}{c|}{\textbf{Local}} & \multicolumn{1}{c|}{\textbf{Remote}} & \multicolumn{1}{c|}{\textbf{All}} & \multicolumn{1}{c|}{\textbf{Local}} & \multicolumn{1}{c|}{\textbf{Remote}} & \multicolumn{1}{c|}{\textbf{All}} & \multicolumn{1}{c|}{\textbf{Local}} & \multicolumn{1}{c|}{\textbf{Remote}} & \multicolumn{1}{c|}{\textbf{All}} \\
    \hline
    MultiChat & 14.0 & 4.5 & \textcolor[rgb]{ .439,  .188,  .627}{18.5} & \cellcolor[rgb]{ .949,  .949,  .949}100\% & \cellcolor[rgb]{ .949,  .949,  .949}96.28\% & \cellcolor[rgb]{ .949,  .949,  .949}\textcolor[rgb]{ .439,  .188,  .627}{97.12\%} & 36.85\% & 100\% & \textcolor[rgb]{ .439,  .188,  .627}{49.27\%} & \cellcolor[rgb]{ .949,  .949,  .949}19.21\% & \cellcolor[rgb]{ .949,  .949,  .949}100\% & \cellcolor[rgb]{ .949,  .949,  .949}\textcolor[rgb]{ .439,  .188,  .627}{34.77\%} \\
    \hline
    NioEcho & 11.4 & 11.3 & \textcolor[rgb]{ .439,  .188,  .627}{21.7} & \cellcolor[rgb]{ .949,  .949,  .949}100\% & \cellcolor[rgb]{ .949,  .949,  .949}93.35\% & \cellcolor[rgb]{ .949,  .949,  .949}\textcolor[rgb]{ .439,  .188,  .627}{96.59\%} & 38.69\% & 100\% & \textcolor[rgb]{ .439,  .188,  .627}{66.20\%} & \cellcolor[rgb]{ .949,  .949,  .949}38.07\% & \cellcolor[rgb]{ .949,  .949,  .949}100\% & \cellcolor[rgb]{ .949,  .949,  .949}\textcolor[rgb]{ .439,  .188,  .627}{65.85\%} \\
    \hline
    xSocket & 203.2 & 281.1 & \textcolor[rgb]{ .439,  .188,  .627}{356.3} & \cellcolor[rgb]{ .949,  .949,  .949}100\% & \cellcolor[rgb]{ .949,  .949,  .949}38.62\% & \cellcolor[rgb]{ .949,  .949,  .949}\textcolor[rgb]{ .439,  .188,  .627}{84.30\%} & 12.45\% & 100\% & \textcolor[rgb]{ .439,  .188,  .627}{49.40\%} & \cellcolor[rgb]{ .949,  .949,  .949}11.46\% & \cellcolor[rgb]{ .949,  .949,  .949}100\% & \cellcolor[rgb]{ .949,  .949,  .949}\textcolor[rgb]{ .439,  .188,  .627}{49.08\%} \\
    \hline
    Thrift & 80.7 & 102.1 & \textcolor[rgb]{ .439,  .188,  .627}{128.0} & \cellcolor[rgb]{ .949,  .949,  .949}100\% & \cellcolor[rgb]{ .949,  .949,  .949}60.60\% & \cellcolor[rgb]{ .949,  .949,  .949}\textcolor[rgb]{ .439,  .188,  .627}{80.80\%} & 33.59\% & 100\% & \textcolor[rgb]{ .439,  .188,  .627}{56.76\%} & \cellcolor[rgb]{ .949,  .949,  .949}31.17\% & \cellcolor[rgb]{ .949,  .949,  .949}100\% & \cellcolor[rgb]{ .949,  .949,  .949}\textcolor[rgb]{ .439,  .188,  .627}{55.50\%} \\
    \hline
    Open Chord & 248.0 & 245.9 & \textcolor[rgb]{ .439,  .188,  .627}{276.0} & \cellcolor[rgb]{ .949,  .949,  .949}100\% & \cellcolor[rgb]{ .949,  .949,  .949}93.36\% & \cellcolor[rgb]{ .949,  .949,  .949}\textcolor[rgb]{ .439,  .188,  .627}{98.95\%} & 48.32\% & 100\% & \textcolor[rgb]{ .439,  .188,  .627}{54.76\%} & \cellcolor[rgb]{ .949,  .949,  .949}0.61\% & \cellcolor[rgb]{ .949,  .949,  .949}100\% & \cellcolor[rgb]{ .949,  .949,  .949}\textcolor[rgb]{ .439,  .188,  .627}{14.60\%} \\
    \hline
    ZooKeeper-integration & 284.1 & 265.5 & \textcolor[rgb]{ .439,  .188,  .627}{492.1} & \cellcolor[rgb]{ .949,  .949,  .949}100\% & \cellcolor[rgb]{ .949,  .949,  .949}94.46\% & \cellcolor[rgb]{ .949,  .949,  .949}\textcolor[rgb]{ .439,  .188,  .627}{96.96\%} & 29.77\% & 100\% & \textcolor[rgb]{ .439,  .188,  .627}{70.03\%} & \cellcolor[rgb]{ .949,  .949,  .949}0.51\% & \cellcolor[rgb]{ .949,  .949,  .949}100\% & \cellcolor[rgb]{ .949,  .949,  .949}\textcolor[rgb]{ .439,  .188,  .627}{62.96\%} \\
    \hline
    ZooKeeper-system & 687.1 & 641.3 & \textcolor[rgb]{ .439,  .188,  .627}{929.5} & \cellcolor[rgb]{ .949,  .949,  .949}100\% & \cellcolor[rgb]{ .949,  .949,  .949}95.12\% & \cellcolor[rgb]{ .949,  .949,  .949}\textcolor[rgb]{ .439,  .188,  .627}{97.23\%} & 55.31\% & 100\% & \textcolor[rgb]{ .439,  .188,  .627}{70.64\%} & \cellcolor[rgb]{ .949,  .949,  .949}18.13\% & \cellcolor[rgb]{ .949,  .949,  .949}100\% & \cellcolor[rgb]{ .949,  .949,  .949}\textcolor[rgb]{ .439,  .188,  .627}{44.49\%} \\
    \hline
    ZooKeeper-load & 577.1 & 560.7 & \textcolor[rgb]{ .439,  .188,  .627}{837.8} & \cellcolor[rgb]{ .949,  .949,  .949}100\% & \cellcolor[rgb]{ .949,  .949,  .949}66.65\% & \cellcolor[rgb]{ .949,  .949,  .949}\textcolor[rgb]{ .439,  .188,  .627}{82.73\%} & 15.50\% & 100\% & \textcolor[rgb]{ .439,  .188,  .627}{48.58\%} & \cellcolor[rgb]{ .949,  .949,  .949}7.25\% & \cellcolor[rgb]{ .949,  .949,  .949}100\% & \cellcolor[rgb]{ .949,  .949,  .949}\textcolor[rgb]{ .439,  .188,  .627}{47.79\%} \\
    \hline
    Voldemort-integration & 654.6 & 578.1 & \textcolor[rgb]{ .439,  .188,  .627}{1052.7} & \cellcolor[rgb]{ .949,  .949,  .949}100\% & \cellcolor[rgb]{ .949,  .949,  .949}89.13\% & \cellcolor[rgb]{ .949,  .949,  .949}\textcolor[rgb]{ .439,  .188,  .627}{95.69\%} & \multicolumn{1}{c|}{-} & \multicolumn{1}{c|}{-} & \multicolumn{1}{c||}{\textcolor[rgb]{ .439,  .188,  .627}{-}} & \multicolumn{1}{c|}{\cellcolor[rgb]{ .949,  .949,  .949}-} & \multicolumn{1}{c|}{\cellcolor[rgb]{ .949,  .949,  .949}-} & \multicolumn{1}{c|}{\cellcolor[rgb]{ .949,  .949,  .949}\textcolor[rgb]{ .439,  .188,  .627}{-}} \\
    \hline
    Voldemort-system & 522.1 & 478.6 & \textcolor[rgb]{ .439,  .188,  .627}{842.9} & \cellcolor[rgb]{ .949,  .949,  .949}100\% & \cellcolor[rgb]{ .949,  .949,  .949}86.97\% & \cellcolor[rgb]{ .949,  .949,  .949}\textcolor[rgb]{ .439,  .188,  .627}{94.01\%} & \multicolumn{1}{c|}{-} & \multicolumn{1}{c|}{-} & \multicolumn{1}{c||}{\textcolor[rgb]{ .439,  .188,  .627}{-}} & \multicolumn{1}{c|}{\cellcolor[rgb]{ .949,  .949,  .949}-} & \multicolumn{1}{c|}{\cellcolor[rgb]{ .949,  .949,  .949}-} & \multicolumn{1}{c|}{\cellcolor[rgb]{ .949,  .949,  .949}\textcolor[rgb]{ .439,  .188,  .627}{-}} \\
    \hline
    Voldemort-load & 286.9 & 709.9 & \textcolor[rgb]{ .439,  .188,  .627}{967.6} & \cellcolor[rgb]{ .949,  .949,  .949}100\% & \cellcolor[rgb]{ .949,  .949,  .949}4.42\% & \cellcolor[rgb]{ .949,  .949,  .949}\textcolor[rgb]{ .439,  .188,  .627}{47.91\%} & \multicolumn{1}{c|}{-} & \multicolumn{1}{c|}{-} & \multicolumn{1}{c||}{\textcolor[rgb]{ .439,  .188,  .627}{-}} & \multicolumn{1}{c|}{\cellcolor[rgb]{ .949,  .949,  .949}-} & \multicolumn{1}{c|}{\cellcolor[rgb]{ .949,  .949,  .949}-} & \multicolumn{1}{c|}{\cellcolor[rgb]{ .949,  .949,  .949}\textcolor[rgb]{ .439,  .188,  .627}{-}} \\
    \hline
    Freenet & 2787.8 & 2780.0 & \textcolor[rgb]{ .439,  .188,  .627}{2956.6} & \cellcolor[rgb]{ .949,  .949,  .949}100\% & \cellcolor[rgb]{ .949,  .949,  .949}3.11\% & \cellcolor[rgb]{ .949,  .949,  .949}\textcolor[rgb]{ .439,  .188,  .627}{93.24\%} & \multicolumn{1}{c|}{-} & \multicolumn{1}{c|}{-} & \multicolumn{1}{c||}{\textcolor[rgb]{ .439,  .188,  .627}{-}} & \multicolumn{1}{c|}{\cellcolor[rgb]{ .949,  .949,  .949}-} & \multicolumn{1}{c|}{\cellcolor[rgb]{ .949,  .949,  .949}-} & \multicolumn{1}{c|}{\cellcolor[rgb]{ .949,  .949,  .949}\textcolor[rgb]{ .439,  .188,  .627}{-}} \\
    \hline
    \textbf{weighted average} & \textbf{1191.6} & \textbf{1251.8} & \textcolor[rgb]{ 1,  0,  0}{\textbf{1494.7}} & \cellcolor[rgb]{ .949,  .949,  .949}\textbf{100\%} & \cellcolor[rgb]{ .949,  .949,  .949}\textbf{41.06\%} & \cellcolor[rgb]{ .949,  .949,  .949}\textcolor[rgb]{ 1,  0,  0}{\textbf{84.85\%}} & \textbf{31.86\%} & \textbf{100\%} & \textcolor[rgb]{ 1,  0,  0}{\textbf{58.75\%}} & \cellcolor[rgb]{ .949,  .949,  .949}\textbf{9.95\%} & \cellcolor[rgb]{ .949,  .949,  .949}\textbf{100\%} & \cellcolor[rgb]{ .949,  .949,  .949}\textcolor[rgb]{ 1,  0,  0}{\textbf{46.02\%}} \\
    \hline
    \end{tabular}%
  \label{tab:meaneff}%
  \vspace{-12pt}
\end{table*}%

The impact-set ratio distribution of {\bf Csd+} and {\bf Scov+} (i.e., the two new framework instantiations first introduced in this paper) % versions
shows their substantial advantage over both the baseline
and the {\bf Msg+} version.
In all of the subjects to which these two new versions were applied, referring to local static dependencies largely cut off
the baseline local impact sets, leading to drastic reduction of the overall baseline impact sets by 20\% to over 50\% for
the majority of the queries (per the 75\% quartiles).
As expected, leveraging statement coverage data moved further along in pruning the local impact sets.
The results demonstrated substantial improvements of {\bf Scov+} over {\bf Csd+} for most queries in most subjects,
which is particularly true of larger subjects (e.g., Open Chord and ZooKeeper) compared to smaller ones (e.g., xSocket and Thrift).

Complementary to Figure~\ref{fig:effdist}, Table~\ref{tab:meaneff}
gives the mean effectiveness results (computed from the same individual data points which the figure depicts the distribution for).
Results of the two new {\distea} versions against
Voldemort and Freenet were missing because of the reason mentioned above.
On average, {\bf Msg+} was not always effective, with a worse case of reporting 99\% of the baseline impact set for Open Chord.
Although it attained the highest reduction of over 50\% %of baseline impact set
(for Voldemort-load),
the reduction was less than 20\% in all other cases.
In comparison, {\bf Csd+} pruned baseline impact sets by mostly by 40\% or more, and 30\% in the worse case.
The highest effectiveness was achieved by {\bf Scov+} on Open Chord, with an over 85\% reduction of baseline results on average.
Our manual inspection of the code of this subject in comparison to other subjects revealed that it contains much more extensive control structures exercised during the analyzed executions, which justifies the best effectiveness of statement-coverage-based pruning for this
subject.

The last row of the table shows the weighted (by the number of queries) averages of per-subject effectiveness measures.
Given the differences in the subjects' code and executions, these overall averages may not well represent an
average-case situation across the subjects. Nevertheless, these numbers provide a summary metric to facilitate comparisons.
In all, {\bf Msg+}, {\bf Csd+}, and {\bf Scov+} pruned 15\%, 41\%, and 54\% of baseline results, respectively, on average across all applicable cases.
Given the safety of all resulting impact sets produced by any of the four compared techniques (Section~\ref{sec:inference}),
these reduction ratios are translated to precision improvements by 17.6\%, 69.5\%, 117.4\% (in terms of ratios rather than
absolute precision values)\footnote{An impact-set size ratio $r$ corresponds to a precision increase by (1-$r$)/$r$.}, respectively.

From a perspective of practical application, these effectiveness improvements imply that developers can save the time that would
be spent on inspecting up to half of the impacts/dependencies given by the baseline approach.
The first four columns of Table~\ref{tab:meaneff} list the baseline impact set sizes, which
inform the implications and significance of these inspection-effort savings.
For instance, for ZooKeeper-load, {\bf Csd+} would save developers effort on examining 430 (false-positive) methods.
On overall average, {\bf Msg+}, {\bf Csd+}, and {\bf Scov+} would save such efforts for inspecting 226, 617, and 807 methods that are
false positives, respectively.
The numbers for the local and remote impact sets further validated our expectations on where (local/remote)
each advanced version would prune the baseline results the most.
The {\bf Msg+} local impact sets are always the same (100\%) as the baseline's, so are the remote impact sets from {\bf Csd+} and {\bf Scov+}.
This consistency further validated the analysis algorithms underlying these three advanced {\distea} versions.

%Overall, the {\bf Basic} {\distea} reports on average only 69\% of the impacts produced by \emph{MCov}.
%%In particular, the reductions in remote impact sets are even higher, by 38\% on average and well above 35\% mostly for the three largest subjects.
%This implies that, relative to the baseline, %s like \emph{MCov},
%developers can save the time that would
%be spent on inspecting over a third of the impacts given by the baseline approach.
%%propagated to remote processes.
%Together with the impact size ratios, the sizes of baseline ({\em mCov}) impact sets (left four columns)
%indicates how many false positives (methods) from the baseline were removed, further suggesting
%the inspection efforts saved by our approach.
%The two advanced versions of {\tech} offered even greater savings,
%with 43\% impact-set reduction on average over all the eight subjects
%and 64\% reduction for all but the two largest subjects.
%The numbers for the local and remote impacts further validated our expectations on where (local/remote)
%each advanced version would prune the {\bf Basic}-version results the most.

\vspace{6pt}\noindent\fbox{\parbox{\dimexpr\linewidth-2\fboxsep-\fboxrule\relax}{
\textbf{RQ1}:
The three advanced instances of {\distea} achieved substantial effectiveness (precision) improvements over
{\distia}, with an impact-set size reduction of 15\% (with {\bf Msg+}), 41\% (with {\bf Csd+}), and 54\% (with {\bf Scov+}).
The implication of these improvements is to save developers' effort of examining 226 to 807 false-positive
methods in the baseline impact sets.
}}
\vspace{2pt}

\subsubsection{RQ2: Contributing Factors}
\begin{table*}[htbp]
  \centering
  \vspace{0pt}
  \caption{Wilcoxon $p$-values and Cliff-Delta effect sizes (in parentheses) with respect to impact-set sizes between all pairs among the four {\distea} techniques}  \vspace{-8pt}
    \begin{tabular}{|l|r|c|c|c|c|c|}
    \hline
    \textbf{Subject \& test input} & \multicolumn{1}{c|}{\textbf{baseline:{\bf Msg+}}} & \textbf{baseline:{\bf Csd+}} & \textbf{baseline:{\bf Scov+}} & \textbf{{\bf Msg+}:{\bf Csd+}} & \textbf{{\bf Msg+}:{\bf Scov+}} & \textbf{{\bf Csd+}:{\bf Scov+}} \\
    \hline
    MultiChat & 5.62E-01 (.08) & \multicolumn{1}{r|}{\textbf{6.20E-05} (.68)} & \multicolumn{1}{r|}{\textbf{2.62E-06} (.68)} & \multicolumn{1}{r|}{\textbf{1.17E-04} (.60)} & \multicolumn{1}{r|}{\textbf{4.55E-06} (.68)} & \multicolumn{1}{r|}{\textbf{6.30E-04} (.80)} \\
    \hline
    NioEcho & 2.92E-01 (.33) & \multicolumn{1}{r|}{\textbf{2.31E-09} (.93)} & \multicolumn{1}{r|}{\textbf{7.97E-10} (.96)} & \multicolumn{1}{r|}{\textbf{2.48E-09} (.93)} & \multicolumn{1}{r|}{\textbf{7.37E-10} (.96)} & \multicolumn{1}{r|}{9.58E-01 (.04)} \\
    \hline
    xSocket & \textbf{5.29E-38} (.71) & \multicolumn{1}{r|}{\textbf{1.12E-181} (.86)} & \multicolumn{1}{r|}{\textbf{1.08E-181} (.86)} & \multicolumn{1}{r|}{\textbf{1.11E-133} (.70)} & \multicolumn{1}{r|}{\textbf{1.08E-133} (.70)} & \multicolumn{1}{r|}{1.00E+00 (.00)} \\
    \hline
    Thrift & \textbf{3.58E-18} (.56) & \multicolumn{1}{r|}{\textbf{1.22E-63} (.77)} & \multicolumn{1}{r|}{\textbf{1.67E-67} (.80)} & \multicolumn{1}{r|}{\textbf{2.17E-24} (.50)} & \multicolumn{1}{r|}{\textbf{6.35E-27} (.53)} & \multicolumn{1}{r|}{6.34E-01 (.10)}  \\
    \hline
    Open Chord & 6.33E-02 (.25) & \multicolumn{1}{r|}{\textbf{3.47E-158} (.81)} & \multicolumn{1}{r|}{\textbf{3.25E-243} (.99)} & \multicolumn{1}{r|}{\textbf{2.93E-157} (.80)} & \multicolumn{1}{r|}{\textbf{3.52E-243} (.99)} & \multicolumn{1}{r|}{\textbf{6.62E-87} (.79)} \\
    \hline
    ZooKeeper-integration & \textbf{5.34E-03} (.28) & \multicolumn{1}{r|}{\textbf{1.01E-40} (.40)} & \multicolumn{1}{r|}{\textbf{4.68E-35} (.36)} & \multicolumn{1}{r|}{\textbf{5.35E-50} (.42)} & \multicolumn{1}{r|}{\textbf{3.88E-48} (.42)} & \multicolumn{1}{r|}{\textbf{2.01E-12} (.55)} \\
    \hline
    ZooKeeper-system & \textbf{1.49E-13} (.64) & \multicolumn{1}{r|}{\textbf{7.97E-88} (.63)} & \multicolumn{1}{r|}{\textbf{2.31E-221} (.94)} & \multicolumn{1}{r|}{\textbf{5.15E-57} (.47)} & \multicolumn{1}{r|}{\textbf{9.35E-221} (.94)} & \multicolumn{1}{r|}{\textbf{5.26E-48} (.63)} \\
    \hline
    ZooKeeper-load & \textbf{9.83E-106} (.56) & \multicolumn{1}{r|}{\textbf{5.50E-163} (.80)} & \multicolumn{1}{r|}{\textbf{7.31E-154} (.78)} & \multicolumn{1}{r|}{\textbf{1.05E-69} (.47)} & \multicolumn{1}{r|}{\textbf{1.01E-64} (.45)} & \multicolumn{1}{r|}{1.20E-01 (.47)} \\
    \hline
    Voldemort-integration & \textbf{4.95E-12} (.18) & - & - & - & - & - \\
    \hline
    Voldemort-system & \textbf{8.34E-15} (.20) & - & - & - & - & - \\
    \hline
    Voldemort-load & \textbf{0.00E+00} (.58) & - & - & - & - & - \\
    \hline
    Freenet & \textbf{5.96E-168} (.49) & - & - & - & - & - \\
    \hline
    \end{tabular}%
  \label{tab:pvalueseffsizes}%
  \vspace{-12pt}
\end{table*}%

%Our framework optionally utilizes different kinds of program information for varied levels of precision.
To understand the contributing factors in the effectiveness improvements of {\tech} over the baseline,
we examined the effectiveness results again but with a focus on the contrast among the three advanced {\tech} versions
evaluated against the baseline: {\bf Msg+}, {\bf Csd+}, and {\bf Scov+}. The motivation for understanding these contributing factors is to
gain knowledge on the pros and cons of varied design decisions (in terms of data use) in dynamic dependence abstraction of distributed programs.

As %discussed above and
shown in Table~\ref{tab:meaneff}, the two versions that utilize static dependencies ({\bf Csd+} and {\bf Scov+}) are both considerably
more effective (achieving 37\% and 26\% smaller impact sets on average, respectively) than {\bf Msg+}.
In comparison, {\bf Msg+} is more effective than the baseline, with much smaller magnitude of improvements though
(14\% smaller impact sets on average).
Thus, in terms of average-case precision, it appeared that message-passing semantics contributed less than static dependencies.
This is intuitively because message-passing semantics itself is a very coarse form of (component/process-level) information.
Also, normally message passing occurs among distributed processes fairly often and bidirectionally, as we
observed in our subjects' executions. As a result, the constraint (Equation~\ref{eq_constraint}) is generally easy to satisfy.
%Meanwhile,
Comparing between {\bf Csd+} and {\bf Scov+} indicates that adding more program information (i.e., statement coverage) led to
further precision improvements. On the other hand, the gain of 11\% is only half of that (26\%)
achieved by {\bf Csd+} over {\bf Msg+}. Thus, static dependencies seem to contribute more than statement coverage too.
%Further, %the impact-set ratio distribution of
Not only were such contrasts among the three advanced versions observed in an average case,
Figure~\ref{fig:effdist} revealed similar contrasts for the majority of individual
queries in all relevant subjects individually.
%
%Not only were such contrasts among the three {\tech} versions observed in an average case,
%Figure~\ref{fig:effdist} demonstrated that the improvements
%of {\bf Csd+} over {\bf Msg+} and that of {\bf Msg+} over the {\bf Basic} version were actually noticeable
%for the majority of queries {\em individually}.

%\noindent
%{\bf Statistical validation.}
Figure~\ref{fig:effdist} also enables statistical comparisons of the contributions made by varied forms of
program information used in {\tech}.
Comparing the medians (circled dots) across the three advanced versions for each subject and test type further
corroborates the considerable effectiveness advantage of {\bf Scov+} over {\bf Csd+}, and the even greater
improvements of {\bf Csd+} over {\bf Msg+}.
%
%To statistically examine these effectiveness differences, we also showed in each boxplot
%a pair of triangular marks, which indicate the comparison interval of the median.
%The statistical meaning of these intervals is that the medians of any two groups are significantly different
%at the 5\% significance level if their intervals do not overlap.
These comparison intervals indicate that, in a median case, (1) {\bf Csd+} was {\em significantly} more effective than {\bf Msg+} in
7 out of the 8 applicable cases (with the only exception of ZooKeeper-system), and (2) {\bf Scov+} was {\em significantly} more
effective than {\bf Csd+} for 3 out of the 8 cases (MultiChat, ZooKeeper-system, and ZooKeeper-load).

%{\bf Msg+} was {\em significantly} more effective than the baseline in six out of 12 cases (i.e., xSocket, Thrift, ZooKeeper and Voldemort with system and load tests), while (2) {\bf Csd+} was {\em significantly} more effective than both the {\bf Basic} and {\bf Msg+} versions in {\em all} of the eight cases.
%%The gaps between these intervals confirmed the greater improvement of
%%
%These comparative results suggest that,
%with partially-ordered method events as the bedrock information,
%message-passing semantics generally contributed much less significantly than static intra-component dependencies to
%the effectiveness improvement in {\tech}.
%Our results also revealed that message-passing semantics contributed more for certain subjects but less for others (without
%consistent correlation with subject sizes), while the static dependencies contributed almost equally to all subjects.
%On the other hand, %however,
%incorporating static dependencies may not be always feasible (especially
%for very-large subjects, due to its much greater overheads than exploiting message-passing semantics).

Beyond the average- and median-case comparisons, we further conducted two statistical analyses:
(1) paired Wilcoxon signed-rank tests~\cite{walpole11jan} to assess the statistical significance (at the $0.95$ confidence level)
of effectiveness differences among the four {\tech} versions, and
(2) effect sizes in terms of Cliff's Delta~\cite{cliff1996ordinal} (in a paired setting with $\alpha=.05$) to assess the
magnitude of the effectiveness differences.
In both analyses, the two groups were the impact-set sizes given by each pair of techniques compared.
Both analyses are nonparametric, allowing us to lift the assumption about the normality of the distribution of
underlying data points.
Table~\ref{tab:pvalueseffsizes} lists the Wilcoxon $p$ values along with corresponding effect sizes (in parentheses) for
all relevant comparison groups for all applicable subjects and test input types. Statistically significant results are
highlighted in boldface.
We differentiate four levels of effect strength as per Cliff's Delta (d) values~\cite{romano2006exploring}:
{\em negligible} ($|d|$$\le$0.147), {\em small} (0.147$<$$|d|$$\le$0.33), {\em medium} (0.33$<$$|d|$$\le$0.474), and {\em large} ($|d|$$>$0.474).

The results (3rd to 6th columns of Table~\ref{tab:pvalueseffsizes}) indicate that the two new {\distea} versions ({\bf Csd+} and {\bf Scov+}) were
strongly significantly more effective, with mostly very strong (large) effect sizes, than both the baseline and {\bf Msg+}.
Meanwhile, {\bf Msg+} impact set sizes were significantly different from (smaller than) those of the baseline for all cases but the two smallest
subjects and Open Chord. Also, the majority of these significant cases came with a large effect size. These observations with {\bf Msg+}
and the baseline were similar to those between {\bf Csd+} and {\bf Scov+}.
In all, results of the statistical analyses corroborated what we observed from the earlier comparisons on medians and averages about the
strengths of the effects of various program information on {\distea}'s effectiveness.
On the other hand, the numbers of Table~\ref{tab:pvalueseffsizes} indicate significant (in terms of $p$ values)
and large (in terms of effect sizes) improvements of the three advanced versions of {\distea} over the basic {\distia}.

\vspace{6pt}\noindent\fbox{\parbox{\dimexpr\linewidth-2\fboxsep-\fboxrule\relax}{
\textbf{RQ2}:
Among the various forms of program information used by {\distea}, static dependencies combined with statement coverage
contributed the most to its effectiveness, followed by static dependencies alone and then by message-passing semantics.
The intra-component dependencies are the most contributing, single form of data, implying that incorporating
static dependence analysis may benefit greatly to the precision of dynamic dependence abstraction of distributed programs.
}}
\vspace{4pt}

\subsubsection{RQ3: Efficiency}
\setlength{\tabcolsep}{.45pt}
\renewcommand{\arraystretch}{0.97}

\begin{table}[htbp]
  \centering
  \caption{Efficiency results of the baseline analysis}\vspace{-8pt}
    \begin{tabular}{l|r|r|r|r|r|r}
    \hline
    \multirow{2}[4]{*}{\textbf{Subject \& input}} & \multicolumn{1}{c|}{\multirow{2}[4]{*}{\textbf{\shortstack{Original\\run (ms)}}}} & \multicolumn{4}{c|}{\textbf{time costs in milliseconds (ms)}} & \multicolumn{1}{c}{\multirow{2}[4]{*}{\textbf{\shortstack{space\\(KB)}}}} \\
\cline{3-6}    \multicolumn{1}{l|}{} &   & \multicolumn{1}{c|}{\textbf{\shortstack{Static\\analysis}}} & \multicolumn{1}{c|}{\textbf{\shortstack{Instr.\\run}}} & \multicolumn{1}{c|}{\textbf{\shortstack{Runtime\\overhead}}} & \textbf{\shortstack{Querying\\(stdev)}} &  \\
    \hline
    MultiChat & 5,461 & \textcolor[rgb]{ .439,  .188,  .627}{12,817} & \textcolor[rgb]{ .439,  .188,  .627}{5,735} & \textcolor[rgb]{ .439,  .188,  .627}{5.02\%} & \textcolor[rgb]{ .439,  .188,  .627}{4 (2)} & \textcolor[rgb]{ .439,  .188,  .627}{7} \\
    \hline
    NioEcho & 3,213 & \textcolor[rgb]{ .439,  .188,  .627}{13,365} & \textcolor[rgb]{ .439,  .188,  .627}{3,619} & \textcolor[rgb]{ .439,  .188,  .627}{12.64\%} & \textcolor[rgb]{ .439,  .188,  .627}{4 (2)} & \textcolor[rgb]{ .439,  .188,  .627}{5} \\
    \hline
    xSocket & 7,753 & \textcolor[rgb]{ .439,  .188,  .627}{24,842} & \textcolor[rgb]{ .439,  .188,  .627}{8,470} & \textcolor[rgb]{ .439,  .188,  .627}{9.25\%} & \textcolor[rgb]{ .439,  .188,  .627}{6 (2)} & \textcolor[rgb]{ .439,  .188,  .627}{85} \\
    \hline
    Thrift & 9,751 & \textcolor[rgb]{ .439,  .188,  .627}{23,143} & \textcolor[rgb]{ .439,  .188,  .627}{10,241} & \textcolor[rgb]{ .439,  .188,  .627}{5.03\%} & \textcolor[rgb]{ .439,  .188,  .627}{7 (1)} & \textcolor[rgb]{ .439,  .188,  .627}{18} \\
    \hline
    Open Chord & 4,856 & \textcolor[rgb]{ .439,  .188,  .627}{14,533} & \textcolor[rgb]{ .439,  .188,  .627}{4,931} & \textcolor[rgb]{ .439,  .188,  .627}{1.54\%} & \textcolor[rgb]{ .439,  .188,  .627}{8 (5)} & \textcolor[rgb]{ .439,  .188,  .627}{73} \\
    \hline
    ZooKeeper-integration & 37,239 & \multirow{3}[2]{*}{\textcolor[rgb]{ .439,  .188,  .627}{39,124}} & \textcolor[rgb]{ .439,  .188,  .627}{38,396} & \textcolor[rgb]{ .439,  .188,  .627}{3.11\%} & \textcolor[rgb]{ .439,  .188,  .627}{10 (2)} & \textcolor[rgb]{ .439,  .188,  .627}{94} \\
\cline{1-2}\cline{4-7}    ZooKeeper-system & 15,385 &   & \textcolor[rgb]{ .439,  .188,  .627}{18,565} & \textcolor[rgb]{ .439,  .188,  .627}{20.67\%} & \textcolor[rgb]{ .439,  .188,  .627}{24 (6)} & \textcolor[rgb]{ .439,  .188,  .627}{132} \\
\cline{1-2}\cline{4-7}    ZooKeeper-load & 94,187 &   & \textcolor[rgb]{ .439,  .188,  .627}{98,891} & \textcolor[rgb]{ .439,  .188,  .627}{4.99\%} & \textcolor[rgb]{ .439,  .188,  .627}{22 (5)} & \textcolor[rgb]{ .439,  .188,  .627}{142} \\
    \hline
    Voldemort-integration & 17,755 & \multirow{3}[2]{*}{\textcolor[rgb]{ .439,  .188,  .627}{132,536}} & \textcolor[rgb]{ .439,  .188,  .627}{18,662} & \textcolor[rgb]{ .439,  .188,  .627}{5.11\%} & \textcolor[rgb]{ .439,  .188,  .627}{22 (7)} & \textcolor[rgb]{ .439,  .188,  .627}{315} \\
\cline{1-2}\cline{4-7}    Voldemort-system & 11,136 &   & \textcolor[rgb]{ .439,  .188,  .627}{12,232} & \textcolor[rgb]{ .439,  .188,  .627}{9.84\%} & \textcolor[rgb]{ .439,  .188,  .627}{19 (4)} & \textcolor[rgb]{ .439,  .188,  .627}{197} \\
\cline{1-2}\cline{4-7}    Voldemort-load & 21,066 &   & \textcolor[rgb]{ .439,  .188,  .627}{21,198} & \textcolor[rgb]{ .439,  .188,  .627}{0.63\%} & \textcolor[rgb]{ .439,  .188,  .627}{29 (5)} & \textcolor[rgb]{ .439,  .188,  .627}{782} \\
    \hline
    Freenet & 54,794 & \textcolor[rgb]{ .439,  .188,  .627}{165,174} & \textcolor[rgb]{ .439,  .188,  .627}{61,876} & \textcolor[rgb]{ .439,  .188,  .627}{12.92\%} & \textcolor[rgb]{ .439,  .188,  .627}{114 (17)} & \textcolor[rgb]{ .439,  .188,  .627}{527} \\
    \hline
    \textbf{Overall average} & \textbf{33,213.7} & \textcolor[rgb]{ .439,  .188,  .627}{\textbf{73,854.9}} & \textcolor[rgb]{ .439,  .188,  .627}{\textbf{36,273.6}} & \textcolor[rgb]{ .439,  .188,  .627}{\textbf{8.07\%}} & \textcolor[rgb]{ .439,  .188,  .627}{\textbf{49.3 (44.9)}} & \textcolor[rgb]{ .439,  .188,  .627}{\textbf{395.7}} \\
    \hline
    \end{tabular}%
  \label{tab:costbaseline}%
\end{table}%

% Table generated by Excel2LaTeX from sheet 'costs-threeversion-in-paper'
\setlength{\tabcolsep}{.1pt}
\begin{table*}[htbp]
  \centering
  \caption{Time cost breakdown (in milliseconds) and storage cost (in KB) of the three advanced versions of {\tech}}\vspace{-6pt}
    \begin{tabular}{|l|r|r|r|r|r||r|r|r|r|r||r|}
    \hline
    \multirow{2}[4]{*}{\textbf{Subject \& input}} & \multirow{2}[4]{*}{\textbf{\shortstack{{\bf Msg+}\\querying\\(stdev)}}} & \multicolumn{4}{c|}{\textbf{{\bf Csd+} time costs in milliseconds}} & \multicolumn{1}{c|}{\multirow{2}[4]{*}{\textbf{\shortstack{{\bf Csd+}\\ Storage\\costs (KB)}}}} & \multicolumn{4}{c|}{\textbf{{\bf Scov+} time costs in milliseconds}} & \multicolumn{1}{c|}{\multirow{2}[4]{*}{\textbf{\shortstack{{\bf Scov+}\\Storage\\costs (KB)}}}} \\
\cline{3-6}\cline{8-11}    \multicolumn{1}{|l|}{} & \multicolumn{1}{c|}{} & \multicolumn{1}{c|}{\textbf{\shortstack{Static\\analysis}}} & \multicolumn{1}{c|}{\textbf{\shortstack{Instr.\\run}}} & \multicolumn{1}{c|}{\textbf{\shortstack{Runtime\\overhead}}} & \textbf{\shortstack{Querying\\(stdev)}} &   & \multicolumn{1}{c|}{\textbf{\shortstack{Static\\analysis}}} & \multicolumn{1}{c|}{\textbf{\shortstack{Instr.\\run}}} & \multicolumn{1}{c|}{\textbf{\shortstack{Runtime\\overhead}}} & \textbf{\shortstack{Querying\\(stdev)}} &  \\
    \hline
    MultiChat & 4 (2) & \textcolor[rgb]{ .439,  .188,  .627}{175,474} & \textcolor[rgb]{ .439,  .188,  .627}{25,298} & \textcolor[rgb]{ .439,  .188,  .627}{363.25\%} & \textcolor[rgb]{ .439,  .188,  .627}{46 (21)} & \textcolor[rgb]{ .439,  .188,  .627}{84} & 221,839 & 26,246 & 380.61\% & \multicolumn{1}{r||}{19 (7)} & 85 \\
    \hline
    NioEcho & 5 (2) & \textcolor[rgb]{ .439,  .188,  .627}{277,903} & \textcolor[rgb]{ .439,  .188,  .627}{8,147} & \textcolor[rgb]{ .439,  .188,  .627}{153.56\%} & \textcolor[rgb]{ .439,  .188,  .627}{52 (19)} & \textcolor[rgb]{ .439,  .188,  .627}{133} & 367,268 & 9,716 & 202.40\% & \multicolumn{1}{r||}{34 (12)} & 136 \\
    \hline
    xSocket & 8 (2) & \textcolor[rgb]{ .439,  .188,  .627}{781,925} & \textcolor[rgb]{ .439,  .188,  .627}{19,156} & \textcolor[rgb]{ .439,  .188,  .627}{147.08\%} & \textcolor[rgb]{ .439,  .188,  .627}{394 (114)} & \textcolor[rgb]{ .439,  .188,  .627}{6,332} & 859,158 & 21,683 & 179.67\% & \multicolumn{1}{r||}{256 (39)} & 6,540 \\
    \hline
    Thrift & 9 (7) & \textcolor[rgb]{ .439,  .188,  .627}{401,621} & \textcolor[rgb]{ .439,  .188,  .627}{21,397} & \textcolor[rgb]{ .439,  .188,  .627}{119.43\%} & \textcolor[rgb]{ .439,  .188,  .627}{775 (48)} & \textcolor[rgb]{ .439,  .188,  .627}{4,407} & 462,480 & 23,924 & 145.35\% & \multicolumn{1}{r||}{337 (17)} & 4,438 \\
    \hline
    Open Chord & 10 (6) & \textcolor[rgb]{ .439,  .188,  .627}{196,958} & \textcolor[rgb]{ .439,  .188,  .627}{29,078} & \textcolor[rgb]{ .439,  .188,  .627}{498.81\%} & \textcolor[rgb]{ .439,  .188,  .627}{21,643 (16,291)} & \textcolor[rgb]{ .439,  .188,  .627}{4,109} & 264,532 & 37,319 & 668.51\% & \multicolumn{1}{r||}{100 (256)} & 4,173 \\
    \hline
    ZooKeeper-integration & 12 (2) & \multirow{3}[2]{*}{\textcolor[rgb]{ .439,  .188,  .627}{3,188,103}} & \textcolor[rgb]{ .439,  .188,  .627}{84,063} & \textcolor[rgb]{ .439,  .188,  .627}{125.74\%} & \textcolor[rgb]{ .439,  .188,  .627}{8,855 (15)} & \textcolor[rgb]{ .439,  .188,  .627}{15,689} & \multirow{3}[2]{*}{3,374,107} & 95,520 & 156.51\% & \multicolumn{1}{r||}{8,850 (15)} & 15,708 \\
\cline{1-2}\cline{4-7}\cline{9-12}    ZooKeeper-system & 28 (5) &   & \textcolor[rgb]{ .439,  .188,  .627}{37,704} & \textcolor[rgb]{ .439,  .188,  .627}{145.07\%} & \textcolor[rgb]{ .439,  .188,  .627}{21,737 (7,208)} & \textcolor[rgb]{ .439,  .188,  .627}{28,402} &   & 48,119 & 212.77\% & \multicolumn{1}{r||}{23,233 (216)} & 36,804 \\
\cline{1-2}\cline{4-7}\cline{9-12}    ZooKeeper-load & 28 (8) &   & \textcolor[rgb]{ .439,  .188,  .627}{303,009} & \textcolor[rgb]{ .439,  .188,  .627}{221.71\%} & \textcolor[rgb]{ .439,  .188,  .627}{16,313 (939)} & \textcolor[rgb]{ .439,  .188,  .627}{28,412} &   & 412,846 & 338.33\% & \multicolumn{1}{r||}{13,988 (1,773)} & 35,058 \\
    \hline
    Voldemort-integration & 25 (7) & \multicolumn{1}{c|}{\multirow{3}[2]{*}{\textcolor[rgb]{ .439,  .188,  .627}{-}}} & \multicolumn{1}{c|}{\textcolor[rgb]{ .439,  .188,  .627}{-}} & \multicolumn{1}{c|}{\textcolor[rgb]{ .439,  .188,  .627}{-}} & \multicolumn{1}{c||}{\textcolor[rgb]{ .439,  .188,  .627}{-}} & \multicolumn{1}{c|}{\textcolor[rgb]{ .439,  .188,  .627}{-}} & \multicolumn{1}{c|}{\multirow{3}[2]{*}{-}} & \multicolumn{1}{c|}{-} & \multicolumn{1}{c|}{-} & \multicolumn{1}{c||}{-} & \multicolumn{1}{c|}{-} \\
\cline{1-2}\cline{4-7}\cline{9-12}    Voldemort-system & 22 (5) &   & \multicolumn{1}{c|}{\textcolor[rgb]{ .439,  .188,  .627}{-}} & \multicolumn{1}{c|}{\textcolor[rgb]{ .439,  .188,  .627}{-}} & \multicolumn{1}{c||}{\textcolor[rgb]{ .439,  .188,  .627}{-}} & \multicolumn{1}{c|}{\textcolor[rgb]{ .439,  .188,  .627}{-}} &   & \multicolumn{1}{c|}{-} & \multicolumn{1}{c|}{-} & \multicolumn{1}{c||}{-} & \multicolumn{1}{c|}{-} \\
\cline{1-2}\cline{4-7}\cline{9-12}    Voldemort-load & 42 (13) &   & \multicolumn{1}{c|}{\textcolor[rgb]{ .439,  .188,  .627}{-}} & \multicolumn{1}{c|}{\textcolor[rgb]{ .439,  .188,  .627}{-}} & \multicolumn{1}{c||}{\textcolor[rgb]{ .439,  .188,  .627}{-}} & \multicolumn{1}{c|}{\textcolor[rgb]{ .439,  .188,  .627}{-}} &   & \multicolumn{1}{c|}{-} & \multicolumn{1}{c|}{-} & \multicolumn{1}{c||}{-} & \multicolumn{1}{c|}{-} \\
    \hline
    Freenet & 141 (24) & \multicolumn{1}{c|}{\textcolor[rgb]{ .439,  .188,  .627}{-}} & \multicolumn{1}{c|}{\textcolor[rgb]{ .439,  .188,  .627}{-}} & \multicolumn{1}{c|}{\textcolor[rgb]{ .439,  .188,  .627}{-}} & \multicolumn{1}{c||}{\textcolor[rgb]{ .439,  .188,  .627}{-}} & \multicolumn{1}{c|}{\textcolor[rgb]{ .439,  .188,  .627}{-}} & \multicolumn{1}{c|}{-} & \multicolumn{1}{c|}{-} & \multicolumn{1}{c|}{-} & \multicolumn{1}{c||}{-} & \multicolumn{1}{c|}{-} \\
    \hline
    \textbf{Overall average} & \textbf{61.3 (56.1)} & \textcolor[rgb]{ .439,  .188,  .627}{\textbf{846,958.6}} & \textcolor[rgb]{ .439,  .188,  .627}{\textbf{88,705.8}} & \textcolor[rgb]{ .439,  .188,  .627}{\textbf{209.42\%}} & \textcolor[rgb]{ .439,  .188,  .627}{\textbf{12,411.8 (11,619.7)}} & \textcolor[rgb]{ .439,  .188,  .627}{\textbf{15,599.8}} & \textbf{914,834.8} & \textbf{114,809.6} & \textbf{283.91\%} & \textbf{4,897.6 (5,444.4)} & \textbf{18,378.6} \\
    \hline
    \end{tabular}%
  \label{tab:threeversioncost}%
  \vspace{-10pt}
\end{table*}%

%
%Table~\ref{tab:costs} %(from the fifth column)
%lists all relevant costs of {\distea} in this study,
%%that {\distea} incurred for producing the impact sets analyzed above,
%including the runtime of the static analyzer (\emph{Static analysis}),
%overhead of the runtime monitors (\emph{Runtime overhead}) measured as ratios of the runtime of original
%program (\emph{Normal run}) over the instrumented one (\emph{Instrumented run}), and runtime of
%the post-processor (\emph{Querying}).

Tables~\ref{tab:costbaseline} and~\ref{tab:threeversioncost} list all relevant costs of the baseline and the three
advanced versions of {\distea}, respectively.
The costs reported include the time cost of static analysis, run-time overhead measured as ratios of the execution time of the original
program (\emph{Original run}) over the execution time of the instrumented one (\emph{Instr. run}), and impact querying time.
%For the {\bf Basic} and {\bf Msg+} version, the static analysis includes (bytecode) instrumentation.
The time costs are reported in milliseconds (ms), and storage costs in KB which include
the disk space taken by serialized MDGs (only for {\bf Csd+} and {\bf Scov+}) and execution traces (for all of the four instantiations).

With the {\bf Basic} version (baseline), the static analysis generally took longer for larger subjects, as expected, yet
still below 3 minutes even on the largest system Freenet.
Run-time and querying costs are consistently correlated to subject sizes as well as the type of
inputs and the size of execution traces (the seventh column), with the worst case seen by ZooKeeper and Freenet, respectively.
%on its unit-test input set, which is by far the largest
%among all subjects and inputs studied.
Nevertheless, the run-time overhead was at worst 21\% and the longest querying time was just a tenth of %no more than
one second.
The last row (of Tables~\ref{tab:costbaseline}) gives the average costs weighted by the number of queries studied in
each subject and input type. Overall, the baseline needed only 74 seconds for static analysis and
incurred only 8\% runtime overhead. Both the querying and storage costs were negligible.

The {\bf Msg+} version shares the same instrumentation as the {\bf Basic} version (second column of Table~\ref{tab:threeversioncost}).
The additional per-process message-receiving maps
incurred negligible overheads. Thus, we did not show the cost breakdown but only the querying time for this framework instantiation.
Our results show the additional data utilized only caused very small increase in the querying costs, which remained
well below one second per query.
The efficiency results of both the {\bf Basic} and {\bf Msg+} versions were highly consistent with those obtained in our
preliminary studies~\cite{cai2016distea}.

The {\bf Csd+} version incurred much higher costs in any of the three phrases of {\tech} than the baseline and {\bf Msg+}.
Since static dependencies are not just affected by the size of a program but more by the its logic complexity,
the static-analysis costs were not linearly correlated to subject sizes. Moreover, very-large systems (Voldemort and Freenet)
were not successfully analyzed by this version as mentioned earlier, thus we missed the results accordingly.
Among the six remaining subjects, the largest took the longest time of about 53 minutes. This should be readily
acceptable for systems at such a scale. Also, this is a one-time cost (for the single program version analyzed by {\distea}) as the instrumented code and the constructed MDG can be reused for any inputs and for computing any queries afterwards.
The runtime overheads were considerably larger too, almost 5x with Open Chord,
since {\bf Csd+} requires full sequences of method-execution events. Further, traversing the MDG and longer traces led to
much higher querying costs. Nevertheless, even as the worse case, querying the impact set of a method in ZooKeeper
against the system test took 22 seconds, which is reasonably affordable.
On overall (weighted) average, {\bf Csd+} took 14 minutes for static analysis and 12 seconds for
answering an impact-set query with 2x run-time overhead, over the 8 cases it was applied to.

The {\bf Scov+} version adds the collection and use of statement coverage data to {\bf Csd+}, thus it is supposedly the
most heavyweight instance of our framework. Our results, however, show that this additional data did not
incur much additional static analysis costs or run-time slowdown (68 seconds and 0.7x, respectively).
%Our current implementation of {\distea} profiles statement coverage in a separate pass (independent of the main analysis pass),
%which causes excessive overhead. The run-time slowdown would be reduced if the profiling is incorporated in the main analysis pass.
Noticeably, not only did the statement coverage contribute significantly to the effectiveness of our framework (see RQ2),
the additional static analysis and run-time overheads are also paid off by the savings of querying costs.
The statement coverage data helped prune the static dependencies before they were used for dynamic dependence
abstraction (see Lines 6 and 14 of Algorithm~\ref{algo2}). As a result, the querying costs incurred by {\bf Scov+}
were only 40\% of those by {\bf Csd+} on overall average.
As for {\bf Csd+}, storage costs of {\bf Scov+} were also negligible (37MB in the worse case, and 18.4MB on overall average).
%over 1.5 hours, which may or may not be affordable for developers.
%For other successfully analyzed subjects, however, the querying costs were at most 2.5 minutes, which should be
%reasonably practical.
%
%For both the {\bf Basic} and {\bf Msg+} versions, storage costs are also closely connected to the type of inputs and subject sizes, of
%which the largest was less than 1MB for Voldemort against the load test.
%In other cases, this cost was at most 0.5MB.
%The storage costs of {\bf Csd+} were considerably higher, yet remained within 28MB. %at worst.

In sum, the {\bf Basic} and {\bf Msg+} versions were highly efficient in both time and space dimensions, thus they are readily
scalable to large distributed systems.
%the {\bf Basic} {\distea} and its {\bf Msg+} version were
%highly efficient in both time and space dimensions, and thus they appears to be readily scalable to large systems:
%on average, they cost one minute for instrumentation %static analysis
%and 50ms for computing the impact set per query, with mean run-time overhead of below 8\%.
%The slightly additional costs of {\bf Msg+}, together with its noticeably higher effectiveness, make it
%a much more cost-effective option than the {\bf Basic} version.
The {\bf Csd+} and {\bf Scov+} versions were considerably more expensive, mainly because of the substantial cost of static dependence
analysis. %as the cost of higher precision.
%The {\bf Csd+} version was by far the most expensive technique within our {\tech} framework.
Nevertheless, %given their much superior precision than the two efficient versions,
for systems of small to medium levels of size and complexity,
they still provide compelling options, offering much higher effectiveness at reasonable costs.
To very large and complex systems, more efficient static dependence analysis would be necessary to scale up these more
precise {\distea} versions.
%
%According to our results, for systems of small to medium levels of size and complexity, these two versions could be a comparable or
%superior option to {\bf Msg+}, due to its even much higher effectiveness at reasonable costs.
%Meanwhile, the static dependence analysis within {\bf Csd+} can be a bottleneck that limits its scalability.

%Tracing the first message-receiving events and utilizing them in addition, the enhanced version is expected to
%incur higher time and space costs. Yet, our results (not shown here) reveal that the increases are quite marginal and
%thus negligible:
%on overall average, the run-time and space overhead is less than 1\% higher, and the querying time is 9.8ms longer.
%Since both versions share the same instrumenter, %static analyzer, the static analysis
%the instrumentation
%cost remains the same.
%Thus, in all, the enhanced version tends to be much more cost-effective.
%%of {\tech} tends to a much more cost-effective option than its {\bf Basic} version.

\vspace{4pt}\noindent\fbox{\parbox{\dimexpr\linewidth-2\fboxsep-\fboxrule\relax}{
\textbf{RQ3}:
The {\bf Basic} and {\bf Msg+} versions of {\distea} were highly efficient and scalable to large distributed systems, taking
less than 1.5 minutes for static analysis and less than 1 second for querying, with 8\% run-time overhead.
The {\bf Csd+} and {\bf Scov+} versions were significantly more expensive, yet their costs remain reasonable and practically
acceptable for systems of small to medium levels of size and complexity, with 15 minutes for static analysis, 12 seconds for answering a query, and 2.8x overhead on average. Storage costs were negligible for the entire framework.
}}
\vspace{4pt}

\subsubsection{RQ4: Cost-Effectiveness Tradeoffs}
In practice, developers need to consider both the cost and effectiveness of an analysis tool (i.e., the balance between these two factors) to make their decisions on tool selection~\cite{latoza2006maintaining,de2008empirical,lindvall1998well,do2017just,cai2016diapro}.
To investigate the capabilities of our framework in offering variable cost-effectiveness, we
put together the effectiveness improvements and overhead increases of the three advanced versions over the baseline.
Specifically, we compute the cost-effectiveness as the ratio of improvement percentage of mean precision (from Table~\ref{tab:meaneff}) to the increase
%percentage,
factor of average costs (from Tables~\ref{tab:costbaseline} and~\ref{tab:threeversioncost}), relative to the baseline.
We consider two different classes of cost separately: (1) the per-query cost for answering each impact-set query, which is
different from query to query, and (2) the one-time cost for static analysis and profiling, which is incurred once for all
queries. Storage costs are not considered here since they were all quite trivial in all cases with any of the four {\distea} versions.

\begin{figure*}[tp]
 \centering
 \setlength{\tabcolsep}{1pt}
\begin{tabular}{c}
 \subfloat{
  \includegraphics[width=1.0\textwidth]{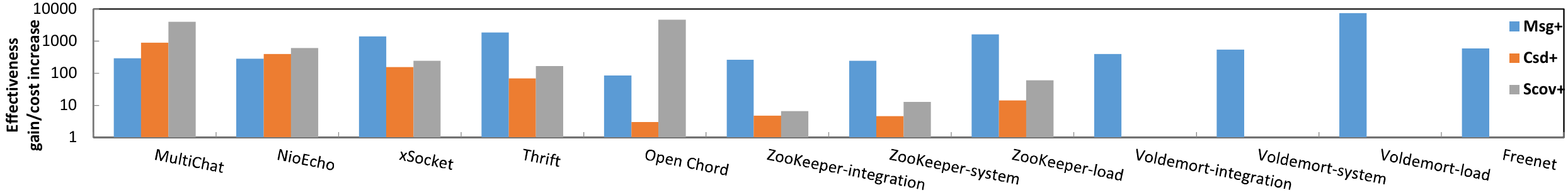}
   \label{fig:effcostquery}
   } \\
 \subfloat{
  \includegraphics[width=1.0\textwidth]{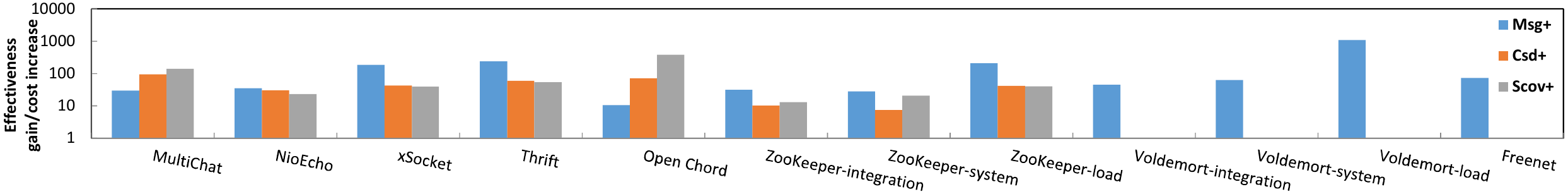}
   \label{fig:effcostother}
   }
\end{tabular}
 \vspace{-8pt}
 \caption{Cost-effectiveness of the three advanced versions of {\distea} expressed as the ratios of their mean effectiveness gain to the factor of increase in the average querying cost (top) and total cost of the first two phases (bottom), both against the baseline {\distia}. The higher the ratio, the better (more cost-effective).}
 \label{fig:effcostall}
 \vspace{-10pt}
\end{figure*}

Figure~\ref{fig:effcostall} shows the cost-effectiveness ($y$ axis) comparisons among the three advanced versions for
all the cases each version was applied to (as listed on the $x$ axis), for the querying cost (top) and one-time costs combined (bottom).
To better differentiate the results of the three techniques compared, the cost-effectiveness is shown in a logarithmic scale (with a base 10) after the effectiveness improvement percentage was enlarged by 1,000 times.
As shown, in terms of querying costs, {\bf Msg+} appeared to be the most cost-effective version for all cases but the two smallest subjects and
Open Chord (for which {\bf Scov+} was the most cost-effective). This is mainly because of the highly lightweight nature of {\bf Msg+}. As a result, while the costs of the two new {\distea} versions were not very high in an absolute term, the ratios of these costs to those of {\bf Msg+}
were large, outweighing their precision improvement percentages over {\bf Msg+}.
Among the two new versions, however, {\bf Scov+} was consistently most cost-effective than {\bf Csd+}, due to the substantially higher precision
yet relatively small cost increases of {\bf Scov+} over {\bf Csd+}.

Concerning the costs of the first two phases combined, the general contrasts among the three advanced versions were similar to what we observed
in the cases in which only querying costs are considered in the cost-effectiveness measure.
Except for Open Chord, {\bf Msg+} always outperformed the other two versions. The reason mainly lies in the costs of static dependence analysis
incurred by {\bf Csd+} and {\bf Scov+} being substantially larger than even the total cost of {\bf Msg+}.
On the other hand, the advantage of {\bf Scov+} over {\bf Csd+} was of a lesser extent (albeit still noticeable) in contrast to their comparisons
when only querying costs were concerned. This is because {\bf Scov+} was always more expensive than {\bf Csd+} for the first two phases.

Overall, {\bf Msg+} achieved the highest cost-effectiveness among the three advanced versions due to its high efficiency and scalability, along with
significant precision advantage over the baseline. The two new versions, {\bf Csd+} and {\bf Scov+},
remain practically cost-effective options, especially for software systems of medium or smaller sizes and complexity or for users who
need much higher precision even at the cost of considerably higher overheads. In particular, {\bf Scov+} was consistently more cost-effective than {\bf Csd+}. Thus, where having less false negatives are of a higher priority, {\bf Scov+} would generally be a better option.
%Finally, we regard all the cost-effectiveness balances provided by the three versions as practical

\vspace{4pt}\noindent\fbox{\parbox{\dimexpr\linewidth-2\fboxsep-\fboxrule\relax}{
\textbf{RQ4}:
{\distea} provides variable and flexible cost-effectiveness balance options to accommodate diverse needs.
The {\bf Basic} and {\bf Msg+} versions provide a relatively rough but rapid solution, between which {\bf Msg+} offers the best cost-effectiveness
when high precision is not a priority; otherwise, {\bf Scov+} is the most cost-effective technique, especially for systems to which
it scales. These tradeoffs provide guidance for developers to choose which {\distea} version to use in varied situations.
}}
\vspace{4pt}

\subsection{Threats to Validity}\label{sec:validitythreats}
The main threat to \emph{internal} validity lies
in possible errors in our {\distea} implementation and experiment scripts.
To reduce this threat, we did a careful code review for our tools and
%with the inputs
manually validated the correctness of their functionalities and analysis results against
the two smallest subjects.
%Also, the low-level bytecode instrumentation
%is largely handled by Soot, a framework that has matured over many years; and the runtime
%monitors reused the code from tools developed in several previous work.
An additional such threat concerns possible missing (remote) impacts due to
network I/Os that were not monitored at runtime. However, we checked the code
of all subjects and confirmed that they only used the most common message-passing APIs monitored
by our tool when executing the program inputs we utilized.
In general, for arbitrary distributed systems, the soundiness of {\distea} relies on the
identification of all such API calls used in the system.
Yet another threat is that there might be hidden dependencies between methods induced by external storage I/Os: for example, a method reading a disk file is hidden-dependent on a method that writes the same file. Similar hidden dependencies can also induced by database accesses. Currently, {\tech} does not consider such hidden dependencies; instead, it focuses on independencies among code entities due to memory accesses and network communications.

%%Also, for Schedule1, we used a version translated from C to Java,
%%%For Schedule1 for which we used a translated Java version from C, we verified that the outcomes of all test cases remained the same.

The main threat to \emph{external} validity is that our study results may not generalize to
all other distributed programs and input sets.
%In this study, we considered only limited
%number of subjects, which may not represent all real-world distributed systems, and only subsets of inputs,
%which do not necessarily represent all behaviours of the studied systems.
To reduce this threat, we have chosen subject programs of various sizes and application domains, including
the six industry-scale systems in different domains. In addition, we considered
different types of inputs, including integration, system, and load tests.
Many of these tests came as part of the subjects, except for the integration tests which
we created according to the official online documentation (quick-start guide) of these systems.
%their system-level functionalities as documented on their official websites.

The main threat to \emph{construct} validity concerns the metrics used for the evaluation.
Without directly comparable peer techniques in the literature,
%we assumed that developers would use coverage-based approach like \emph{MCov}, as a possible %representative
%alternative to {\distea}.
we used the {\bf Basic} version of our framework as the baseline approach to assess the improvements made by
the advanced versions of the same framework.
While it is intuitive to compare {\tech} with a technique that it extends, this choice of baseline
might cause potential biases.
%
%, to narrow down the search space of potential impacts in
%the context of distributed %multiprocess
%executions.
%
%To mitigate this threat, we examined the composition of each impact set and analyzed
%the effectiveness with respect to its local and remote subsets in addition to that
%of the holistic impact set to help demonstrate the usefulness of {\distea}.
%
Also, we did not have the ground-truth impact sets to compute precision and recall metrics in absolute terms.
A precise statement-level forward dynamic slicer would be able to generate the ground-truth needed, yet such
a slicer is not currently available to us.
Thus, we used the impact-set size ratio as a {\em relative} effectiveness measure.
Yet, referring to the {\em same} baseline enabled our comparison of the three advanced versions of {\tech}.
While not ideal, such relative comparisons and measures suffice for our goal of understanding how the varied design factors
in the dynamic dependence abstraction of distributed programs affect the cost-effectiveness tradeoffs.

%is %in our experimental design.
%Without any additional knowledge, we gave the same weight to the impact sets of every method.
%However, in practice, developers may find some methods more important than others, and thus
%the reported average precision improvements of {\tool} might not represent the actual precisions
%that developers will experience.
%Also,
%that we used
%impact set sizes as inverse indicators of precision
%assuming that recall is not affected.
%This is the safe for program understanding, but for predicting the impacts that \emph{actual} changes will have,
%recall might be less than perfect if those changes modify the control flow of the program
%to execute methods not reported by {\tech}.
%
%the assumption about the safety of impact sets of our analyses. With respect to actual
%impact sets for concrete changes, however, the recall may not be perfect especially when those changes modify control flows at runtime. Nevertheless, our analyses are safe relative to
%the execution data utilized for the single program version available to them.

Finally, a \emph{conclusion} threat concerns the data points analyzed:
We applied the statistical analyses only to methods for which
impact sets could be queried (i.e., methods executed at least once).
Also, the present study only considered potential changes in single methods for each query, while in practice
developers may plan for changes in multiple methods at a time, which may lead to different %study
results.
To minimize this threat, we adopted the strategy for all experiments and calculated
%experimental
the metrics for every possible query.
%We also considered every such method as a query in
%our study.
%Investigating multiple-method queries, and real changes, is among future work. %, among others.
To reduce possible biases in our statistical analyses, we chose non-parametric hypothesis testing and effect size measures that
do not rely on the normality of underlying data's distribution.

%% file: discussion.tex
%!TEX root = paper.tex
%  \vspace{-6pt}
\section{Discussion}\label{sec:discussion}
In this section, we discuss two additional pertinent issues with our technical approach:
validation of {\tech}'s analysis results and selection of {\tech} instantiations for practical use.

\subsection{Result Validation}
As discussed in Section~\ref{subsubsec:impactcomp}, the safety of the analysis results produced by any of the instantiations of {\tech} is established through the soundiness of the respective analysis technique. In principle, this safety {\em implies} perfect recall---although as the safety comes from soundiness, the perfection holds only for queries that do not involve dependencies induced by dynamic language constructs (i.e., reflection and JNI in our case of Java).
Also, as discussed in Sections~\ref{sec:validitythreats}, due to the lack of ground truth, we could not compute recall and precision in absolute terms. Instead, we gained confidence about recall of each advanced version of {\tech} as per the conservative nature of its dependence pruning relative to the baseline, and used the impact-set size ratios, also relative to the baseline, to measure precision indirectly. These strategies suffice for the purpose of studying the improvements of these advanced versions and their underlying techniques over the baseline, which is indeed our main goal with this work.

Nevertheless, it is still important to understand the actual analysis accuracy of the dependence sets produced by (any instantiation of) our framework. Unfortunately, we are not aware of any existing automated tool that can compute the ground-truth dependence set of any given query against the distributed programs and their executions used in our evaluation; neither is it possible to produce all such ground truth manually.
Thus, we extended the manual inspection adopted for the preliminary version of this work~\cite{cai2016distea}, following a prior methodology aiming at a similar purpose~\cite{cai2018hybrid} as summarized as follows.
For each of our studied subjects and executions, we randomly chose ten queries for which none of the {\tech} instantiations produced a dependence set that includes more than 50 methods---we had to limit the scale of this manual study because of its tedious and heavyweight nature.
Then, for each chosen query, we manually produced the ground-truth dynamic dependence set
by in-depth code review and step-through tracking of the executions (like step-over debugging) while leveraging available documentation of respective subject systems.
Using these ground-truth dependencies, we computed the precision and recall for each of the chosen queries.

With these cases, the average (over all the subject executions and the ten queries per execution) precision of {\bf Basic}, {\bf Msg+}, {\bf Csd+}, and {\bf Scov+} was 55.1\%, 64.8\%, 93.4\%, and 98.6\%, respectively. Importantly, for all these cases, the recall of any instantiation was 100\%---indeed, the dependence set of each of these sample queries did not include any dependencies induced by reflective or JNI calls. This confirmed that the pruning carried out by the three advanced versions only removed false-positive dependencies.

\subsection{Instantiation Selection}
As shown, our framework offers flexible options of cost-effectiveness tradeoffs through the four instantiations each providing a distinct level of such tradeoffs. For a user with a particular task that relies on the dynamic dependencies computed by {\tech}, a practical problem is to how to choose the right instantiation (tool) out of the varied alternatives. Thus, we make recommendations in this regard as guidelines as follows.

Figure~\ref{fig:effcostall} suggests that the most cost-effective option varied both with different subject systems (of varying code size and complexity) and with different executions (of varying execution complexity) of the same systems.
Yet for systems and executions of a small to medium size and complexity, the user might want to choose {\bf Scov+} given that it tended to offer the highest cost-effectiveness in most cases in our evaluation study. %more likely than others did.
For large-scale and highly-complex distributed systems and executions, though, the user would be most likely recommended to opt for {\bf Msg+} given the highest level of cost-effectiveness it offered.

However, there are also situations in which the user's best option may not be the most cost-effective one.
For instance, the user would choose the instantiation that offers higher precision for the system and execution on hand as long as the added costs are still affordable, even though the extra costs are not best paid off. For another example, the fastest instantiation might be the best choice to
the user when the small efficiency advantages (over the second fastest) matter, although this fastest tool is not as cost-effective as other options.
In these situations, the user should choose the instantiation that satisfies the cost or effectiveness priority with respect to the particular task.
Note that the presence of these situations justifies not only offering the most cost-effective analysis techniques but also providing those of other levels of cost-effectiveness, just as {\tech} did.

%% file: related.tex
%!TEX root = paper.tex
\vspace{-0pt}
\section{Related Work}\label{sec:related}
\vspace{-2pt}
%The most related papers, ideally grouped by category or topic. How our technique and evaluation is better/worse/differs.
% 1. dynamic impact analysis
Four main categories of previous work are most related to ours: impact analysis, dependence analysis of
distributed programs, logging for distributed systems, and dynamic partial order reduction.
\subsection{Impact Analysis}
%The execute-after-sequences ({\EAS})
The EAS approach~\cite{apiwattanapong05may} which partially inspired {\distea} %utilized by {\distea}
%for impact prediction
is a performance optimization of its predecessor {\PI}%introduced by Law and Rothermel
~\cite{law03may}.
Many other dynamic impact analysis techniques also exist~\cite{Li2013ASC,cai2020reflection}, aiming at
improving precision~\cite{cai14diver,cai14sensa,caiprioritizing},
%~\cite{Breech2006IIM,Huang2007PDI,Hattori2008OTP,cai14diver},
recall~\cite{maia2010hybrid},
efficiency~\cite{Breech2005ACO},
%~\cite{Breech2004OIA,Breech2005ACO},
and cost-effectiveness~\cite{cai15diverplus,cai2016diapro} over {\PI} and {\EAS}.
However, these techniques did not address distributed or multiprocess programs that we focus on in this work.
%In this work, we leveraged the order of method executions for efficient impact analysis as well as these previous
%approaches did. However, none of these techniques can fully work with distributed multiprocess programs, computing
%impacts both with single processes and across multiple asynchronous processes, as {\distea} is devoted to doing.
%
% 8. relation to our recent work
%Impact analysis purely based on execute-after relation inference has been known to be imprecise~\cite{cai14mar}. It can
%also suffer from imperfect recall relative to ground-truth impacts of actual changes~\cite{cai15jss}, but it is safe
%with respect to the concrete executions utilized by the analysis.
Two recent advances in dynamic impact analysis,
{\diver}~\cite{cai14diver} and the unified framework in~\cite{cai15diverplus,cai2016diapro},
utilize hybrid program analysis to achieve higher precision and more flexible cost-effectiveness options over {\EAS}-based
approaches, but still target single-process programs only.
On the other hand, the {\bf Csd+} version of {\tech} was initially motivated by these prior works to
incorporate the static intra-component dependencies for effectiveness improvement.
%
%As a first step, {\distea} sacrifices precision for high efficiency. However, it would be interesting to %of great interest to
%adopt hybrid approaches for distributed systems too.
%For instance, %among other improvements, one may be to immediately gain
%one may immediately gain
%better analysis precision by first using static dependencies to prune false-positive impacts
%\emph{within} each process, as {\diver} did, and then propagating impacts across process boundaries by means of the
%%Lamport timestamps as
%analysis algorithms
%used in {\distea}. More aggressive pruning may further be obtained by leveraging
%more and/or finer-grained interprocess dependencies, such as communication dependencies and synchronization dependencies~\cite{kamkar1995dynamic,mohapatra2006distributed}, as already exploited by distributed-program slicing techniques~\cite{barpanda2011dynamic}.
%
In preliminary work~\cite{cai2016distia,cai2016distea}, we developed an early prototype of {\tech} that includes the
the {\bf Basic} and {\bf Msg+} versions. This paper extends that prior work both technically by adding {\bf Csd+} and {\bf Scov+}, and
empirically through a larger-scale and more extensive evaluation.

% 5. impact analysis for distributed systems
An impact analysis for distributed systems, Helios~\cite{popescu2012impact}
can predict impacts of potential changes to support evolution tasks for DEBS.
%This technique,
However, it relies on particular message-type filtering and manual annotations %by developers
in addition to a few other constraints. Although these limitations are largely lifted
by its successor Eos~\cite{garcia2013identifying}, both approaches are %the analysis is
\emph{static} and limited to DEBS only, as is another
 technique~\cite{Tragatschnig2014IAE} which identifies impacts based on change-type classification yet
 ignores intra-component dependencies hence provides incomplete results. %, similar to ~\cite{sun2010change,ren04oct}.
%This approach relies on sink/source matching among components
%according to event type correlation, with ignorance of intra-component dependencies and limited applicability (to DEBS only).
While sharing similar goals, {\distea} targets a broader range of distributed systems than DEBS using \emph{dynamic} analysis and without relying on special source-code information (e.g., interface patterns) as those techniques do.
%
%automatically finds message types and message flow that involve relevant statements, thus seemingly capable of retrieving
%code-based impacts, yet is designed exclusively for DEBS as the performance-wise optimizations in~\cite{jayaram2011program}.
%
Unlike our dependence-based approach, a traceability-based solution~\cite{bohner96jun} is presented in~\cite{sneed2001impact} which relies on a well-curated repository of various software models. The dynamic
impact analysis for component-based software in~\cite{feng2006applying} works at architecture level, different
from ours working for distributed programs at code level.

\subsection{Dependence Analysis of Distributed Programs}
%%% why dependency analysis
Historically, dependence analysis has underlain a wide range of code-based software engineering tasks and
associated techniques~\cite{podgurski90sep,horwitz1992use}. In particular, a large body and variety of dependence-based
approaches have been proposed over the past few decades to support development and maintenance in general ~\cite{gallagher1991using,loyall1993using,Li2013ASC,orso04apr,petrenko2009variable,cai15tracer}, and fault diagnosis~\cite{baah10jul,bates93jan,faultlocsurvey2016,xqfu20fsetool} and security defense in particular~\cite{dalton2007raksha,attariyan2010automating,kemerlis2012libdft,xqfu19fsetool,fu2020scaling,fu2021flowdist}.
For instance, for maintaining and evolving a software system, it is essential to assess the influence of given program entries of interest on the rest of the program~\cite{bohner96jun,Li2013ASC} for change planning (deciding whether to realize changes at
candidate change locations) and fulfillment (deciding where to realize the changes)~\cite{bohner96jun,rovegard2008empirical,tao2012software}.

% 2. fine-grained dependence analysis of concurrent (multithreaded but centralized) programs
Using fine-grained %code-based
dependency analysis,
a large body of work attempted to extend traditional slicing algorithms to concurrent
programs~\cite{krinke2003context,xiao2005improved,Nanda2006ISM,giffhorn2009precise,xu2005brief} yet
mostly focused on centralized, and primarily multithreaded, ones.
%either multithreaded or multiprocess but \emph{centralized} software.
For those programs, traditional dependence analysis was extended to handle additional
%data and control
dependencies due to shared variable accesses, synchronization, and communication
between threads and/or processes (e.g.,~\cite{krinke2003context,Nanda2006ISM}).
While {\distea} also handles multithreaded programs, %executions as {\EAS} did,
it %mainly
targets %the needs for analyzing
multiprocess ones running on distributed machines, and aims at method-level dependence abstraction
%lightweight predictive impact analysis
instead of
fine-grained slicing.
%of them. their evolution and maintenance.
%
% 3. slicing of distributed (multiprocess) systems
For systems running in multiple processes where interprocess communication is realized
via socket-based message passing,
%(e.g., the client/sever model using socket-based messaging),
an approximation for static slicing was discussed in~\cite{krinke2003context}.
Various dynamic slicing algorithms have been proposed too, earlier for procedural programs only ~\cite{korel1992dynamic,cheng1997dependence,goswami2000dynamic, duesterwald1993distributed,kamkar1995dynamic} and
recently for object-oriented software also~\cite{mohapatra2006distributed,barpanda2011dynamic,pani2012slicing,xqfu20tosem}.
%Specially for Java programs based on RMI (remote method invocation), a points-to analysis algorithm was developed
%to support slicing (and other static analysis) of RMI-based distributed applications~\cite{sharp2006static}.
And a more complete and detailed summary of slicing techniques for distributed programs can be found in~\cite{xu2005brief}
 and~\cite{barpanda2011dynamic}.
%Although these slicing algorithms were rarely evaluated against large real-world distributed systems, it can be anticipated
%that they would face scalability issues with large systems based on the limited
%empirical results they reported and the heavyweight nature of their technical design.

%In contrast to these fine-grained (statement-level) analysis,
%{\distea} aims at a highly efficient
%method-level dynamic impact analysis
%that can readily scale to large distributed systems.
A few other static analysis algorithms for distributed systems exist as well but focus on
 other (special) types of systems, such as RMI-based Java programs~\cite{sharp2006static},
 %and Android applications~\cite{octeau2013effective},
 different from the common type of distributed systems~\cite{Coulouris2011DSC} {\distea} addresses.
%
%In addition, a few static analysis and its supporting alogrithms (e.g., points-to analysis)
%have been developed for (particular types of) distributed software as well (e.g.,~\cite{sharp2006static,octeau2013effective}).
%
%{\distea} is a dynamic analysis focusing on distributed systems where inter-component communications are realized via socket-based network I/Os.
%
%Also, in general dynamic slicing can be too expensive for a method-level impact analysis~\cite{apiwattanapong05may,Law2003IDI}.
%
%However, like other such slicing techniques, these approaches build dependence graphs based on whole-program
%analysis and then perform slicing by simple graph-reachability analysis, which thus suffer from
%significant imprecision~\cite{giffhorn2009precise} and/or performance issues~\cite{goswami2000dynamic}
%\footnote{The technique creates a dynamic dependence graph that is unbounded as its size grows with the
%execution trace length~\cite{Agrawal1990DPS}}.
%
% 4. coarser-level dependence analysis of distributed systems in contrast to these prohibitive code-based dependence analyses,
At coarser levels, researchers resolve dependencies in % analysis of
distributed systems too but for different purposes such as %the purpose of
enhancing parallelization~\cite{psarris2004experimental}, system configuration~\cite{kon2000dependence},
%security analysis~\cite{octeau2013effective},
and high-level system %understanding% of system behaviour
modeling~\cite{abrahamson2014shedding,beschastnikh2014inferring}, or limited to static analysis~\cite{murphy1996lightweight,popescu2012impact,garcia2013identifying}.
%A static analysis, LSME%(lexical source model extraction)
%~\cite{murphy1996lightweight} extracts %also computes
%inter-component
%dependencies due to implicit invocations, but it is both %has been shown both
%imprecise and unsound~\cite{popescu2012impact,garcia2013identifying}.
%In~\cite{garcia2013identifying}, another static analysis is proposed to infer
%inter-component dependencies based on messaging-interface matching.
%Compared to these techniques,
In contrast,
{\distea} performs code-based analysis while providing more focused dependencies (impacts) relative to concrete program executions than
static-analysis approaches.

\subsection{Logging for Distributed Systems}
% 6. logging and/or log mining in distributed systems
%To facilitate
%For high-level comprehension and modeling of distributed systems,
%To facilitate
Targeting high-level understanding of distributed systems, %and executions,
techniques like logging and mining run-time logs~\cite{beschastnikh2014inferring,lou2010mining}
%has been explored recently. Such techniques
%usually utilize textual analysis and log message correlations to infer interdependencies among system components,
infer inter-component interactions using textual analysis of system logs,
relying on the availability of particular data such as informative logs and/or patterns in them.
{\distea} utilizes similar information (i.e., the Lamport timestamps) %transferred among processes)
but infers the happens-before relation between method-execution events mainly for code-level dependence analysis.
%The timestamps we utilized is related to the logs employed by those techniques
Also, {\distea} automatically generates such information it requires rather than
relying on existing information in the original programs. %under analysis
%(e.g., logging statements).

% 7. time-stamping distributed events
%In terms of partially ordering distributed events,
The Lamport timestamp used is related to vector clocks~\cite{fidge1988timestamps,mattern1989virtual} used by
other tools, such as ShiVector~\cite{abrahamson2014shedding} for ordering distributed logs and Poet~\cite{kunz1997poet}
for visualizing distributed systems executions.
While we could utilize vector clocks also, %in {\distea} as well,
we chose the Lamport timestamp as it is lighter-weight yet suffices for {\tech}.
%---vector clocks may be preferred for a more elaborate design (e.g., a more
% precise analysis leveraging message-passing semantics as discussed in Section~\ref{sec:discuss}). %for {\distea}. %in our case.
In addition, unlike ShiVector, which requires accesses to source code and recompilation using
the AspectJ compiler, {\distea} does not have such constraints as it works on bytecode. %as it directly works on Java bytecode.

Tracing message-passing systems was explored before but for different purposes, such as
overcoming non-determinacy/race detection~\cite{netzer1995optimal} and reproducing buggy
executions~\cite{konuru2000deterministic}.
Tools like RoadRunner~\cite{flanagan2010roadrunner} and
ThreadSanitizer~\cite{serebryany2009threadsanitizer} targeted multithreaded, centralized programs.
None of these solutions work
%can be readily incorporated/adopted for our analysis working
for large, diverse distributed systems without perturbations as {\tech} did.

%In contrast, inferring inter-component dependencies through key-pair correlation of text messages when mining
%distributed execution logs~\cite{lou2010mining}
%is applicable to a larger class of distributed system, yet such techniques tend to be opportunistic as they are subject to
%the availability of \emph{informative} logs.
%%
%Recently, a dedicated impact analysis called Helios has been presented~\cite{popescu2012impact} but specially only for
%distributed \emph{event-based} systems (DEBS)~\cite{muhl2006distributed}. In addition, like Helios and Eos,
%provide only coarse dependencies between components (e.g.,classes).

%\textbf{Dynamic Partial Order Reduction.}
%Introduced for multithreaded-program verification, dynamic partial order reduction
%(DPOR) has been used to reduce the size of state space to be searched by model checking
%algorithms.

\subsection{Dynamic Partial Order Reduction (DPOR)}
DPOR has been used to improve the performance of model checking concurrent software by avoiding
the examination of independent transitions
%, utilizing run-time program information to compute minimal persistent sets
~\cite{Flanagan2005DPR}.
Sharing the spirit of DPOR, especially distributed DPOR~\cite{yang2007distributed}, {\tech} synchronizes %only the first and last
method-execution events for efficient computation of partial ordering of methods {\em within} processes, and exploits message-passing semantics to avoid partial-ordering independent methods {\em across} processes.
However, unlike existing DPOR techniques which target multi-threaded programs, {\tech} focuses on multi-process, distributed programs.
On the other hand, DPOR may be adopted for dynamic dependence analysis of distributed programs with
extensions/adaptations. % and/or accommodations.
First, the execution conditions of methods can be modeled as method-level states (versus statement-level states in the original DPOR). Also, to represent state transitions across processes, the identifiers of parent processes need to be incorporated in thread identifiers. Finally, partial ordering algorithms like LTS may be employed to determine transition dependence at process level.

%% file: concl.tex
\vspace{-2pt}
\section{Conclusion}\label{sec:conclusion} %and Future Work}
\vspace{-2pt}
%Summary of what the paper presented and what the paper means for science and practice. Main things to do next (what this work enables that wasn't possible before) -- can be a bit speculative and can include things we don't necessarily plan to do, but others might want to do to expand this work.
%Components in distributed systems usually run concurrently in separated processes and
%communicate via socket-based message passing without explicitly invoking or
%referencing each other. In consequence,
%existing dynamic impact analysis, which relies on
%explicit invocations (dependencies), tends to be either entirely inapplicable or
%at best quite ineffective to use for those systems.

Modeling and reasoning about program dependencies has long been a fundamental approach to
many advanced techniques and tools for various software engineering tasks, ranging from testing and debugging to
refactoring and optimizations. However, traditional
dependence models and analysis algorithms, which assume explicit references among code entities,
cannot be readily applied to distributed systems with full
potential, because the architectural design of these systems encourages implicit references among decoupled components
via networking facilities such as socket.

To unleash the power of dependence analysis for distributed software, we presented {\distea},
a framework for dynamic dependence abstraction that safely approximates run-time code
dependencies at method level both within and across process boundaries.
{\tech} offers cost-effective dependence abstraction by partially ordering method-execution
events, exploiting message-passing semantics, incorporating static intra-component dependencies, and leveraging
whole-system statement coverage data.
Blending multiple forms of program information,
{\tech} offers flexible cost-effectiveness balances via four instantiations, to
accommodate varying time and other resource budgets in diverse task scenarios in distributed software development and maintenance.

We motivated and described the design of this framework in its application for dynamic impact
analysis, and evaluated its effectiveness and efficiency also in the context of impact prediction.
Our empirical results with large real-world distributed programs have shown
the superior effectiveness of advanced {\distea} versions over its basic, purely control-flow based
version as a baseline %existing options
by safely reducing false positives of the baseline %of the baseline
by 15\% to 54\% at the cost of varied but reasonable analysis overheads.
There are many applications of {\distea} beyond impact analysis. As immediately next steps, we plan to
apply the framework for performance diagnosis and security defense of distributed systems.